\documentclass[doublecol]{epl2} 

\usepackage{graphicx,color}
\graphicspath{{./}{./figures/}}
\usepackage{dcolumn}
\usepackage{bm}
\usepackage{amsmath}
\usepackage{amssymb}
\usepackage{hyperref}
\usepackage{commath}
\usepackage{makecell}
\usepackage{amsthm}
\usepackage{subfig}

\usepackage{color}

\hypersetup{
    colorlinks=true,       
    linkcolor=blue,          
    citecolor=magenta,        
    filecolor=red,      
    urlcolor=cyan           
}
 
\definecolor{labelkey}{cmyk}{.4,.2,0,0}

\newcommand{\be}{\begin{equation}}
\newcommand{\ee}{\end{equation}}
\newcommand{\bea}{\begin{eqnarray}}
\newcommand{\eea}{\end{eqnarray}}

\newcommand{\rmd}{\mathrm{d}}

\def\Xint#1{\mathchoice {\XXint\displaystyle\textstyle{#1}}%
{\XXint\textstyle\scriptstyle{#1}}%
{\XXint\scriptstyle \scriptscriptstyle{#1}}%
{\XXint\scriptscriptstyle \scriptscriptstyle{#1}}%
\!\int} \def\XXint#1#2#3{{ \setbox0=\hbox{$#1{#2#3}{\int}$} \vcenter{\hbox{$#2#3$}}\kern-.5\wd0}} 
\def\dashint{\Xint-}

\newcommand{\doidoi}[2]{\href{http://dx.doi.org/#1}{#2}}

\shorttitle{} 
\title{Linear statistics and pushed Coulomb gas at the edge \\ of $\beta$-random matrices: four paths to large deviations}

\author{Alexandre Krajenbrink and Pierre Le Doussal}
\shortauthor{A.~Krajenbrink and P.~Le~Doussal}

\institute{Laboratoire de Physique Th\'eorique de l'Ecole Normale Sup\'erieure,\\
PSL University, CNRS, Sorbonne Universit\'es, 
24 rue Lhomond, 75231 Paris, France.
}

\pacs{02.10.Yn}{Matrix theory}

\abstract{The Airy$_\beta$ point process, $a_i \equiv N^{2/3} (\lambda_i-2)$, describes
the eigenvalues $\lambda_i$ at the edge of the Gaussian $\beta$ ensembles of random matrices
for large matrix size $N \to \infty$.
 We study the probability distribution function (PDF) of linear statistics
${\sf L}= \sum_i t \varphi(t^{-2/3} a_i)$ for large parameter $t$. We show the large deviation forms 
$\mathbb{E}_{{\rm Airy},\beta}[\exp(-{\sf L})] \sim \exp(- t^2 \Sigma[\varphi])$ 
and $P({\sf L}) \sim \exp(- t^2 G(L/t^2))$ for the cumulant generating function
and the PDF.
We obtain the exact rate function $\Sigma[\varphi]$ using {\it four} apparently different
methods \textit{(i)} the electrostatics of a
Coulomb gas \textit{(ii)} a random Schr{\"o}dinger problem, i.e. the stochastic Airy operator
\textit{(iii)} a cumulant expansion \textit{(iv)} a non-local non-linear differential Painlev\'e type equation. Each method was independently introduced previously to obtain the lower tail of the Kardar-Parisi-Zhang equation \cite{sasorov2017large,JointLetter,krajenbrink2018systematic,tsai2018exact}. Here
we show their equivalence in a more general framework.
Our results are obtained for a class of functions $\varphi$, \textit{the monotonous
soft walls}, containing the monomials $\varphi(x)=(u+x)_+^\gamma$
and the exponential $\varphi(x)=e^{u+x}$ 
and equivalently 
describe the response of a Coulomb gas \textit{pushed at its edge}.
The small $u$ behavior of
the excess energy $\Sigma[\varphi]$ exhibits a change at $\gamma=3/2$ 
between a non-perturbative hard wall like regime for $\gamma<3/2$ (third order free-to-pushed transition)
and a perturbative deformation of the edge for $\gamma>3/2$ (higher order transition).
Applications are given, among them: \textit{(i)} truncated linear statistics such as $\sum_{i=1}^{N_1} a_i$, leading to a formula for the
PDF of the ground state energy of $N_1 \gg 1$ noninteracting fermions in a linear plus random
potential \textit{(ii)} $\sim (\beta-2)/r^2$ interacting spinless fermions in a trap 
at the edge of a Fermi gas \textit{(iii)} traces of large powers of random matrices.}

\begin{document}

\maketitle

Random matrix theory \cite{mehta,ForresterBook,eynard2015random,JohanssonReviewRMT} has an enormous range of current applications
e.g. for quantum chaos, transport  and entanglement \cite{beenakker1997random, sommers2007statistics, vivo2008distributions,nadal2010phase, texier2013wigner, AltlandBagrets2017,deLucaChalker2018,BertiniProsen2018}, Anderson localization \cite{mirlin1996transition},
string theory \cite{brezin,CotlerShenker2017}, data analysis \cite{bun2017cleaning}, fluctuating
interfaces and interacting Brownians \cite{nadal2009nonintersecting}, 
stochastic growth \cite{PrahoferSpohn2000,BorodinFerrari},
combinatorics such
as tilings, dimers and random permutations \cite{BDJ}, 
trapped fermions \cite{eisler_prl, marino_prl, marino_pre,castillo, 
CDM14,dean2016noninteracting,multicriticalfermions},
A classical problem, called linear statistics, amounts to study the probability distribution
function (PDF) of sums ${\sf L} = \sum_{i=1}^N f(\lambda_i)$
over eigenvalues $\lambda_i$ of a size $N$ random matrix.
Varying the function $f$ and the ensemble, it describes e.g.
conductance, shot noise, Renyi entropies, interfaces center of mass,
particule number fluctuations.
At large $N$, central limit theorems, universality, and connections to the Gaussian free field
were shown \cite{johansson2015gaussian, johanssonLinStatDysonBM,BWZ, 
BorodinFerrari,borodinclt,DumitriuPaquette,DimitriuPaquette2012,cunden2014universal,moments,krajenbrink2018systematic} for typical fluctuations of ${\sf L}$
in the bulk of the spectrum. Large deviations were also 
studied in the bulk \cite{dean2006large, dean02, majumdar2009index, majumdar2011many, AGZ}, from the 
Coulomb gas representation, and recently for truncated linear statistics \cite{grabsch2017truncated,grabsch2017truncated2},
showing interesting phase transitions.

In this Letter we study the edge of the spectrum where fluctuations are stronger and much fewer results exist \cite{basor1999determinants}. 
For the classical random matrix ensembles, an array of methods 
exists to study 
spectral correlations \cite{mehta,ForresterBook,eynard2015random,JohanssonReviewRMT}, 
such as the Coulomb gas, resolvant, 
orthogonal polynomials, Selberg integrals,
determinantal processes, Painlev\'e equations,
the Dimitriu-Edelman tridiagonal representation \cite{DumitriuEdelman2002} and the stochastic operators \cite{EdelmanSutton2007,RamirezRiderVirag2011,ViragReview}. These methods
however often appear 
disconnected: in this Letter we unveil
relations between some of them, valid at the edge. Strongly perturbed Coulomb gas are interesting correlated systems by themselves, extensively studied recently \cite{cunden2018universality}, but not much at their edge \cite{dhar2018extreme}.

We focus on the Gaussian $\beta$ ensemble of random matrices \cite{DumitriuEdelman2002} for which the joint probability distribution function (JPDF) of the eigenvalues $\lambda_i$, 
has the form
\be \label{PDF}
P[\lambda] \sim e^{ \beta \sum_{1 \leq i < j \leq N} \log |\lambda_i-\lambda_j|
- \frac{\beta N}{4} \sum_{i=1}^N \lambda_i^2} 
\ee
also identical for $\beta>1$ to the quantum JPDF for the ground state of $N$ spinless fermions in an harmonic trap with mutual $(\beta-2)/r^2$ interactions and to the stationary measure of the $\beta$ Dyson Brownian motion. Eq.~\eqref{PDF} leads to 
the celebrated semi-circle eigenvalue density of support $[-2,2]$.
The JPDF \eqref{PDF} can be seen as the Gibbs measure 
of a Coulomb gas (CG) with logarithmic repulsion between the eigenvalues,
which, at large $N$, can be described by a continuous density. We 
study here the eigenvalues located near the \textit{right} edge of this CG, 
in a window of width
$\sim N^{-2/3}$. In that window for $N \to \infty$,
the scaled eigenvalues $a_i \equiv N^{2/3} (\lambda_i-2)$ 
define the 
Airy$_\beta$ point process (APP). We consider the linear statistics of the APP, i.e. the sum 
\be
{\sf L}= t \sum_{i=1}^{+\infty} \phi(u+t^{-2/3} a_i)
\ee
where $u$ is a control parameter and $\phi$ the shape function.
We study a restricted set of functions $\phi$, denoted $\Omega_0$ and defined below, which is a subset of all continuous positive increasing functions such that $\phi(x)=0$ for $x \leq 0$.
The parameter $t$ thus controls how many eigenvalues contribute to the sum, indeed
since the ordered eigenvalues behave as 
$a_i \simeq_{i \to +\infty}- (3 \pi/2)^{2/3} i^{2/3}$,
only $K \simeq \frac{2 u^{3/2}}{3 \pi} t$ eigenvalues {\it typically} contribute to the sum. Hence for a function $\phi$ of order 1, the sum ${\sf L}$ is of order $t^2$ at large $t$. It is natural to define the scaled eigenvalue empirical density of the APP,
$\rho(b)=t^{-1} \sum_i \delta(b+ t^{-2/3} a_i)$, so that\footnote{For convenience the density is defined for the reversed APP $-a_i$.}
\be
{\sf L} = t^2 \int_{-\infty}^{+\infty} \rmd b \, \rho(b) \phi(u-b)
\ee
Since the mean density
of the APP 
converges for $a_i \to -\infty$ to the semi-circle, at large $t$ the mean value of ${\sf L}$
is 
\be \label{mean}
\mathbb{E}_{\beta}[ {\sf L}] \simeq t^2 \int_{-\infty}^{+\infty}  \rmd b \, \rho_{\rm Ai}(b) \phi(u-b) \quad , \quad
\rho_{\rm Ai}(b) := \frac{ \sqrt{(b)_+} }{\pi}
\ee
$\mathbb{E}_{\beta}$ denoting an average over the APP, and $(b)_+:= \max(b,0)$. We are interested in the large fluctuations of ${\sf L}$, and calculate the large deviation function $\Sigma_\phi(u)$ 
\be \label{defsig}
 \Sigma_\phi(u) := \lim_{t \to +\infty} t^{-2} \log Q_t(u) \quad , \quad 
Q_t(u) := \mathbb{E}_{\beta} [ e^{ - {\sf L} } ]
\ee
We show that the PDF of ${\sf L}$, $P({\sf L})$,  becomes  at large $t$
\be \label{proba}
P({\sf L}) \simeq e^{- t^2 G(\ell) }, \quad \ell = {\sf L}/\mathbb{E}_{\beta}[ {\sf L}]
\ee
and obtain its expression for $\ell \leq 1$ from a Legendre transform involving $\Sigma_\phi(u)$. 
We interpret $Q_t(u)$ as the Gibbs measure of the Coulomb gas 
upon a perturbation by a \textit{soft wall} external potential 
described by 
$\phi$. For $\phi$ in $\Omega_0$,
the external force $\phi' \geq 0$ pushes the charges towards
the bulk which defines the pushed CG problem, well studied in the
bulk. The novelty here is to study a Coulomb gas {\it pushed at its edge} and probe its rigidity.
The rate function $\Sigma_\phi(u)$ is then the excess total energy resulting from its
reorganization 
measured by the deviation of the equilibrium pushed density,
denoted $\rho_*(b)$, from the unperturbed one $\rho_{\rm Ai}(b)$. One outstanding 
question is the nature of the phase transition at $u=0^+$ between pushed and free, usually third order in the bulk \cite{majumdar2014top}. We find here transitions with continuously varying exponent larger than three. 

This problem was studied before for $\phi(x)= \phi_{\rm KPZ}(x)=(x)_+$ to obtain 
the lower tail of the Kardar-Parisi-Zhang equation
(Refs.~\cite{sasorov2017large,JointLetter,krajenbrink2018systematic,tsai2018exact,corwin2018lower })
 for $\beta=2$ (full-space KPZ), $\beta=1$ (half-space KPZ)
and arbitrary $\beta$ (extended polymer of Ref.~\cite{gorin2018kpz}).
No less than {\it four methods} were devised to treat that case: \textit{(i)} the electrostatics of a
Coulomb gas \textit{(ii)} a random Schr{\"o}dinger problem known as the stochastic Airy operator
\textit{(iii)} a cumulant expansion \textit{(iv)} a non-local non-linear differential Painlev\'e type equation
(the last two for $\beta=2$ only).
Although apparently unrelated, 
they 
lead to 
the same result for $\Sigma_\phi(u)$
for this KPZ problem.
The aim of this Letter is to unveil the connections between these methods,
make explicit the 
underlying structure and apply them
to 
more general functions $\phi$ beyond $\phi_{\rm KPZ}$. Two saddle point equations,
denoted {\sf SP1} and {\sf SP2}, shown to be dual to each other,
play an important role in the large $t$ limit, and appear alternatively in each
method as the genuine saddle point equation. 

{\it SAO/WKB method}. We start with the method based on the stochastic Airy
operator (SAO) \cite{EdelmanSutton2007},
introduced by Tsai for the KPZ problem in Ref.~\cite{tsai2018exact}.
It is known \cite{RamirezRiderVirag2011} that the APP can be generated as
$- a_i = \varepsilon_i$ where $\varepsilon_i$ are the eigenvalues of the 
following Schr{\"o}dinger problem on the half-line $y>0$, defined by the Hamiltonian
\be \label{HSAO} 
\mathcal{H}_{\rm SAO} = - \partial_y^2 + y + \frac{2}{\sqrt{\beta}} V(y) 
\ee
where $V(y)$ is a unit white noise $V(y) = B'(y)$ and the wave functions
vanish at $y=0$. Since we are interested
in energy levels of order $t^{2/3}$ we can rescale $y=t^{2/3} x$, $V(y)=t^{2/3} \frac{\sqrt{\beta}}{2} v(x)$,
$\mathcal{H}'_{\rm SAO} = t^{-2/3} \mathcal{H}_{\rm SAO}$ with energy levels 
$b_i = t^{-2/3} \varepsilon_i=- t^{-2/3} a_i$ and obtain
\be \label{Hprime} 
\mathcal{H}'_{\rm SAO}  = - t^{-2} \partial_x^2 + x + v(x) 
\ee 
This corresponds to a Schr{\"o}dinger problem for a particle of 
mass $m=1/2$ with $\hbar=1/t$. In the large $t$ limit it is natural 
to use the WKB semi-classical approximation for the density
of energy levels of \eqref{Hprime}, $\hat \rho(b)=\sum_i \delta(b-b_i)$
as $\hat \rho(b) \simeq t \rho(b)$ with \cite{SM}
\be \label{rho} 
\rho(b) = \frac{1}{\pi} \frac{\rmd}{\rmd b} \int_0^{+\infty} \rmd x\,  \sqrt{(b- x - v(x))_+}
\ee
The average over the APP in Eq.~\eqref{defsig} is
an average 
over the white noise
$V(y)$, 
of measure 
$ \sim e^{-\frac{1}{2} \int \mathrm{d}y \,  V(y)^2 }
=e^{-\frac{\beta}{8} t^2 \int \mathrm{d} x v(x)^2}$, hence we obtain that $\Sigma_\phi(u)$ is
the solution of the following variational problem for $x>0$
\be \label{sigmasao}
\Sigma_\phi(u)  = \min_{v(x)} \big[ \int_{-\infty}^{+\infty} \mathrm{d}b \, \rho(b) \phi(u-b) + \frac{\beta}{8} \int_0^{+\infty}   \mathrm{d}x\, v(x)^2 \big] 
\ee 
where $\rho(b)$ is defined in Eq.~\eqref{rho}.
This approach was made rigorous in the case $\phi_{KPZ}$ in Ref.~
\cite{tsai2018exact} using explosions in the Riccati formulation of the
SAO. 

We now display the resulting saddle point equation, which we denote {\sf SP1}, for a more general
$\phi$. From Eq.~\eqref{sigmasao} for $x>0$, the optimal $v(x)=v_*(x)$ is solution of
\be \label{saddle1} 
 \frac{\beta}{4} v_*(x) = \frac{1}{2 \pi} \int_{-\infty}^{+\infty} \frac{\mathrm{d}b}{\sqrt{(b)_+}} \phi'(u-b-x - v_*(x)) 
\ee
where everywhere $\frac{1}{\sqrt{(b)_+}}:=\frac{\theta(b)}{\sqrt{b}}$.
Choosing
$\phi$ in $\Omega_0$ we show that:
\textit{(i)} there
is a unique solution $v_*(x)$ to \eqref{saddle1} \textit{(ii}) $\phi$ is increasing 
which implies $v_*(x) \geq 0$, \textit{(iii)}
$\phi(x \leq 0)=0$ hence 
$v_*(x\geq  u)=0$,
\textit{(iv)} from simple manipulations, see Ref.~\cite{SM}, $\Sigma_\phi(u)$ can be rewritten simply as
\be \label{SigmaSAO} 
\Sigma_\phi(u) = \frac{\beta}{4} \int_0^{+\infty} \rmd x\,  x v_*(x) 
\ee

{\it Cumulant method}. Another method was recently developed for $\beta=2$
when the APP has a determinantal structure. A systematic time expansion, i.e. in $t$, on the Fredholm determinant representation, Eq.~\eqref{FD}, of the average in Eq.~\eqref{defsig}, led to the following series formula, shown in Ref.~\cite{krajenbrink2018systematic} for $\beta=2$, and conjectured here for any $\beta$
\be \label{series}
\Sigma_\phi(u) = - \sum_{n \geq 1} \frac{\tilde \kappa_n(u)}{n!} ~ , ~
\tilde \kappa_n(u) =  \frac{\beta}{4}  (-\frac{4}{\beta \pi})^n
\partial_u^{n-3} f(u)^n 
\ee
It yields the cumulants of ${\sf L}$, i.e.
$\mathbb{E}_\beta[{\sf L}^n]^c~=~t^2 (-1)^n \tilde \kappa_n(u)$,
with $\tilde \kappa_2(u)=\frac{4}{\beta\pi^2} \int_0^u \rmd u' f(u')^2$ and the mean value was given in Eq.~\eqref{mean}.
%
%
We have defined
\be \label{deff} 
f(u) := \frac{1}{2} 
\int_{-\infty}^{+\infty} \frac{\rmd b}{\sqrt{(b)_+} } \phi'(u-b) 
\ee
which, for $\phi$ in $\Omega_0$, is positive with $f(u\leq 0)=0$.
It is possible to sum up the series of cumulants of Eq.~\eqref{series}. We show in Ref.~\cite{SM} that if the following equation
\be \label{equation}  
 f(u - \frac{4}{\beta \pi} w(u))= w(u)
\ee
admits a unique positive increasing solution $w(u)$ for all $u\geq0$, with $w(u\leq 0)=0$, then 
\bea \label{m22} 
\Sigma_\phi(u) = \frac{1}{\pi} \int_0^u \rmd u'  (u-u') w(u')
\eea 
The uniqueness of the solution of \eqref{equation} is equivalent to
$z\mapsto h(z)= f(z) + \frac{4}{\beta \pi} z$ being strictly increasing, in which case $w(u)=\frac{\beta \pi}{4} (u- h^{-1}(u))$.
We call $\Omega_2$ the set of $\phi$
such that the associated $f$ satisfies this condition and further restrict to the subset
$\Omega_0 \subset \Omega_2$ 
such that
$f$ 
is positive, increasing, with $f(z \leq 0)=0$, leading to \eqref{equation}.
From Eq.~\eqref{m22}
we see that $\Sigma_\phi(u)$ is positive (as required) and  since
$\Sigma_\phi''(u)=\frac{1}{\pi}w(u)$ is also positive, it is also convex (as required). This method 
gives a simpler parametric representation than the SAO/WKB method and one can ask about the connection between the two.

Comparing \eqref{saddle1} and \eqref{equation} we note
that we can find the solution of the SAO/WKB {\sf SP1} saddle point equation
as 
\be \label{vw} 
v_*(x) = \frac{4}{\beta \pi} w(u-x) \quad 0 \leq x \leq u
\ee
with $v_*(x)=0$ for $x \geq u$. Hence Eqs.~\eqref{saddle1} and \eqref{equation} 
are the same {\sf SP1} equation. Using \eqref{vw} we see that
formulae \eqref{SigmaSAO} and \eqref{m22} for $\Sigma_\phi(u)$,
also coincide. 
Conversely, going from the SAO/WKB to the cumulant expansion
amounts to use the Lagrange inversion formula on Eq.~\eqref{saddle1},
or equivalently on Eq.~\eqref{equation} (see Ref.~\cite{SM}). 
This coincidence for any $\beta$ confirms our conjecture 
for the $\beta$ dependence of the series. This concludes the equivalence
between the two methods for $\phi$ in the class $\Omega_0$.

It is useful to invert the (convolution) relation \eqref{deff} between 
$\phi$ and $f$. It is 
invertible as a convolution\cite{SM}
\bea \label{inversion} 
\phi(u) = \frac{2}{\pi}  \int_{-\infty}^{+\infty} \frac{{\rmd}b}{\sqrt{(b)_+}} f(u - b) 
\eea 

{\it Painlev\'e/WKB method}. This method, introduced
in the context of large deviations in \cite{LargeDevUs}, (based on \cite{ACQ11}), was developed in 
\cite{sasorov2017large} for $\phi=\phi_{\rm KPZ}$. 
For $\beta=2$ the $\lbrace a_i \rbrace$ form the usual Airy point process
and one rewrites \eqref{defsig} for all $t$ as a Fredholm determinant (FD)
\bea \label{FD} 
Q_t(u) = {\rm Det}[ I - \sigma_t K_{{\rm Ai}}] 
\eea 
where $\sigma_t(a)=1- e^{- t  \phi(u+t^{-2/3} a)}$ and 
$K_{{\rm Ai}}(a,a')$ is the standard Airy kernel, see Ref.~\cite{SM}.
This FD is shown \cite{ACQ11} to obey the 
following equation, with $s=-ut^{2/3}$ 
\bea
&& \log Q_t(u) = \int_{s}^{+\infty} \rmd r (s-r) \Psi_t(r) \\
&& \Psi_t(r)  = - \int_{-\infty}^{+\infty} \rmd v (q_t(r,v))^2 \frac{\rmd}{\rmd v} e^{- t \phi(v t^{-2/3})} \label{Psi} \\
&& \partial_r^2 q_t(r,v) = [v+r+ 2 \Psi_t(r)] q_t(r,v)  \label{3} 
\eea 
with $q_t(r,v) \simeq_{r \to +\infty} {\rm Ai}(r+v)$. Following Ref. \cite{sasorov2017large}
and introducing the scaled variables $r = t^{2/3} X$ and
$v = t^{2/3} V$ one shows (see Ref.~\cite{SM}) that at large $t$, Eq. \eqref{3} can be solved by the WKB
method in the form
$\Psi_t(r) \simeq t^{2/3} g(X)$, under the condition that 
$g(X)$ satisfies {\it simultaneously} a pair of equations: while Eq.~\eqref{eq1} is shown to identify to
{\sf SP1}, the second one Eq.~\eqref{eq2} is new and denoted {\sf SP2}
\bea
\hspace*{-1.4cm} && \frac{\beta}{2} g(X) = \frac{1}{2 \pi} \int_{-\infty}^{+\infty} \frac{\rmd V \phi'(V)}{\sqrt{ (- V - X - 2 g(X))_+}}  \label{eq1} \\
\hspace*{-1.4cm}&& \phi(V) = \beta \int_{-\infty}^0 \rmd X' \sqrt{(V+ X' + 2 g(X'))_+} - \frac{2 \beta}{3} (V)_+^{\frac{3}{2}}  
\label{eq2} 
\eea 
These generalize Eqs.~(28) and (29) of Ref.~\cite{sasorov2017large}, where the compatibility
of these two equations was qualified as a {\it miracle}. 
We have extended the above pair of equations
to any $\beta$, by consistency with the other methods. In addition
\be \label{sigpainleve}
\Sigma_\phi(u) = \frac{\beta}{2} \int_{-u}^0 \rmd X (X+u) g(X) 
\ee 
We now unveil the connection to the other methods, and explain the
miracle. First, we see that Eq.~\eqref{eq1} reduces to our previous saddle point 
equation {\sf SP1} (in the equivalent forms of Eqs.~\eqref{saddle1} and  \eqref{deff}-\eqref{equation}) 
upon the identification 
\be \label{24} 
u \equiv - X \quad , \quad 
w(u) \equiv \frac{\beta}{2} \pi g(-u)   
\ee
From it, 
we see that formula \eqref{sigpainleve} for $\Sigma_\phi(u)$ 
becomes equal to the one obtained with the other methods, e.g.
Eq.~\eqref{m22}.

For full consistency we now show that, within the class of $\phi$ studied here,
Eqs. \eqref{eq1} and \eqref{eq2} are equivalent, proving that {\sf SP1} and
{\sf SP2} are dual forms of the same condition. Let us show that {\sf SP1} 
implies {\sf SP2}. Denoting $\mathcal{I}$ the r.h.s. of 
Eq.~\eqref{eq2}, using Eq.~\eqref{24} we can rewrite it as
\bea \label{I}
\mathcal{I} = \beta \int_0^{+\infty} \rmd u' \sqrt{(V - u' + \frac{4}{\beta \pi} w(u'))_+} - \frac{2 \beta}{3} (V)_+^{3/2} 
\eea 
We use the change of variable $z = u' - \frac{4}{\beta \pi} w(u')$, $f(z) = w(u')$. If $f(u)$ is positive and increasing then $z$ is an increasing function of $u'$. In addition 
since $u'= f^{-1}(w(u')) + \frac{4}{\beta \pi} w(u')$, we also have $u'=z + \frac{4}{\beta \pi} f(z)$ hence
\bea
&& \mathcal{I} = \beta \int_0^{+\infty} \rmd z (1+ \frac{4}{\beta \pi} f'(z)) \sqrt{(V-z)_+}  - \frac{2 \beta}{3} (V)_+^{3/2} \nonumber  \\
&& 
 = \frac{2}{\pi} \int_{-\infty}^{+\infty} \rmd z \frac{f(z)}{\sqrt{(V-z)_+}} = \phi(V)
\eea
In the last equality we used the inversion formula \eqref{inversion} and the miracle is explained. It is also simple to show the converse,
i.e. {\sf SP2} implies {\sf SP1}, see Ref.~\cite{SM}. Hence, the Painlev\'e/WKB method is equivalent
to the two others. 

{\it Electrostatic Coulomb gas method}. In Ref.~\cite{JointLetter} the edge limit of the standard Coulomb gas describing the bulk eigenvalues of the GUE was taken, 
and applied to study the large deviations for ${\phi=}\phi_{\rm KPZ}$. 
For general $\phi$, the function $\Sigma_\phi(u)$ is given \cite{footnote0} by the minimization problem
\bea
\hspace*{-1cm} &&\Sigma_\phi(u) = \min_{\rho} \big[ \int_{-\infty}^{+\infty} \rmd b\,  \rho(b) \phi(u-b)+ \mathcal{J}(\rho) + U(\rho)\big] \label{CG1} \\
\hspace*{-1cm}  && \mathcal{J}(\rho) = - \frac{\beta}{2} \iint_{-\infty}^{+\infty} \log|b_1-b_2| \prod_{i=1}^2 \mathrm{d}b_i (\rho(b_i)-\rho_\text{Ai}(b_i)) \nonumber 
\eea
where $U(\rho) = \frac{2 \beta}{3\pi}\int_{-\infty}^0 \mathrm{d}b\, |b|^{\frac32} \rho(b)$ 
is irrelevant below.
The minimum is over mass conserving measures $\rho(b)$
such that $\int_\mathbb{R}  \rmd b (\rho(b)-\rho_\text{Ai}(b))=0$,
where $ \rho_\text{Ai}(b) = \frac{1}{\pi}\sqrt{(b)_+}$. The variational
equation determines the optimal density $\rho^*(b)$ as the unique solution such that
\bea \label{28}
\phi(u-b) - \beta \int_{-\infty}^{+\infty} \rmd b' \log|b-b'| (\rho_*(b')-\rho_{\rm Ai}(b'))  \geq c 
\eea 
with equality on the support of $\rho^*$. We assume, and verify later, that 
for $\phi \in \Omega_0$,
the support is an interval $[u_0>0,+\infty[$. 
Taking a derivative, we have for $b \in [u_0,+\infty[$
\bea \label{29}
- \phi'(u-b) - \beta \dashint_{-\infty}^{+\infty}  \frac{\rmd b' }{b-b'} (\rho_*(b')-\rho_{\rm Ai}(b')) = 0 
\eea 
In Ref.~\cite{JointLetter} $\rho_*(b)$ and $\Sigma_\phi(u)$ were calculated
for ${\phi=}\phi_{\rm KPZ}$ and here we 
obtain these quantities for a
general $\phi$ in $\Omega_0$.

We now unveil the connection between the Coulomb gas and the other methods.
First note that Eq.~\eqref{rho} provides a parametrization of the
density $\rho(b)$ in terms of a function $v(x)$ (
at this stage
\textit{arbitrary}, i.e. not necessarily solution of a saddle point). 
This parametrization has some remarkable properties. The first
is that 
we can exactly identify the electrostatic energy 
of the Coulomb gas, ${\cal J}(\rho)$, with the Brownian
weight function appearing in the SAO/WKB method, i.e. the second term in Eq.~\eqref{sigmasao}, as 
\be \label{J} 
{\cal J}(\rho) = \frac{\beta}{8} \int_0^{+\infty} \rmd x \, v(x)^2 
\ee
This is shown in Ref.~\cite{SM} under the condition that $x+v(x)$ is an increasing function for $x \geq 0$. This condition is in particular realized at the saddle point {\sf SP1} for
$\phi$ in  $\Omega_0$. 

Consider now the solution $v^*(x)$ of the saddle point {\sf SP1} of the
SAO/WKB method, and define $\rho_1(b)$ its associated
density under the parametrization Eq. \eqref{rho}. We show in Ref.~\cite{SM} that
$\rho_1(b)=\rho_*(b)$, i.e. the unique minimizing density for the Coulomb gas. Indeed, the Hilbert
transform 
can be explicitly calculated with this parametrization and the variational
condition Eq.~\eqref{29} becomes, for $b > u_0$ 
\be \label{30}
\phi'(u-b) =  \frac{\beta}{2} \int_0^{+\infty} \!\!\!\! \rmd x [  \frac{1}{\sqrt{(-b + x + v_*(x))_+}} -  \frac{1}{\sqrt{(-b + x)_+}} ] 
\ee
Given that $v_*(x)$ is solution to the saddle point {\sf SP1}, we notice that 
Eq.~\eqref{30} is exactly the derivative of Eq.~\eqref{eq2} (equivalently
Eq.~\eqref{I}) upon the identification of the {\sf SP2} equation in terms of 
$v_*$ (Eqs.~\eqref{vw} and \eqref{24}). 
This identification can be
extended to Eq.~\eqref{28}, see Ref.~\cite{SM}. Hence we find that the Coulomb gas saddle point equation matches
the saddle point equations of the other methods. Furthermore, 
since the first term in Eqs.~\eqref{sigmasao} and \eqref{CG1} are also the same,
we obtain by inserting $v(x)=v^*(x)$ into the Coulomb energy \eqref{J} 
that $\Sigma_\phi(u)$ given
by the Coulomb gas method coincide with the one of the SAO/WKB method.

Finally, two equivalent explicit formulae for the optimal density $\rho^*$ 
for both the SAO/WKB and 
Coulomb gas methods are obtained
using the saddle point {\sf SP1} and read
\bea
&& \rho^*(b) = \frac{1}{\pi} \sqrt{(b-u_0)_+} + \delta \rho(b) \label{trico}\\
&& \delta \rho(b) = \frac{2}{\beta \pi^2} \int_0^{w(u)} \frac{\rmd w'}{\sqrt{(b-u+f^{-1}(w))_+}} \nonumber \\
&& \delta \rho(b) = \frac{1}{\beta \pi^2} 
\dashint_{-\infty}^{\infty} \rmd b' \frac{\phi'(b')}{b+b'-u} \frac{\sqrt{(b-u_0)_+}}{\sqrt{(u-u_0-b')_+}}  \nonumber 
\eea
where the lower edge of the support is $u_0=\frac{4}{\beta \pi} w(u)$ and the first expression does not involve principal parts\footnote{Although we used a different route, the second is akin to a 
Tricomi inversion formula.}.
Remarkably, the sole knowledge of the edge $u_0$ as
a function of $u$, obtained solving Eq.~\eqref{equation}, determines completely the energy $\Sigma_\phi(u)$,
indeed $\Sigma_\phi''(u) = \frac{1}{\pi} w(u)=\frac{\beta}{4} u_0$, and
the effective restoring force, i.e. the pressure of the gas is (see Ref.~\cite{SM})
$\Sigma'_\phi(u) = \frac{\beta}{4} \int_0^{+\infty} \rmd x\, v(x) = \frac{1}{\pi} \int_0^u \rmd u' w(u')$.

{\it Calculation of the PDF $P({\sf L})$.}  
Requiring that
$\mathbb{E}_{\beta} [ e^{ - B {\sf L} } ] = \int \rmd {\sf L} P({\sf L}) e^{ - B {\sf L} }
\sim e^{- t^2 \Sigma_{B \phi}(u)}$ and inserting 
Eq.~\eqref{proba},
yields, 
upon Legendre inversion,
\be \label{Gl} 
G(\ell) = \max_{B} [ \Sigma_{B \phi}(u) - A B \ell] 
\ee 
with $A=\mathbb{E}_{\beta}[ {\sf L}]/t^2=\partial_B\Sigma_{B \phi}(u)|_{B=0^+}$
given by Eq.~\eqref{mean}. 
We are able to probe only the \textit{pushed} Coulomb gas, i.e. $B \geq0$ and $\ell \leq 1$. The side
$\ell >1$ corresponds to $B<0$, a \textit{pulled} Coulomb gas in which case $B \phi$ does not belong to $\Omega_0$. The phenomenon found in \cite{grabsch2017truncated} {\it in the bulk} that for $\ell>1$ 
the support of the optimal density splits, leading to a distinct phase, is likely to carry to the edge.


We now apply these methods to calculate $\Sigma_\phi(u)$ (the excess energy),
$\rho_*(b)$ (the equilibrium density) and $G(\ell)$ (the PDF) for some examples.

{\it Monomial soft walls}. For $\phi(x) = (x)_+^\gamma$, see Fig.~\ref{gamma15} for instance, the associated $f(u) = C_\gamma (u)_+^{\gamma-\frac{1}{2}}$ with 
$C_\gamma = \frac{\sqrt{\pi}}{2} \frac{\Gamma(\gamma+1)}{\Gamma(\gamma+ \frac{1}{2})}$.
Hence $\phi$ is in $\Omega_0$ iff $\gamma>1/2$, to which we restrict. The energy
is a simple polynomial in $u,w$
\be 
\Sigma_\phi(u)= a_\gamma u^2 w + b_\gamma u w^2 + c_\gamma w^3
\ee
with $w$ is the unique positive solution of the trinomial equation
$u= \frac{4}{\beta \pi} w + (\frac{w}{C_\gamma})^{\frac{1}{\gamma-1/2}}$, and 
$a_\gamma=\frac{4 }{\pi  (2 \gamma +1) (2 \gamma +3)}$
$b_\gamma=\frac{(2 \gamma -3) (6 \gamma +1)}{4 \pi  \beta  \gamma } a_\gamma$
$c_\gamma= \frac{(2 \gamma -3)^2 (2 \gamma +1)}{3 \pi ^2 \beta ^2 \gamma } a_\gamma$.
More explicit forms exist for some values of $\gamma$, see Ref.~\cite{SM}. 
To comment this result it is 
useful to compare with the 
infinite hard wall, $\lim_{B \to +\infty} \Sigma_{B \phi}(u)=\frac{\beta}{24} u^3$,
a standard result related to the cubic tail of the Tracy-Widom distribution
(see Refs.~\cite{dean2006large,perret2015density, perret2014near, KrajLedou2018})
which is an upper bound for $\Sigma_\phi$ (see Ref.~\cite{SM}). Combined with the first cumulant (Jensen) bound, $-\tilde \kappa_1(u)$, we obtain for all $u$
\begin{equation} \label{bound} 
\Sigma_\phi(u)\leq \min\big(\frac{\beta}{24}u^3 , \frac{\Gamma(\gamma+1)}{\sqrt{4\pi}\Gamma(\frac{5}{2}+\gamma)}u^{\gamma+\frac{3}{2}} \big)
\end{equation}
It turns out that this inequality is saturated at small and large $u$ for all $\gamma \neq 3/2$, i.e. it gives the exact asymptotics
(prefactor included) in both limits. Comparing the exponents $3$ and $\gamma+3/2$, we see that $\Sigma_\phi(u)$ is cubic (and $\phi$ acts as a hard wall breaking the edge) for small $u$ for $\gamma<3/2$, and
for large $u$ for $\gamma>3/2$. 
The other limiting behavior, $u^{\gamma+\frac{3}{2}}$, given by the first cumulant bound,
appears as a weak, perturbative, response of the edge to the potential $\phi$. 
This change at $\gamma=3/2$ is also seen on the density.

\begin{figure}[ht!]
  \centering
  \subfloat[$u=25$]{\label{fig:edge-a}\includegraphics[scale=0.35]{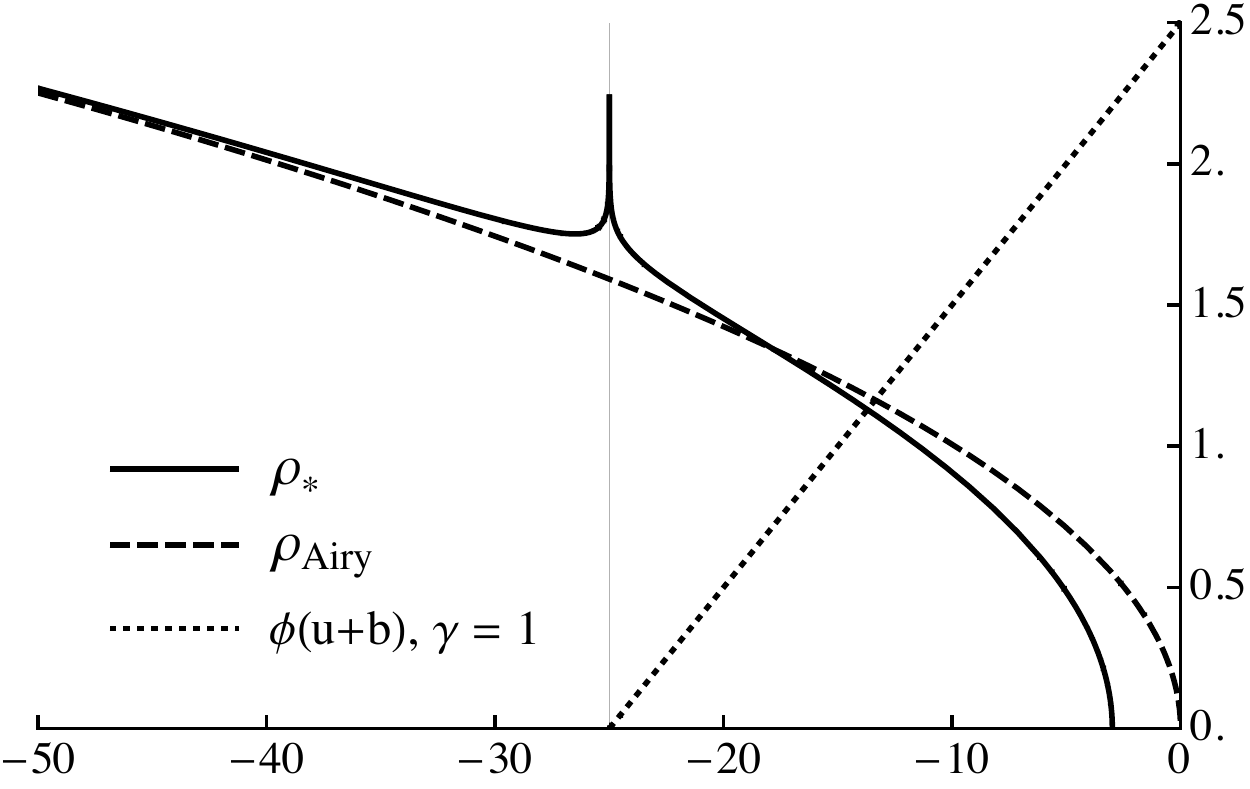}}
  \hspace{-5pt}
  \subfloat[$u=0.05$]{\label{fig:contour-b}\includegraphics[scale=0.35]{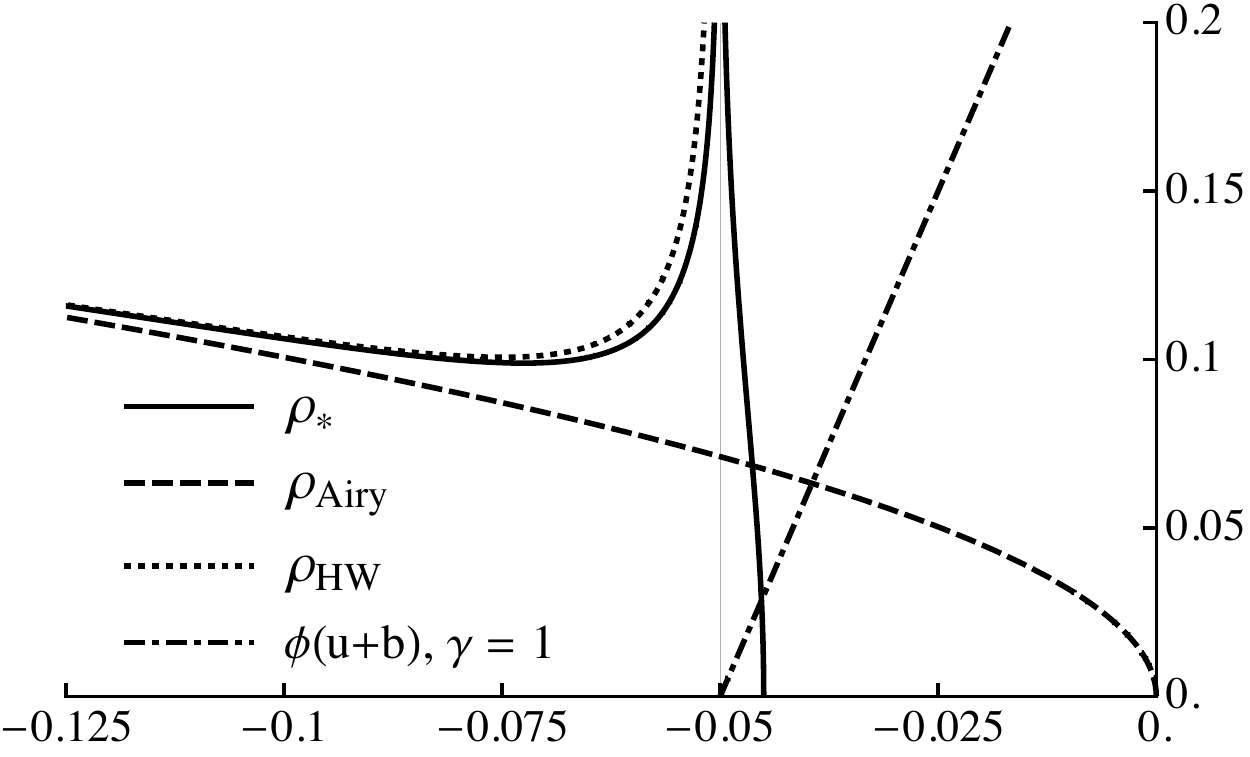}}
  \caption{Optimal density $\rho_*(-b)$ for $\beta=2$ and the linear wall $\gamma=1$ (solid line), compared to
  the semi-circle density $\rho_{\rm Ai}(-b)$
  (dashed line), the potential $\phi(u+b)$, 
  and to the infinite hard-wall $\rho_{HW}(-b)$.}
  \label{fig:gamma1}
\end{figure}

\begin{figure}[ht!]
  \centering
  \subfloat[$u=100$]{\label{fig:edge-a}\includegraphics[scale=0.35]{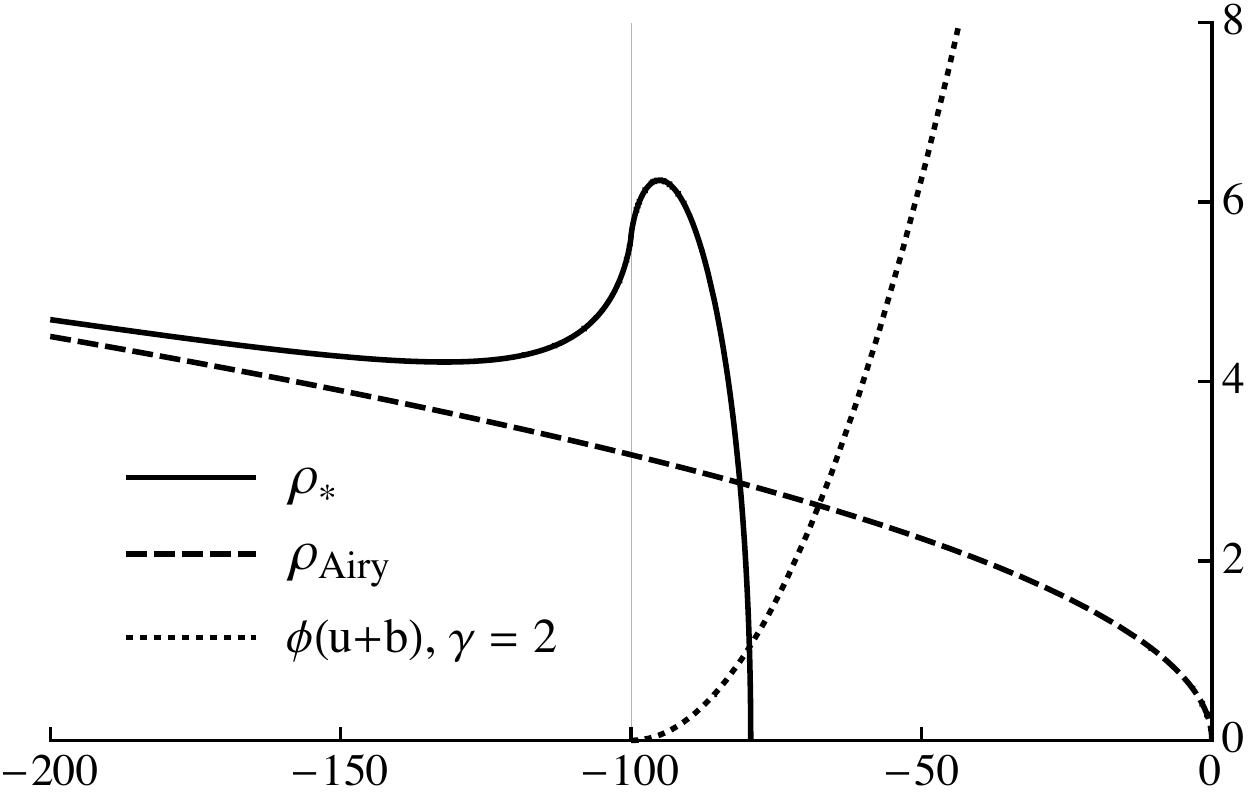}}
  \hspace{-5pt}
  \subfloat[$u=1$]{\label{fig:contour-b}\includegraphics[scale=0.35]{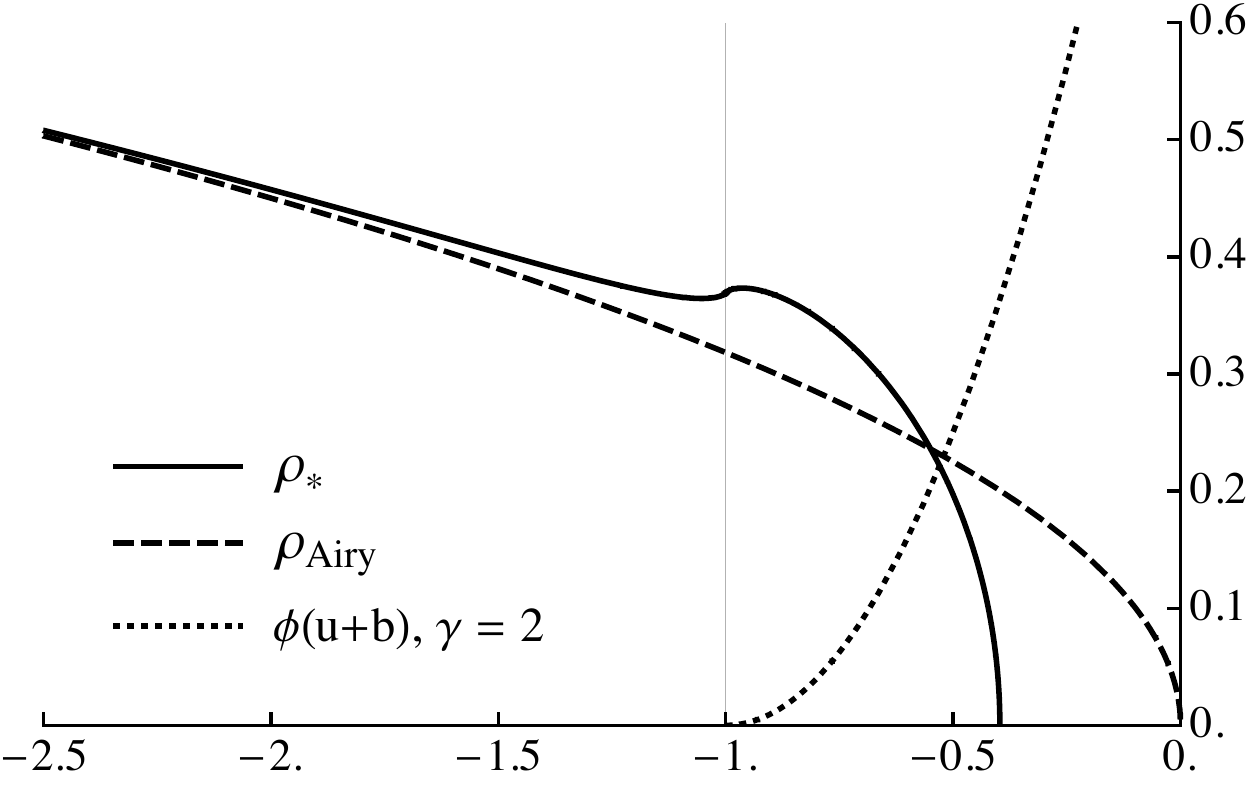}}
  \caption{Same as Fig. \ref{fig:gamma1} for the quadratic wall $\gamma=2$.}
  \label{fig:gamma2}
\end{figure}

\begin{figure}[h!]
\centerline{\includegraphics[scale=0.6]{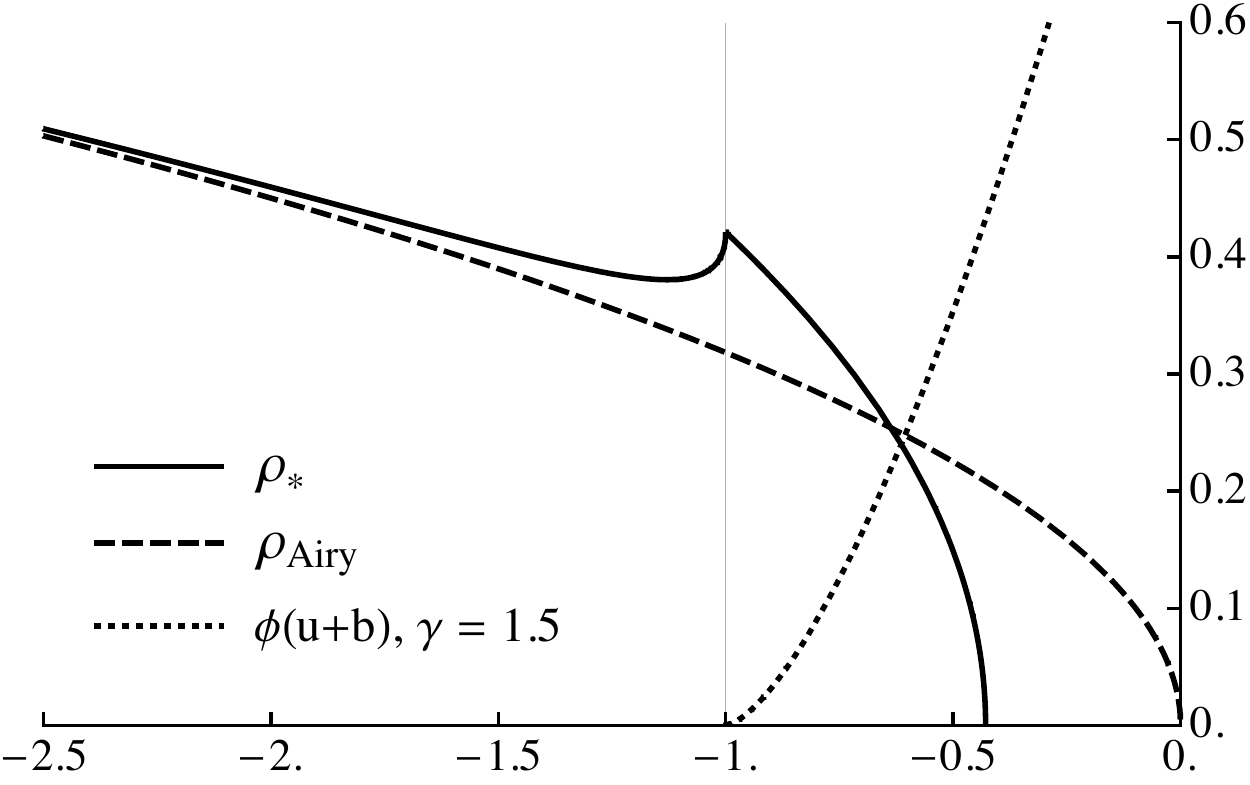}}
\caption{Same as Fig. \ref{fig:gamma1} for the critical case $\gamma=3/2$, $B=1$ and 
$u=1$.}
\label{gamma15}
\end{figure}

The optimal density $\rho_*(b)$ is obtained as a hypergeometric function $_2F_1$
for any $\gamma$ \cite{SM}.
It is smooth except at \textit{(i)} $b=u$, with singularity $|b - u|^{\gamma-1}$ for non-integer $\gamma$, and $
(b-u)^{\gamma-1} \log|b-u|$ for integer $\gamma$ \textit{(ii)} at
$b=u_0$, the lower edge, 
always of semi-circle type. 
It is plotted 
in Fig. \ref{fig:gamma1} for $\gamma=1$ (linear wall) and in
Fig. \ref{fig:gamma2} for $\gamma=2$ (quadratic wall) for respectively large and 
small $u$. We see that for $\gamma=1$ the rearrangement 
of the CG is weak for large $u$, consistent with the (first cumulant) perturbative result 
$\Sigma_\phi(u) \sim u^{5/2}$. 
For small $u$ the rearrangement is strong and the density converges 
to the known infinite hard wall 
optimal density $\rho_{HW}(b) = \frac{2 b - u}{2 \pi \sqrt{(b-u)_+}}$, see Ref.~\cite{KrajLedou2018},
plotted in 
Fig.~\ref{fig:gamma1}b) for comparison, 
consistent with the cubic $\Sigma_\phi(u) \simeq \frac{\beta}{24} u^3$ behavior. 
that the same holds in Fig. \ref{fig:gamma2} for $\gamma=2$ but with small
and large $u$ behaviors exchanged.

\vspace*{-0.8cm}


%

To explore the critical case $\gamma=3/2$, we study 
$B \phi(x) = B (x)_+^\gamma$. The saddle point 
equation {\sf SP1} then admits the simple solution $v_*(x)=\frac{3B}{3B+2\beta}(u-x)_+$ leading to
\be
\Sigma_{B \phi}(u) =\frac{\beta}{24}\frac{3B}{3B+2\beta} u^3
\ee
remarkably, a simple cubic for any $u$. The prefactor exhibits a smooth crossover between the 
perturbative 
($B$ small) and the hard wall regime ($B$ large). The optimal density 
has a remarkably simple form (plotted in Fig.~\ref{gamma15})
\be
\rho_*(b)= \frac{1}{\pi}\sqrt{(b-u_0)_+} + 
 \frac{3 B}{2 \beta \pi} ( \sqrt{(b-u_0)_+} - \sqrt{(b-u)_+} ) 
\ee
with $u_0=\frac{3B}{3B+2\beta} u$ which recovers $\rho_{HW}(b)$ for $B \to +\infty$. Finally, from Eq.~\eqref{Gl},
we find that the PDF for all monomial
walls, i.e for all $\gamma>1/2$, takes the form
\be
P({\sf L}) \sim e^{- t^2 \frac{\beta}{24} u^3 q_\gamma(\ell) } , \quad \ell = {\sf L}/\mathbb{E}_{\beta}[ {\sf L}]
\ee
where $q_\gamma(\ell)$ is independent of $u$ and $\beta$, see Ref.\cite{SM},
and with $q_{3/2}(\ell) = (1 - \sqrt{\ell})^2$ for $\ell \leq 1$. Note that 
$q_\gamma(0)=1$ 
is always true
since it corresponds to all $a_i<-u$, i.e. an infinite hard-wall.


In summary 
for the monomial walls 
the pushed-free phase transition at $u=0$ in the excess energy is third order for $\gamma \leq 3/2$ and order $\gamma+3/2$ 
for $\gamma \geq 3/2$. This change of behavior indicates a critical rigidity for the edge of the Coulomb gas 
in its sensitivity to a perturbation.
%

{\it Exponential wall}. An
interesting case is $\phi(x)=e^{x}$,
for which $f$ is also exponential $f(u) =\frac{\sqrt{\pi}}{2} e^{u}$. Since $f(u\leq0)>0$
it belongs to a larger class {$\Omega_1 \supset \Omega_0$} for which uniqueness holds 
and the above formulae still hold with minor modifications \cite{SM}.
Denoting $W=W_0(\frac{2}{\beta \sqrt{\pi}} e^{u})$ 
the standard branch of the Lambert function\cite{corless1996lambertw}, i.e. the solution $W(x)$ of $W e^{W}=x$,
we find $w(u)= \frac{\beta \pi}{4} W$,
and using Eq.~\eqref{m22} with the lower bound at $u=-\infty$ we obtain
\be
 \Sigma_\phi(u) =  \frac{\beta}{48} (2 W^3+9 W^2+12 W)
\ee
The limiting behaviors of the energy are $\Sigma_\phi(u) \simeq  \frac{\beta}{24} u^3$ for large positive $u$
and $\Sigma_\phi(u) \simeq  \frac{1}{2 \sqrt{\pi} } e^{u}$ for large negative $u$ and the
optimal density is, see Ref.~\cite{SM}
\be \rho_*(b)  = \frac{1}{ \pi}  \sqrt{(b - u_0 )_+  }
+  \frac{u_0}{2 \sqrt{\pi}}   e^{u_0-  b} \text{Erfi}\left(\sqrt{b
   -u_0}\right)
\ee
The probability is given by Eq.~\eqref{proba} with
\be  G(\ell) = - \frac{\beta}{48} (2 + \tilde W)^2 (1+ 2 \tilde W)
~ , ~ \tilde W=W_{-1}(-2 \ell e^{-2})
\ee
in terms of the second real branch of the Lambert function.

{\it Inverse monomial walls,} of the form 
$\phi(x)=(- x)^{-\delta}$
for $x<0$, $\phi(x>0)=+\infty$, for $\delta>3/2$ are
another example in $\Omega_1$, 
which penetrate strongly, as power laws, into the Coulomb gas. 
Explicit results are displayed in Ref.\cite{SM}.

An important set of applications concern {\it truncated linear statistics}. 
A sum over the $N_1$ largest eigenvalues of the Laguerre ensemble (LE),
${\cal L}=\sum_{i=1}^{N_1} f(\lambda_i)$ were
studied by CG methods at large $N$ in \cite{grabsch2017truncated}
in the bulk, i.e. for $\kappa=N_1/N$ fixed. For $f(\lambda)=\sqrt{\lambda}$ it was shown that the PDF of
the scaled variable $s=N^{-3/2} {\cal L}$ takes the form $\exp(- N^2 \Phi_\kappa(s))$. We 
have shown \cite{SM} that the $\kappa \to 0$ limit of these results match
our edge results {\it for the linear wall $\gamma=1$}.
Using universality at the soft edge (of the LE), both can be related
to truncated linear statistics of the APP of the type
\be \label{sum1} 
L_{N_1} = - \sum_{i=1}^{N_1} a_i = \sum_{i=1}^{N_1} \varepsilon_i
\ee 
Since $N_1$ is fixed, $u$ must be determined self-consistently,
$u=u_*(N_1/t)$ by the condition that in the optimal density 
$\rho_{*,u_*}(b)$ there are $N_1$ eigenvalues below level $u_*$. 
This leads to the PDF, for $1 \ll N_1 \ll N$, 
\be \label{probFermions} 
P(L_{N_1}=L) \sim \exp\left(- \frac{\beta N_1^2 }{2} 
\frac{2 \pi^2}{3}  \bar \Phi(a \frac{\mathbb{E}_\beta[L]-L}{N_1^{5/3}})\right)
\ee
with $a=(\frac{3}{2 \pi^2})^{\frac{1}{3}}$, $\bar \Phi(S)$ being
given parametrically as \cite{SM}
\be \label{PhiS0}
 \bar \Phi(S)= \frac{y^6}{12}+\frac{y^3}{2}+\frac{2}{3 y^3}-\frac{5}{4} 
~~,~~ S = \frac{y^5}{10} + y^2 - \frac{2}{y} + \frac{9}{10} 
\ee
for $S \in ]-\infty,0]$ corresponding to $y \in ]0,1]$. $\bar \Phi(S)$ has
a cubic tail at large negative $S$. Since 
$L_{N_1}=E_0(N_1)$ is {\it the ground state energy of $N_1$ non-interacting fermions
in a linear plus random potential} described by ${\cal H}_{SAO}$ in \eqref{HSAO},
Eq. \eqref{probFermions}  is also the PDF for this problem (studied recently in 
\cite{HartmannMajumdar2018} without a linear potential).
Other applications of 
\eqref{sum1} include the center of mass
of the $N_1$ rightmost fermions in an harmonic trap with $\sim \beta(\beta-2)/r^2$
mutual interactions, or of fluctuating interfaces \cite{SM}. Applications
of the exponential wall include traces of
large powers ($\sim (N/t)^{2/3}$) of random matrices,
as in \cite{gorin2018kpz},
as well as a directed polymer or a quantum particle
at high temperature, in presence of linear plus random (static)
disorder \cite{SM}. Finally,
note that any bulk linear statistics
with a function $f(\lambda)$ which is smooth at a soft edge is
universally described by the linear soft wall $\gamma=1$
(i.e. with a log-divergent optimal density).

In conclusion we have unified \textit{four} apparently distinct methods to
study the large deviations for linear statistics at the edge of the
$\beta$ ensemble of random matrices. It equivalently 
describes the response of a logarithmic Coulomb gas pushed \textit{delicately} at its edge,
with various applications to trapped fermions.
Our results raise multiple questions such as the extensions to more general soft potentials $\phi$ leading to non-unique solutions of the saddle points or multiple supports for the optimal density. This direction is currently in progress. Other outstanding questions are extensions of our methods to other random matrix ensembles, or to other types of Coulomb gases.

\acknowledgments
We thank L.C. Tsai for discussions and sharing preliminary findings.
We are grateful to I. Corwin, P. Ghosal, V. Gorin, S. N. Majumdar, S. Prolhac, G. Schehr for interesting discussions and collaborations.  We acknowledge support from ANR grant ANR-17-CE30-0027-01 RaMaTraF,
and KITP hospitality under NSF Grant No. NSF PHY-1748958.


\begin{thebibliography}{99}
\bibitem{footnote0}
We see there that the dependence in $\beta$ is \cite{SM} \\
$\Sigma^{(\beta)}_{\frac{\beta}{2} \phi}(u) = \frac{\beta}{2} \Sigma^{(2)}_{\phi}(u)$
and $G^{(\beta)}(\ell) =  \frac{\beta}{2}  G^{(2)}(\ell)$.
\end{thebibliography}
\end{document}


\title{Supplementary materials of "Linear statistics and pushed Coulomb gas at the edge of $\beta$-random matrices: four paths to large deviations"}

\author{Alexandre Krajenbrink and Pierre Le Doussal}
   
\address{Laboratoire de Physique Th\'eorique de l'Ecole Normale Sup\'erieure,\\
PSL University, CNRS, Sorbonne Universit\'es, 
24 rue Lhomond, 75231 Paris, France.}

\begin{abstract}
We give the details of the derivations described in the Letter. We explain all the connections
between the four methods to study the large deviations. We give all details of the applications
of these methods to various systems. 
\end{abstract}
\maketitle
{\hypersetup{linkcolor=black}
	\tableofcontents
}
\section{Mathematical preliminaries} 
We display here some useful formula for the calculations presented in this Letter.
\subsection{Square root of the Heaviside function}
We recall the notation from the Letter $\frac{1}{\sqrt{(\lambda)_+}}=\frac{\theta(\lambda)}{\sqrt{\lambda}}$ and introduce the following integral
\be
\label{eq:heaviside}
\frac{1}{\pi} \int_{-\infty}^{+\infty} \frac{\rmd V}{\sqrt{(A-V)_+(V-A')_+} } = \theta(A-A')
\ee 
where $A,A'$ are real constants and $\theta$ is the Heaviside function. We further write this relation in a reduced convolution form
\begin{equation}
\frac{1}{\pi} \frac{1}{\sqrt{(\lambda)_+}} * \frac{1}{\sqrt{(\lambda)_+}} = \theta
\end{equation}
From the convolution point of view, the function $1/\sqrt{(\lambda)_+}$ thus acts as a square root of the Heaviside function. Integrating this relation leads to the useful identity
\be
\label{eq:SquareRootIdentity}
\frac{2}{\pi} \int_{-\infty}^{+\infty} \rmd V\frac{\sqrt{(V-A')_+}}{\sqrt{(A-V)_+} } = (A-A')_+
\ee 
\subsection{Hilbert transform}
Let us recall the definition of the Hilbert transform of a function $f$, as the convolution integral
\begin{equation}
H(f)(b)=\frac{1}{\pi}\dashint_{-\infty}^{+\infty} \rmd b' \, \frac{f(b')}{b-b'}
\end{equation}
where $\dashint$ is the Cauchy principal value.
It is an anti-involution as $H(H(f))=-f$. The alternative expression
\begin{equation}
\label{eq:hilbert_log}
H(f)=\frac{1}{\pi}(\log\abs{.} * f)'
\end{equation}
will be useful below. 
From a simple residue calculation, one obtains the Hilbert transform of $1/\sqrt{(\lambda-A)_+}$ for a constant $A$ as
\be
\label{eq:sqrt_hilbert}
\frac{1}{\pi} \dashint_{-\infty}^{+\infty} \rmd \lambda' \frac{1}{\lambda - \lambda'} \frac{1}{\sqrt{(\lambda'-A)_+}} =-\frac{1}{\sqrt{(A-\lambda)_+}}
\ee 
From the anti-involution property of the Hilbert transform, one further has
\be
\frac{1}{\pi}  \dashint_{-\infty}^{+\infty} \rmd\lambda' \frac{1}{\lambda - \lambda'} \frac{1}{\sqrt{(A- \lambda')_+}} =
\frac{1}{\sqrt{(\lambda-A)_+}} 
\ee

\section{Details for the Section "SAO/WKB method"   }
\subsection{Semi-classical density of states}

In the Letter we use the standard WKB argument \cite{Cohen_book, landau} to obtain the semi-classical density of states associated to a Schr{\"o}dinger Hamiltonian describing a quantum particle of mass $m$ in a potential $W(x)$ in one
dimension
\be
{\cal H}(p,x) = \frac{p^2}{2 m} + W(x) \quad , \quad p= \frac{\hbar}{i} \partial_x
\ee 
One considers classical periodic trajectories between two
consecutive turning points $x_{\pm}$ where the classical momentum 
$p(x)=\sqrt{ 2 m (E- W(x))}$ vanishes. In the limit of small $\hbar$, or
for high energy levels, the 
quantification condition for the $n$-th level becomes well approximated by 
$\int_{x_-}^{x_+} \rmd x\,  p(x) =  \pi n \hbar$. Hence the integrated density of states,
i.e. the number of levels below the energy $E$
\be
N(E) = \frac{1}{\hbar \pi} \int \rmd x \, \sqrt{ 2 m (E- W(x)) } 
\ee 
Taking from Eq. (8) of the Letter, ${\cal H}={\cal H}'_{SAO}$, which corresponds to 
$W(x)=x + v(x)$, $m=1/2$, $E=b$ and $\hbar=1/t$, we can apply
for large $t$ this WKB estimate leading to the formula (9) in the Letter.
There we consider that there is an infinite barrier at $x=0$, hence
$x_-=0$ and $x_+$ denotes the first turning point.\\

The highly surprising, and quite non-trivial point is that this can be 
a useful approximation despite the fact that $v(x)$ is not at all smooth.
One way to understand it is to remember that in effect the approximation is used for
describing the {\it optimal} $v(x)$ (or near optimal one)
which is way smoother, as we find.\\

An equivalent way to justify the starting point for the density is to use the Ricatti
equation. Writing first the Schr{\"o}dinger equation ${\cal H}'_{SAO} \psi=b \psi$ 
and introducing $g(x) = \psi'(x)/\psi(x)$ it is well known \cite{virag2018operator, fukushima1977spectra, ramirez2011beta, tsai2018exact} that the 
number of eigenvalues below level $b$, $N(b)=N(b_i < b)$, of ${\cal H}'_{SAO}$ equals the total number of
explosions of the Ricatti equation satisfied by $g$
\be
g'(x) = t^2( x-b +  v(x)) - g(x)^2  \quad  , \quad g(0)=+\infty
\ee
It turns out that in the limit of a large parameter $t$ the blow ups are very densely
spaced in $x$, hence in each blow up interval we can solve this equation assuming 
that ${\sf b}={\sf b}(x):=b - x - v(x) >0$ is constant. The equation is then
\be
g'(x) = - t^2 {\sf b} - g^2 
\ee 
and its solution is $g(x)=- t \sqrt{{\sf b}} \tan (t \sqrt{{\sf b}} x + C)$: if 
$x_i$ denote the $i$-th blow up time the separation in $x$
between two consecutive blow up 
\be
x_{i+1}-x_i = \frac{\pi}{t \sqrt{{\sf b}(x_i)} } \ll 1
\ee 
is indeed small in the semi-classical limit, i.e. for large $t$ 
(note that for ${\sf b}<0$ there is no blow up). 
Taking a continuum limit we can write $\frac{\rmd x}{\rmd i} = \frac{\pi}{t \sqrt{b(x)}}$
leading to 
\be \label{5}
N(\lambda_i < b) = \frac{1}{t} \int_{-\infty}^{b} \rmd b' \,\hat \rho(b')   \simeq \frac{1}{\pi}  \int_0^{+\infty} 
\rmd x\,  \sqrt{(b- x - v(x))_+} 
\ee
which leads to Eq. (9) in the Letter. This argument was sketched to us by L.C. Tsai and
later made rigorous by him in the case of the application to $\phi=\phi_{\rm KPZ}$ in Ref.~\cite{tsai2018exact}. 

\subsection{Simplification of $\Sigma_\phi(u)$ at the saddle point {\sf SP1}} 
Let us derive now  the expression of $\Sigma_\phi(u)$ taken at the saddle point expression Eq.~(12) in the Letter. We have 
\be \label{sigmasao}
\Sigma_\phi(u)  = \int_{-\infty}^{+\infty} \mathrm{d}b \, \rho_*(b) \phi(u-b) + \frac{\beta}{8} \int_0^{+\infty}   \mathrm{d}x\, v_*(x)^2 
\ee 
where $v_*(x)$ is the solution of the saddle point {\sf SP1}  and $\rho_*(b)$ is the optimal density
\be
 \frac{\beta}{4} v_*(x)   =  \frac{1}{2 \pi} \int_0^{+\infty} \frac{\mathrm{d}b}{\sqrt{b}} \phi'(u-b -x - v_*(x)) \quad , \quad 
\rho_*(b) = \frac{1}{2\pi} \int_0^{+\infty} \rmd x\,  \frac{1}{\sqrt{(b- x - v_*(x))_+}}
 \label{rho} 
\ee
Let us transform the first term as 
\begin{equation}
\begin{split}
\int_{-\infty}^{+\infty} \mathrm{d}b \, \rho_*(b) \phi(u-b)&=\frac{1}{2\pi }\int_{-\infty}^{+\infty} \mathrm{d}b \int_0^{+\infty} \rmd x\,  \frac{\phi(u-b)}{\sqrt{(b- x - v_*(x))_+}} \\
&=\frac{1}{2\pi }\int_{0}^{+\infty} \mathrm{d}b \int_0^{+\infty} \rmd x\,  \frac{1}{\sqrt{b}} \phi(u-b-x-v_*(x))\\
&=\frac{1}{2\pi }\int_{0}^{+\infty} \mathrm{d}b \int_0^{+\infty} \rmd x\,x(1+v_*'(x))  \frac{1}{\sqrt{b}} \phi'(u-b-x-v_*(x))\\
&=\frac{\beta}{4} \int_0^{+\infty} \rmd x\,x\big(1+v_*'(x)\big) v_*(x)\\
&=\frac{\beta}{4} \int_0^{+\infty} \rmd x\,\big(xv_*(x)-\frac{v_*(x)^2}{2}\big)
\end{split}
\end{equation}
The transformation from the first line to the second is a shift of $b$ by $x+v_*(x)$.  From the second to the third line we proceeded to an integration by part with respect to $x$. From the third to the fourth, we used the saddle point {\sf SP1} and finally from the fourth to the fifth, we integrated by part the $v_*v_*'$ term. We observe that the quadratic term $v_*^2$ cancels the one from the Brownian measure in $\Sigma_\phi(u)$ therefore leading to Eq.~(12) from the Letter
\begin{equation}
\Sigma_\phi(u)=\frac{\beta}{4} \int_0^{+\infty} \rmd x\,xv_*(x)
\label{eq:sigma*}
\end{equation}
\subsection{Simplification of $\Sigma'_\phi(u)$ at the saddle point {\sf SP1}}
The derivative of the free energy $\Sigma_\phi(u)$ at the saddle point is obtained by taking the explicit derivative with respect to $u$ of Eq.~\eqref{sigmasao}, leading to $\Sigma'_\phi(u)  = \int_{-\infty}^{+\infty} \mathrm{d}b \, \rho_*(b) \phi'(u-b) $. Inserting the WKB parametrization for the density and using the saddle point equation {\sf SP1}, we get
\begin{equation}
\label{eq:sigma'*}
\begin{split}
\int_{-\infty}^{+\infty} \mathrm{d}b \, \rho_*(b) \phi'(u-b)&=\frac{1}{2\pi }\int_{-\infty}^{+\infty} \mathrm{d}b \int_0^{+\infty} \rmd x\,  \frac{ \phi'(u-b)}{\sqrt{(b- x - v_*(x))_+}}\\
&=\frac{1}{2\pi }\int_{0}^{+\infty} \mathrm{d}b \int_0^{+\infty} \rmd x\,  \frac{1}{\sqrt{b}} \phi'(u-b-x-v_*(x))\\
&=\frac{\beta}{4} \int_0^{+\infty} \rmd x\,v_*(x)\\
&= \frac{1}{\pi} \int_0^{u} \rmd u'\,w(u')
\end{split}
\end{equation}
which provides another way to show that $\Sigma_\phi''(u)=\frac{1}{\pi}w(u)$ as stated in the Letter. Note that Eq.~\eqref{eq:sigma'*} can directly be obtained from Eq.~\eqref{eq:sigma*} using the correspondence $v_*(x)=\frac{4}{\beta \pi}w(u-x)$ implying that $\partial_u v_* = -\partial_x v_*$ at the saddle point.

\section{Details for the Section "cumulant method"  }
\subsection{Cumulant expansion for $\Sigma_\phi(u)$}
Although we will insert factors of $\beta/2$ in some formulae here, the 
derivation here is restricted to $\beta=2$. As discussed in the Letter,
comparison with the other methods validates our proposed extension to
arbitrary $\beta$. The expectation $Q_t(u)$ defined in the Letter in Eq.~(5) over the Airy point process for $\beta=2$ can be expressed as a Fredholm determinant (see also Eq.~(19) in the Letter)
\be
\label{eq:fredholm}
Q_t(u) = \mathrm{Det}[ I - \sigma_t K_{{\rm Ai}}]_{\, \mathbb{L}^2(\mathbb{R})}
\ee 
where $I$ is the identity operator,  $\sigma_t(a)=1- e^{- t  \phi(u+t^{-2/3} a)}$ and  $K_{{\rm Ai}}(a,a')$ is the Airy kernel, i.e. $
K_{\mathrm{Ai}}(a,a')=\int_{0}^{\infty} \! \rmd r \; \mathrm{Ai}(r+a) \mathrm{Ai}(r+a')$. Let us recall the results of Ref.~\cite{krajenbrink2018systematic} providing the expansion in cumulants of Fredholm determinants such as \eqref{eq:fredholm} with the choice of functions $\sigma_t(a)=1-\exp[g_t(\sigma e^{at^{1/3}})]$ for a class of functions $g_t$. The expansion in cumulants is defined by the following series

 \begin{equation}
\label{deter_process}
 \log \mathrm{Det}\left[I- (1- \rme^{\alpha \hat g}) K_{{\rm Ai}}\right] = \sum_{n=1}^\infty \frac{\alpha^n \kappa_n}{n!} 
\end{equation}
where we denote $\hat g(a)=g_t(\sigma e^{t^{1/3} a})$, $Q_t(u)$ being obtained by setting the book-keeping parameter $\alpha$ to 1. The first two cumulants are given by $ \kappa_1=\mathrm{Tr}(\hat g K_{{\rm Ai}})$ and  $\kappa_2=\mathrm{Tr}(\hat g^2 K_{{\rm Ai}})-\mathrm{Tr}(\hat g K_{{\rm Ai}} \hat g K_{{\rm Ai}})$, see \cite{krajenbrink2018systematic, KrajLedou2018, krajenbrink2018large} for more details. It was found in Ref.~\cite{krajenbrink2018systematic} that the cumulants can be written as 
\begin{equation}
\label{eq:kappa}
\kappa_n= t^{\frac{n}{2}-1} 2^{n-1} (\sigma\partial_{\sigma})^{n-3}\mathfrak{L}_{1}(\sigma)^{n} +\dots 
\end{equation}
where 
		\begin{equation} 
		\label{eq:l1}
\mathfrak{L}_{1}(\sigma)=\frac{1 }{\pi }(\sigma \partial_\sigma)^{2} \int_{0}^{+\infty}\mathrm{d}x\sqrt{x} g_t(\sigma e^{-x})
		\end{equation}
As discussed in Ref.~\cite{krajenbrink2018systematic}, under some conditions, the term displayed in Eq.~\eqref{eq:kappa} is dominant compared to the (complicated) remainder indicated by the $"\dots"$, this is case  for large time if one chooses $\sigma=-e^{ut}$ and functions such that
\begin{equation}
\label{eq:limit_g}
\lim_{t\to +\infty}-\frac{g_t(-e^{t(u+b)})}{t}=\phi(u+b)
\end{equation}
and $g_t(0)=0$, which is precisely what is needed for the results to apply to the class of functions $\sigma_t(a)=1- e^{- t  \phi(u+t^{-2/3} a)}$. Note that $g(0)=0$ implies that $\phi(x)=0$ for $x\leq 0$. Inserting $\sigma=-e^{ut}$ into Eqs.~\eqref{eq:kappa} and \eqref{eq:l1} and taking the large time limit Eq.~\eqref{eq:limit_g},  the $n$-th cumulant reads (until now for $\beta=2$)
\begin{equation}
\label{eq:kappa_general}
\kappa_n=t^{\frac{n}{2}-1} \frac{2^{n-1}}{\pi^n} (\frac{1}{t}\partial_u)^{n-3}\left(-\frac{1}{t}(\partial_u)^{2}  \int_{0}^{+\infty}\mathrm{d}x\sqrt{x} \phi(u-\frac{x}{t})\right)^{n}
\end{equation}
Regrouping the different factors, one observes that all leading terms of $\kappa_n$ are proportional to $t^2$ and we can then write the summation over the cumulant index $n$, and now insert appropriate factors of $\beta/2$, leading to
\bea \label{Qtsum} 
\log Q_t(u)=\sum_{n\geq 1} \frac{\kappa_n}{n!}=-t^2\Sigma_\phi(u) \quad , \quad \Sigma_\phi(u)= -  \frac{\beta}{4} \sum_{n \geq 1} \frac{1}{n!} (-\frac{4}{\beta \pi})^n
\partial_u^{n-3} f(u)^n 
\eea 
where 
\begin{equation}
\label{eq:def_ff}
f(u) =  \int_\mathbb{R} \rmd b \,\sqrt{(b)_+} \phi''(u-b)
\end{equation}
Upon integration by part, one obtains formulae (13) and (14) as given in the Letter. For $n=1, 2$ we clarify the meaning of the anti-derivative in Eq.~\eqref{Qtsum} as
\begin{equation} \label{kappa12}
\begin{split}
\kappa_1= -\frac{t^2}{\pi}  \int_{0}^{+\infty}\mathrm{d}b\sqrt{b} \, \phi(u-b)=-\frac{t^2}{\pi}  \int_0^u \int_0^{u'} \rmd u'\rmd u'' f(u'') \quad , \quad \kappa_2= \frac{4t^2}{\beta\pi^2}\int_0^u \rmd u' \, f(u')^2
\end{split}
\end{equation}

\subsection{Cumulants of ${\sf L}$} 
The generating function of the cumulants of ${\sf L}$ can be obtained at large $t$ as
\be \label{gener} 
\log \mathbb{E}_\beta[ e^{- B {\sf L}} ] = \sum_{n \geq 1} 
\frac{(-B)^n}{n!}  \mathbb{E}_\beta[{\sf L}^n]^c  \simeq - t^2 \Sigma_{B \phi}(u)
\ee 
i.e. it can be obtained by the multiplication of $\phi$ by an amplitude $B$.
Inserting $\phi \to B \phi$ in \eqref{Qtsum} and identifying order by order
we obtain
\be \label{cumL1} 
\mathbb{E}_\beta[{\sf L}^n]^c = (-1)^n \kappa_n \simeq \frac{\beta}{4} t^2 (\frac{4}{\beta \pi})^n
\partial_u^{n-3} f(u)^n 
\ee 
where $\kappa_1$ and $\kappa_2$ are given explicitly in \eqref{kappa12},
the formula being quite explicit for $n \geq 3$. Note that in the Letter we use 
$\tilde \kappa_n(u)=\kappa_n/t^2$.

\subsection{Resummation of $\Sigma_\phi(u)$}

It is possible and convenient to perform the summation 
of the series representation of $\Sigma_\phi$ by writing its third derivative as
\begin{equation} \label{3der}
\Sigma_\phi'''(u)= -  \frac{\beta}{4} \sum_{n \geq 1} \frac{1}{n!} (-\frac{4}{\beta \pi})^n
(\partial_u)^{n} f(u)^n 
\end{equation}
We have used a Mellin-Barnes summation method presented in the Appendix, 
and displayed in Eq.~\eqref{eq:mellin} with $a=-\frac{4}{\beta \pi}$
and $\Sigma_\phi'''(u)= -  \frac{\beta}{4} \mathcal{S}(u)$. 
The summation is mapped to the problem of solving the following equation for $w=w(u)$
\be 
\label{eq:algebraic}
 f(u - \frac{4}{\beta \pi} w)= w
\ee
We consider for now functions $f$ which are positive, increasing with $f(b)=0$ for $b\leq 0$. There is then a unique solution of Eq.~\eqref{eq:algebraic} which can be written
\begin{equation} \label{uw} 
u:=u(w)=f^{-1}(w)+\frac{4}{\beta \pi}w , \quad \forall w>0 , \quad \text{and} \; u(0)=0
\end{equation}
It is convenient to extend $w$ and $u$ to negative values setting $u(w\leq 0)=0$ and $w(u\leq 0)=0$. Given the uniqueness, from Eq.~\eqref{eq:mellin} and \eqref{eq:Su}, $\Sigma_\phi'''$ is given by
\begin{equation}
\Sigma_\phi'''(u)=-  \frac{\beta}{4}  \left( \frac{1}{1+ \frac{4}{\beta \pi} f'(u - \frac{4}{\beta \pi} w(u))} -1 \right) =\frac{1}{\pi}w'(u)
\end{equation}

We then perform the integrations, noting that for $u \to -\infty$ the Coulomb gas is not affected by the wall
and $\Sigma_\phi(u)$ and its derivatives should vanish. For $\phi$ in $ \Omega_0$, $f$ and $w$ 
vanish for $u \leq 0$, so we can even use that $\Sigma(0)=\Sigma'(0)=\Sigma''(0)=0$.
The first integration gives
\be
 \Sigma_\phi''(u)= \frac{1}{\pi}\int_{-\infty}^u \rmd u' \, w'(u') = \frac{1}{\pi} w(u) 
\ee
The second integration gives 
\be
 \Sigma_\phi'(u)= \frac{1}{\pi} \int_{-\infty}^u \rmd u'\, w(u') = \frac{1}{\pi} \int_{0}^u \rmd u' w(u') 
 = \frac{1}{\pi} \int_{0}^{w(u)} \rmd w' [u-u(w')]
\ee
The third integration gives 
\be
\label{eq:triple_int}
 \Sigma_\phi(u)=\frac{1}{2 \pi}\int_0^{w(u)}\mathrm{d}w'\left[u-u(w')\right]^2=\frac{1}{\pi} \int_0^{u}\mathrm{d}u' \, w(u')\left[u-u'\right]
\ee
The forms as integrals in $w'$ are quite useful in practice when $f^{-1}(w)$ has a simple form, as
in the examples given below, since $u(w)$ is then explicit using \eqref{uw} and
the integral can often be calculated. The second form is given in the Letter in Eq. (16)
and allows easy comparison with the other methods.

\section{Ensembles of functions $\phi$ considered for the linear statistics  }

It is useful to recapitulate the ensembles of functions $\phi$ considered here. The condition that for all $u$ in $A$ there is a unique solution $w=w(u)$ to 
\be
w = f(u - \frac{4}{\beta \pi} w) 
\ee 
is equivalent to the condition that for all $u$ in $A$ there is a 
unique solution $z=z(u)$ to 
\be
u = z + \frac{4}{\beta \pi}  f(z) 
\ee 
with the relation $z(u)=u- \frac{4}{\beta \pi} w(u)$. 
This condition is in turn equivalent to the condition
that $z \to h(z)=z + \frac{4}{\beta \pi}  f(z)$ is strictly monotonous
and has no jump (is continuous) in $h^{-1}(A)$. We call $\Omega_2$ the set of functions $\phi$ such that their associated
$f$ has this property (where monotonous means increasing)
with $A=\mathbb{R}$.
\begin{equation*}
\Omega_2= \lbrace \phi \mid z\mapsto z+\frac{4}{\beta \pi}f(z) \;\;  \text{is strictly increasing and continuous} \rbrace
\end{equation*}
We call $\Omega_0$, a subset of $\Omega_2$ such that 
$f(z)$ itself is increasing, positive, continuous with $f(z\leq 0)=0$. 
\begin{equation*}
\Omega_0= \lbrace \phi \mid z\mapsto f(z) \; \; \text{is increasing positive and continuous}, \; \; f(z\leq 0)=0 \rbrace
\end{equation*}
It implies that $\phi$ is also increasing, positive, continuous with $\phi(z\leq 0)=0$, however not all such functions are in $\Omega_0$ (roughly, $\phi$ has to
grow fast enough on the positive side - e.g. if $\phi'$ has these properties
then $\phi$ is in $\Omega_0$). Since $w(u)=f(z(u))$ we see that if $\phi$ is in $\Omega_2$ but
not in $\Omega_0$ then $w(u)$ may be non monotonous in $u$, or
negative. \\

Finally, we define $\Omega_1$ the set of functions $\phi$ such that $f(z)$ is increasing, positive, continuous, but we do not require $f(z\leq 0)=0$.
\begin{equation*}
\Omega_1=\lbrace \phi \mid   z\mapsto f(z) \; \; \text{is increasing positive and continuous} \rbrace
\end{equation*}
Note that here we further require that $\lim_{b\to -\infty} (-b)^{3/2}\phi(b)=0$ to guarantee a finiteness of the excess energy (see Section on the inverse monomial walls).\\
 We have the ordering relation $\Omega_0 \subset \Omega_1 \subset \Omega_2$. Most of the Letter focuses on the set $\Omega_0$, the \textit{monomial walls} and an extension to $\Omega_1$, the \textit{exponential wall} and the \textit{inverse monomial}, will be presented.\\
 
 For $\phi$ in $\Omega_1$ all formula presented in the Letter hold with the slight modification that $w(u)$ does not vanish for $u<0$ (but remains positive and vanishes at $u=-\infty$) hence all integrations over $u$ 
must start from $u=-\infty$ (while those over $w$ still start at 0). The saddle point $v_*(x)$ now is
non zero for $x>u$ but keeps the same properties (positivity and $x+v(x)$ increasing) and vanishes at $x=+\infty$. The relations (17) and (26) in the Letter between $v_*$,$w$,$g$ remain true, i.e.
\begin{equation}
\begin{split}
&\forall u\in \mathbb{R}, \; w(u)=\frac{\beta \pi }{2}g(-u)\\
& \forall x\geq 0, \;  v_*(x)=\frac{4}{\beta \pi}w(u-x)
\end{split}
\end{equation}
\section{From WKB/SAO to the cumulant expansion }

Let us now study the saddle point equation for the WKB/SAO method. Equation (11) in the Letter can be written as 
\begin{equation}
\begin{split}
\frac{\beta}{4}v_*(x)&=\frac{1}{2\pi} \int_{-\infty}^{+\infty}\frac{\rmd b}{\sqrt{(b)_+}}\phi'(u-b-x-v_*(x))\\
&=\frac{1}{\pi}f(u-x-v_*(x))
\end{split}
\end{equation}
where we have used the definition of Eqs. (14) in the Letter and \eqref{eq:def_ff} in the Supp. Mat. of the function $f$. To make contact with the method of cumulants, we study a solution $v_*(x)$ which has the form
\be \label{vw} 
v_*(x) = \frac{4}{\beta \pi} w(u-x) \quad 0 \leq x \leq u
\ee
with $v_*(x)=0$ for $x \geq u$. With this parametrization, the saddle point equation becomes
\begin{equation}
\label{eq:saddle_beta}
f(u-\frac{4}{\beta \pi}w(u))=w(u)
\end{equation}
This is precisely the equation \eqref{eq:algebraic} encountered in the resummation of the series in the cumulant method.
 In addition, using this parametrization, the resulting equation for $\Sigma_\phi(u)$ within the WKB/SAO method Eq.~(9) reads 
\begin{equation}
\label{eq:sigmaSAOCum}
\Sigma_\phi(u)=\frac{\beta}{4}\int_0^{+\infty} \rmd x \, x \,v_*(x) = \frac{1}{\pi }\int_0^u \rmd u' w(u') [u-u']
\end{equation}
This identifies with the one Eq.~(14) from the cumulant method. We now derive through the WKB/SAO method the series expansion previously obtained from the cumulant method. This is realized by the means of the Lagrange inversion formula. Let us recall that the Lagrange inversion formula states that for a sufficiently nice function $h$, the equation $z=x+yh(z)$ can be inverted as
\begin{equation}
z=x+\sum_{n\geq 1}\frac{y^n}{n!} (\partial_x)^{n-1} h(x)^n
\end{equation}
Identifying $x=u$, $z=u-\frac{4}{\beta \pi} w$, $y=-\frac{4}{\beta \pi}$, $h=f$ leads to a series representation for the solution $w(u)$ of Eq.~\eqref{eq:saddle_beta}
\be
-\frac{4}{\beta \pi} w(u) = \sum_{n \geq 1} \frac{1}{n!}(-\frac{4}{\beta \pi})^n \partial_u^{n-1} f(u)^n 
\ee
From Eq.~\eqref{eq:sigmaSAOCum}, we also have $\Sigma_\phi''(u)=  \frac{1}{\pi} w(u)$ and hence we obtain 
\be
\Sigma_\phi''(u) = - \frac{\beta}{4} \sum_{n \geq 1} \frac{1}{n!}(-\frac{4}{\beta \pi})^n \partial_u^{n-1} f(u)^n 
\ee
which coincides precisely with the second derivative of the series expansion obtained for the cumulant method (valid at $\beta=2$ and generalized there to arbitrary $\beta$).

\section{Details for the Section "Painlev\'e method"  }
\subsection{Analysis of the non-local Painlev\'e equation}

Let us recall here the analysis of 
Ref.~\cite{sasorov2017large} and present its generalization.
To make it easier we stick to the notations of Ref.~\cite{sasorov2017large}.
Starting from the equations (20-22) of the Letter
we introduce as in Ref.~\cite{sasorov2017large} the scaled variables 
$r = t^{2/3} X$, $v = t^{2/3} V$ and make the ansatz $q_t(r,v)=t^{-1/6} \tilde q_t(X,V)$
and $\Psi_t(r) \simeq t^{2/3} g_t(X)$, with $g_t(X)>0$. The remarkable fact is
that the function $g_t(X)$ becomes independent of $t$ at large $t$,
and one denotes $g(X) = \lim_{t \to +\infty} g_t(X)$. Performing the rescaling,
Eq. (21) in the Letter becomes
\be \label{gV} 
 g(X) = \int_{-\infty}^{+\infty} \rmd V [\tilde q_t(X,V)]^2 \phi'(V)
e^{- t \phi(V)} 
\ee 
It is precisely the condition that the r.h.s. does not depend on $t$ at large $t$ which
leads to the two equations {\sf SP1} and {\sf SP2} (Eqs~(23) and (24) in the Letter)
and of the consistency of the ansatz, as we now discuss.\\

Performing the rescaling, the Eq. (22) of the Letter becomes
\be \label{shrod} 
- t^{-2} \partial_X^2 \tilde q_t(X,V) + (V + X + 2 g(X)) \tilde q_t(X,V) =0 
\ee
with the boundary condition $\tilde q_t(X,V) \to_{X \to +\infty} t^{1/6} {\rm Ai}(t (X+V)) \simeq
\frac{\exp(- \frac{2}{3} t (X+V)^{3/2})}{2 \sqrt{\pi} (X+V)^{1/4}}$\footnote{{This boundary condition implies the condition $\phi(+\infty)=+\infty$. If this is not the case, e.g. $\phi(+\infty)=\phi_\infty<\infty$ then the boundary condition becomes  $\tilde q_t(X,V) \to_{X \to +\infty} t^{1/6}\sqrt{1-e^{-t\phi_\infty}} {\rm Ai}(t (X+V)) $, see Ref.~\cite{TW}, hence generalizing Proposition 5.2 of Ref.\cite{ACQ11}.}}. 
It can be interpreted as the Schr{\"o}dinger equation of a quantum particle 
of mass $m=1/2$ at energy $-V$ in the potential $X + 2 g(X)$, 
in the semi-classical limit since $\hbar = 1/t$ is small. 
As in Ref~\cite{sasorov2017large} we consider cases such that $g(X)$ 
is a positive and monotonic decreasing function,
and, as seen below, $g(X)$ vanishes for $X>0$.
The potential $X+ 2 g(X)$ is however an increasing function of $X$. 
Hence there is a unique classical turning point at $X=a$ with $V + a + 2 g(a)=0$. The classical momentum of the particle is $p(X,V)=\sqrt{-V - X - 2 g(X)}$, which is
positive in the the classically allowed region $X<a$, and imaginary for $X>a$,
the forbidden region. The standard WKB method then gives the
following approximation of the wave function for large $t$
\begin{equation}
\begin{split}
 \tilde q_t(X,V) \simeq & \; \frac{C_t(V)}{|V+X+2 g(X)|^{1/4} } \\
&\bigg( \cos\left[ t \int_{-\infty}^X \rmd X' 
\sqrt{(-V - X' - 2 g(X'))_+} - \frac{\pi}{4}\right] \theta(-V-X- 2 g(X)) \nonumber 
\\
& + \frac{1}{2} 
\exp\big[ - t \int_{-\infty}^X \rmd X' \sqrt{(V+X'+2 g(X'))_+} \big]\;  \theta(V+X+2 g(X))
\bigg)
\end{split}
\end{equation}
The boundary condition determines the amplitude $C_t(V)$ as
\bea
C_t(V) = \frac{1}{\sqrt{\pi}} \exp\left( t \int_{-\infty}^{+\infty} \rmd X' \big[ \sqrt{(V+X'+2 g(X'))_+}  - \sqrt{(V+X')_+}\big] \right)
\eea
Inserting $\tilde q_t(X,V)$ into Eq.~\eqref{gV} we obtain a sum of
two contributions
\begin{equation}
\begin{split}
& g(X) = \int_{-\infty}^{+\infty} \rmd V \frac{C_t(V)^2  \phi'(V) e^{- t \phi(V)}}{|V+X+2 g(X)|^{1/2}}\\
&\bigg( \frac{\theta(-V-X- 2 g(X))}{2}  + \frac{\theta(V+X+ 2 g(X))}{4} e^{-2 t \int_{-\infty}^X \rmd X' \sqrt{(V+X'+2 g(X'))_+} }    \bigg)
\end{split}
\end{equation}
The second term can be neglected at large $t$ compared to the first
(see Ref.~\cite{sasorov2017large} for more discussion of the validity of the WKB approximations)
leading to
\begin{equation}
\begin{split}
g(X) =& \frac{1}{2 \pi}  \int_{-\infty}^{-X-2 g(X)}
\frac{\rmd V \phi'(V)}{\sqrt{-V-X-2 g(X)}}\\
& \exp\bigg(t \big(
2 \int_{-\infty}^{+\infty} \rmd X' [ \sqrt{(V+X'+2 g(X'))_+} 
- \sqrt{(V+X')_+}] -\phi(V)\big)\bigg)
\end{split}
\end{equation}
The condition of $t$ independence gives the two equations in the Letter, namely
\bea
&& g(X) = \frac{1}{2 \pi}  \int_{-\infty}^{-X-2 g(X)}
\frac{\rmd V \phi'(V)}{\sqrt{-V-X-2 g(X)}}  \label{sp1n} \\
&& \phi(V) = 2 \int_{-\infty}^{+\infty} \rmd X' \left[ \sqrt{(V+X'+2 g(X'))_+} 
- \sqrt{(V+X')_+}\right] \label{sp2n} 
\eea
If $\phi$ belongs to the class $\Omega_0$, $\phi'(V) \geq 0$ and strictly vanishes for
$V<0$ and one finds that indeed $g(X) \geq 0$ with $g(X)=0$ for $X>0$ as
anticipated. Hence the upper bound of
the integral in the second equation can be chosen to be $X'=0$, recovering the 
Eq. (24) in the Letter.  Finally, performing the rescaling in the Eq. (20) of the Letter leads to
\be \label{finalQ} 
\log Q_t(u)= -  t^2 \int_{-u}^{+\infty} \rmd X (u+X) g(X) 
\ee 
leading to the formula (25) in the Letter whenever $g(X >0)=0$. \\

The important property in the above derivation seems to be the uniqueness of the
turning point, i.e. that $X \to X + g(X)$ is a strictly increasing function. 
Comparison with the other method (see below) suggests that it can be extended to cases where $g(X)$ does not vanish for
$X>0$ but decays sufficiently fast so that the integral in \eqref{sp2n}
converges. \\

We note the amazingly close resemblance to the WKB analysis of the 
SAO operator. The connection is discussed in the Letter, and
combining the equations (16) and (25) there we can identify
\be
v_*(x) = 2 g(x-u)  
\ee 
where $v(x)$ is taken at the saddle point {\sf SP1}.
The property that $v(x> u)=0$ thus maps to the property that
$g(X>0)=0$ (and holds for $\phi \in \Omega_0$). 
Note however that the function $g(X)$ also lives for any $X<u$,
and so does $w(u)$ for any $u>0$, hence there is a 
natural extension of the function $v(x)$ of the SAO method.\\

Note that Eqs.~\eqref{gV}, \eqref{shrod} and \eqref{finalQ} 
are all exact for any $t$ if one replace $g(X) \to g_t(X)$ and equivalent to
the system (20-22) in the Letter, being simply their scaled version. 
Hence there may be a more general connection, for arbitrary $t$.
Each methods evaluate one line of the equality, for any $t$
\bea
\mathbb{E}_v\big[ {\rm Det}[ e^{- t  \phi(u - {\cal H}'_{SAO})} ] \big] &&=
\mathbb{E}_v[ e^{- t {\rm Tr}\,  \phi(u - {\cal H}'_{SAO})} ]\\
&& = {\rm Det}[ I - (1- e^{-t \phi(u + t^{-2/3} a)}) K_{\rm Ai}]
\eea

\subsection{Inversion formula between $f$ and $\phi$}
 We recall the definition of $f(u)$ as
\be
f(u) = \frac{1}{2} \int_{-\infty}^{+\infty} \frac{\rmd b}{\sqrt{(b)_+}} \phi'(u-b) 
=\frac{1}{2} \frac{1}{\sqrt{(b)_+}} * \phi'
\ee
Note that $f(u)$ also vanishes for $u \leq 0$.
We can convolute $f(u)$ and obtain
\be
\frac{2}{\pi} \frac{1}{\sqrt{(b)_+}} * f =  \frac{1}{\pi} \frac{1}{\sqrt{(b)_+}} * (\frac{1}{\sqrt{(b)_+}} * \phi' )
=  \theta * \phi'  = \phi 
\ee
Hence the inversion formula
\be
\phi(u) = \frac{2}{\pi} (\frac{1}{\sqrt{(b)_+}} * f )(u) =
\frac{2}{\pi}  \int_{-\infty}^{+\infty} \frac{\rmd b}{\sqrt{(b)_+}} f(u - b) 
\ee
This inversion formula is now used to proof the \textit{miracle}, i.e. Eq.~(24) of the Letter.
\subsection{Proof of the "miracle" for arbitrary $\beta$ : {\sf SP1} $\Rightarrow$ {\sf SP2}}
For all $\beta$, let us define $\mathcal{I}$ by
\be
\label{eq:I_sp1}
\mathcal{I} = \beta \int_0^{+\infty} \rmd u' \sqrt{(V - u' +\frac{4}{\beta  \pi} w(u'))_+} - \frac{2\beta }{3} V^{3/2} 
\ee
and we now prove that for all $V\geq 0$, $\mathcal{I}=\phi(V)$. Recalling that $w(u)=f(u-\frac{4}{\beta \pi} w(u))$, we perform the change of variable in the integral of Eq.~\eqref{eq:I_sp1}
\bea
z = u' -\frac{4}{\beta \pi} w(u')  \quad , \quad f(z) = w(u') 
\eea 
If $f(u)$ is positive and increasing then $z$ is increasing fonction of $u'$. In addition 
since $u'= f^{-1}(w(u')) +\frac{4}{\beta \pi}w(u')$ we also have $u'=z +\frac{4}{\beta  \pi}f(z)$ hence
\begin{equation}
\begin{split}
 \mathcal{I}&=\beta  \int_0^{+\infty} \rmd z (1+\frac{4}{\beta \pi} f'(z)) \sqrt{(V-z)_+}  - \frac{2\beta }{3} V^{3/2} = \frac{4 }{ \pi} \int_0^{+\infty} \rmd z f'(z) \sqrt{(V-z)_+} \\
&= \frac{4}{ \pi}  \int_0^{+\infty} \rmd z \frac{f(z)}{\sqrt{(V-z)_+}} = \frac{2}{ \pi} \int_{-\infty}^{+\infty} \rmd z \frac{f(z)}{\sqrt{(V-z)_+}}\\
&=\phi(V) 
\end{split}
\end{equation}
\subsection{Proof that for arbitrary $\beta$ : {\sf SP2} $\Rightarrow$ {\sf SP1}} 
We now show the converse implication. Given some function $g(X)$ for $X\leq 0$, calculating $\phi$ from Eq.~(24) of the Letter and inserting it into $\tilde{\mathcal{I}}$ defined by
\begin{equation}
\tilde{\mathcal{I}} = \frac{1}{2 \pi} \int_{-\infty}^{+\infty} \frac{\rmd V \phi'(V)}{\sqrt{ (- V - X - 2 g(X))_+}}  \label{eq1}
\end{equation}
we obtain
\be
\tilde{\mathcal{I}}=\frac{\beta}{4 \pi}  \int_0^{+\infty} \frac{\rmd V}{\sqrt{ (- V - X - 2 g(X))_+}} \left( 
\int_{-\infty}^0 \frac{\rmd X'}{\sqrt{(V+ X' + 2 g(X'))_+} } - 2 \sqrt{(V)_+} \right)
\ee
Interverting the integrals, the first term can be written as 
\be
\frac{\beta}{4 \pi}  \int_{-\infty}^0 \rmd X' \int_0^{+\infty} \rmd V 
 \frac{1}{\sqrt{ (- V - X - 2 g(X))_+} \sqrt{(V+ X' + 2 g(X'))_+} }
 = \frac{\beta}{4 }  \int_{X}^0 \rmd X' = -\frac{\beta}{4 } X
\ee
where we have used the identity of Eq.~\eqref{eq:heaviside} with $A=- X - 2 g(X)$ and $A'=- X' - 2 g(X')$ ad the fact that $X + 2 g(X)<X' + 2 g(X')$ is equivalent to $X<X'$ since $X\mapsto X+2g(X)$ is increasing. The second term is calculated using the identity of Eq.~\eqref{eq:SquareRootIdentity} with $A=- X - 2 g(X)$ and $A'=0$.
\be
-\frac{\beta}{2 \pi}  \int_0^{+\infty} \frac{\rmd V  \sqrt{V} }{\sqrt{ (- V - X - 2 g(X))_+}} 
= -\frac{\beta}{4}   (-X-  2 g(X))_+
\ee 
Summing for $X\leq 0$ both contributions, we obtain $\tilde{\mathcal{I}}=\frac{\beta}{2}g(X)$ which is precisely {\sf SP1}.

\section{Details for the Section "Coulomb gas method"  }
\subsection{Parametrization of the density of the Coulomb gas}
The WKB/SAO method suggests to study a parametrization of the density $\rho(b)$ of a Coulomb gas in terms of a function $v(x)$ for $x>0$ as
\begin{equation}
\rho(b) =\frac{1}{2\pi} \int_0^{+\infty} \rmd x\,  \frac{1}{\sqrt{(b- x - v(x))_+}}
\end{equation}
where $v(x)$ encodes the deviation from the Airy density which is recovered for $v(x)=0$, indeed
\begin{equation}
\rho_{\mathrm{Ai}}(b)=\frac{\sqrt{(b)_+}}{\pi}=\frac{1}{2\pi} \int_0^{+\infty} \rmd x\,  \frac{1}{\sqrt{(b- x )_+}}
\end{equation}
This parametrization of the density verifies the bare condition of mass conservation
\begin{equation}
\label{eq:mass_conservation}
\begin{split}
\int_{-\infty}^{+\infty} \rmd b & \left[ \rho(b)-\rho_{\rm Ai}(b) \right]=\frac{1}{\pi}\int_0^{+\infty} \rmd x\, \left[ \sqrt{(b- x - v(x))_+}- \sqrt{(b- x )_+} \right]_{b=-\infty}^{+\infty}\\
&=-\frac{1}{\pi}\int_0^{+\infty} \rmd x\, v(x) \left[ \frac{1}{\sqrt{(b- x - v(x))_+}+ \sqrt{(b- x )_+}} \right]_{b=-\infty}^{+\infty}=0
\end{split}
\end{equation}
provided weak conditions on $v$, e.g. $\int_0^{+\infty} \rmd x\,  v(x) < \infty$. The following two representations in terms of $v(x)$ of \textit{(i)} the Hilbert transform of the density and \textit{(ii)} the energy of the Coulomb gas are quite general and not assume $v(x)$ is a saddle point. The precise class of functions $v(x)$ which parametrizes the general density remains to be investigated. Here in practice, we consider functions such that $x \mapsto \tilde{v}(x)=x+v(x)$ are increasing and positive (in particular, this is verified by the saddle point). In that case, there exists an inversion formula obtained using Eq.~\eqref{eq:heaviside}
\begin{equation}
\tilde{v}^{-1}(y)=2\int_{-\infty}^{+\infty}  \frac{\rmd b}{\sqrt{(b)_+}}\rho(y-b)
\end{equation}
This inversion procedure is identical to the one relating Eqs.~(14) and (18) in the Letter. For more general functions, note that there a formula
\begin{equation}
\int_0^{+\infty} \rmd x \, \theta(y-x-v(x))=2\int_{-\infty}^{+\infty}  \frac{\rmd b}{\sqrt{(b)_+}}\rho(y-b)
\end{equation}
\subsection{Hilbert transform of the density parametrization}
We calculate the Hilbert transform of the density using the aforementioned parametrization and the Hilbert transform in Eq.~\eqref{eq:sqrt_hilbert}.
\begin{equation}
\label{eq:hilbert_reparametrization}
\begin{split}
H(\rho-\rho_{\rm Ai})(b')&=\frac{1}{\pi}\dashint_{-\infty}^{+\infty} \rmd b \, \frac{\rho(b)-\rho_{\rm Ai}(b)}{b'-b}\\
&=\frac{1}{2\pi^2}\int_0^{+\infty} \rmd x\, \dashint_{-\infty}^{+\infty} \rmd b \, \frac{1}{b'-b}\left[ \frac{1}{\sqrt{(b- x - v(x))_+}} -\frac{1}{\sqrt{(b- x )_+}}\right]\\
&=-\frac{1}{2\pi}\int_0^{+\infty} \rmd x\,\left[ \frac{1}{\sqrt{(-b'+ x + v(x))_+}} -\frac{1}{\sqrt{(-b'+ x )_+}}\right]\\
\end{split}
\end{equation}
%
\subsection{Parametrization of the electrostatic energy of the Coulomb gas}
In terms of the above parametrization $v$, the electrostatic energy of the Coulomb gas adopts the remarkably simple representation which identifies the Brownian weight in the WKB/SAO method
\begin{equation}
\mathcal{J}(\rho)=- \frac{\beta}{2}\iint_{-\infty}^{+\infty} \log|b_1-b_2| \prod_{i=1}^2 \mathrm{d}b_i (\rho(b_i)-\rho_\text{Ai}(b_i))=\frac{\beta}{8}\int_0^{+\infty}\rmd x \, v(x)^2
\end{equation}
To show this equality, we write the electrostatic energy of the Coulomb gas as a convolution
\begin{equation}
\label{eq:entropy_convolution}
\mathcal{J}(\rho)=\frac{\beta}{2}\int_{-\infty}^{+\infty} \rmd b_1  \left[\rho(b_1)-\rho_{\rm Ai}(b_1)\right] \times \log\abs{.}* (\rho-\rho_{\rm Ai})(b_1) 
\end{equation}
From the mass conservation property of Eq.~\eqref{eq:mass_conservation} and using the additional representation of the Hilbert transform of Eq.~\eqref{eq:hilbert_log}, we rewrite the r.h.s of Eq. \eqref{eq:entropy_convolution} as 
\begin{equation}
\begin{split}
&\int_{-\infty}^{+\infty}\! \! \rmd b_1  \left[\rho(b_1)-\rho_{\rm Ai}(b_1)\right] \left[\log\abs{.}* (\rho-\rho_{\rm Airy})(b_1) -\log\abs{.}*(\rho-\rho_{\rm Ai})(\Xi) \right]\\
&=\pi \int_{-\infty}^{+\infty}\! \! \rmd b_1 \left[\rho(b_1)-\rho_{\rm Ai}(b_1)\right] \int_{\Xi}^{b_1}\rmd b'  H(\rho-\rho_{\rm Ai})(b')\\
\end{split}
\end{equation}
where $\Xi$ is arbitrary. Applying now the result of the Hilbert transform  $H(\rho-\rho_{\rm Ai})$ of Eq. \eqref{eq:hilbert_reparametrization} with the choice $\Xi=-\infty$, and applying the parametrization to the first term, $\rho(b_1)-\rho_{\rm Ai}(b_1)$, the electrostatic energy reads
\begin{equation}
\begin{split}
&\mathcal{J}(\rho)=\frac{\beta}{4\pi} \int_0^{+\infty} \rmd x\int_0^{+\infty}  \rmd x'\,\int_{-\infty}^{+\infty}\! \! \rmd b_1 \\
& \times \left[ \sqrt{(-b_1+ x + v(x))_+}-\sqrt{(-b_1+ x )_+}\right]\left[ \frac{1}{\sqrt{(b_1- x' - v(x'))_+}} -\frac{1}{\sqrt{(b_1- x' )_+}}\right]
\end{split}
\end{equation}
One successively applies Eq. \eqref{eq:SquareRootIdentity} to all cross-products to integrate w.r.t $b_1$. The resulting integral reads  
\begin{equation}
\begin{split}
&\mathcal{J}(\rho)= \frac{\beta}{8}\int_0^{+\infty} \rmd x\int_0^{+\infty}  \rmd x'\ \big[(x+v(x)-x'-v(x'))_++(x-x')_+ \\
&\hspace*{5cm}-(x+v(x)-x')_+ - (x-x'-v(x'))_+   \big]
\end{split}
\end{equation}

We first proceed to an integration by part on $x'$ and restrict as above to parametrization such that $x\mapsto x+v(x)$ is an increasing function of $x$. This leads to
\begin{equation}
\begin{split}
\mathcal{J}(\rho)&=\, \frac{\beta}{8}\int_0^{+\infty} \rmd x' \, x'(1+v'(x))\int_0^{+\infty}\rmd x \, \left[ \theta(x'+v(x')<x)-\theta(x'<x)\right]\\
&+\frac{\beta}{8}\int_0^{+\infty} \rmd x \, \int_0^{+\infty} \rmd x' \, x' \left[ \theta(x'<x+v(x))-\theta(x'<x) \right]\\
&=-\frac{\beta}{8}\int_0^{+\infty}\rmd x' \, x' (1+v'(x'))v(x')+\frac{\beta}{8}\int_0^{+\infty} \rmd x \left[ \frac{(x+v(x))^2}{2}-\frac{x^2}{2} \right]
\end{split}
\end{equation}
Grouping the various terms and performing a last integration by part on the $vv' $ term, the electrostatic energy finally reads
\begin{equation}
\mathcal{J}(\rho)=\frac{\beta}{8}\int_0^{+\infty}\rmd x \, v(x)^2
\end{equation}

\section{SAO/WKB to the Coulomb gas}
The optimal density $\rho_*$ for the variational problem associated to the Coulomb gas is the unique solution of the following equations \cite{JointLetter}
\begin{equation}
\label{eq:cg_pair}
\begin{split}
&\phi(u-b)- \beta \int_{-\infty}^{+\infty} \rmd b'\,\log \abs{b-b'} (\rho_*(b')-\rho_{\rm Ai}(b'))= c \qquad \text{for} \;\; b\geq u_0 \\
&\phi(u-b)- \beta \int_{-\infty}^{+\infty} \rmd b'\,\log \abs{b-b'} (\rho_*(b')-\rho_{\rm Ai}(b'))\geq c \qquad  \text{for} \;\; b\leq u_0
\end{split}
\end{equation}
for some constant $c$. We have anticipated here that the optimal density has a single support $[u_0,+\infty[$, which is valid for the class of functions $\phi$ considered here.  We now show that the equation {\sf SP2} of the Painlev\'e/WKB method Eq.~(24) of the Letter identifies with the pair of saddle point equations Eqs.~\eqref{eq:cg_pair} for the Coulomb-gas. In the course of the derivation, we also use properties of {\sf SP1} which is equivalent to {\sf SP2}. The equation {\sf SP2}, generalized to any $\beta$, reads
\bea
&& \phi(V)  = \beta  \int_{-\infty}^0 \rmd X' \sqrt{(V+ X' + 2 g(X'))_+}  - \beta\int_{-\infty}^0 \rmd X' \sqrt{(V+ X')_+} 
\eea
Upon the identification (Eq.~(26) in the Letter) $u-V=b$ and $X'=-u'$ and $\frac{ \beta \pi}{2} g(-u')= w(u')$
\begin{equation}
\label{eq:SP2}
\begin{split}
& \phi(u-b) = \beta  \int_0^{+\infty} \rmd u' \sqrt{(u-b-u' +\frac{4}{\beta \pi} w(u'))_+} - 
\beta \int_0^{+\infty} \rmd u' \sqrt{(u-b-u' )_+}  \\
& = \beta  \int_0^{\max(u,u-b+\frac{4}{\beta \pi}f(u-b))}\hspace*{-0.4cm} \rmd u' \sqrt{(u-b-u' +\frac{4}{\beta \pi} w(u'))_+} - \beta\int_0^{\max(u,u-b)}\hspace*{-0.4cm}  \rmd u' \sqrt{(u-b-u' )_+} ]  
\end{split}
\end{equation}
The upper bound in the second line can be obtained as follows. The domain of integration in the first line is
\begin{equation}
u-b\geq u'-\frac{4}{\beta \pi}w(u') \Leftrightarrow f(u-b)\geq w(u')
\end{equation}
using that $f$ in increasing. Therefore it implies that
\begin{equation}
u-b+\frac{4}{\beta \pi}f(u-b)\geq u'
\end{equation}
To discuss further upper bound we introduce $u_0:=\frac{4}{\beta \pi}w(u)\geq0$ which is the lower edge of the support of the optimal density $\rho_*$ as we show below. When $b=u_0$, $u-b+\frac{4}{\beta \pi}f(u-b)=u$ implying a crossover in the upper bound as $b \mapsto u-b+\frac{4}{\beta \pi}f(u-b)$ is decreasing in $b$.
\begin{itemize}
\item If $b\geq u_0$, Eq.~\eqref{eq:SP2} becomes
\end{itemize}
\begin{equation}
\begin{split}
 \phi(u-b) & = \beta   \int_0^{u} \rmd u' \left[ \sqrt{(u-b-u' +\frac{4}{\beta \pi} w(u'))_+} -  \sqrt{(u-b-u' )_+} \right]
\end{split}
\end{equation}
Using the correspondence of Eq.~(17) of the Letter, $v(x)=\frac{4}{\beta \pi} w(u-x)$ for $0\leq x \leq u$, this is equivalent to 
\begin{equation}
\begin{split}
 \phi(u-b) & = \beta  \int_0^{u} \rmd x \left[ \sqrt{(-b+x+v(x))_+} - \sqrt{(-b+x )_+} \right]\\
 &= \beta  \int_0^{+\infty} \rmd x \left[ \sqrt{(-b+x+v(x))_+} - \sqrt{(-b+x )_+} \right]
\end{split}
\end{equation}
where we have used the fact that $v(x)=0$ for $x\geq u$. Using the expression of the Hilbert transform Eq.~\eqref{eq:hilbert_reparametrization}, we prove the saddle point equation of the Coulomb-gas inside the support. Note that the derivative version of this equation leads to Eq.~(33) of the Letter more generally.
\begin{itemize}
\item If $b\leq u_0$, Eq.~\eqref{eq:SP2} becomes
\end{itemize}
\begin{equation}
 \phi(u-b) = \beta  \int_0^{u-b+\frac{4}{\beta \pi}f(u-b)} \hspace*{-0.4cm}\rmd u' \sqrt{(u-b-u' +\frac{4}{\beta \pi} w(u'))_+} - \beta\int_0^{\max(u,u-b)}\hspace*{-0.4cm} \rmd u' \sqrt{(u-b-u' )_+} 
\end{equation}
Using the correspondence, $v(x)=\frac{4}{\beta \pi}w(u-x)$ for $0\leq x \leq u$, this is equivalent to 
\begin{equation}
\begin{split}
& \phi(u-b)  = \beta \int_0^{u} \rmd x \left[ \sqrt{(-b+x+v(x))_+} -  \sqrt{(-b+x )_+}\right]\\
 & +\beta   \int_u^{u-b+\frac{4}{\beta \pi}f(u-b)} \hspace*{0.2cm}\rmd u' \sqrt{(u-b-u' +\frac{4}{\beta \pi} w(u'))_+}- \beta \int_u^{\max(u,u-b)}\hspace*{-0.4cm} \rmd u' \sqrt{(u-b-u' )_+} 
\end{split}
\end{equation}
It is easy to see that the the sum of the terms on the second line is always positive. If $b\geq 0$, then the very last term is zero, and if $b \leq 0$, we can split the first term up to $u-b$ and use the fact that $w(u')\geq 0$. It implies that 
\begin{equation}
\phi(u-b)- \beta  \int_0^{+\infty} \rmd x\left[  \sqrt{(-b+x+v(x))_+} - \sqrt{(-b+x )_+}\right]\geq 0
\end{equation}
which is an equality for $b\in [u_0,+\infty[$, i.e. $b$ in the support of the optimal density $\rho^*$. Integrating the Hilbert transform of Eq.~\eqref{eq:hilbert_reparametrization} with respect to $b$, we have
\begin{equation}
\label{eq:CG_saddle}
 \int_0^{+\infty} \rmd x \left[\sqrt{(-b+x+v(x))_+} - \sqrt{(-b+x )_+}\right]= \int_{-\infty}^{+\infty} \rmd b'\,\log \abs{b-b'} (\rho_*(b')-\rho_{\rm Ai}(b'))+C
\end{equation}
where $C$ is an integration constant which we now show to be 0. Indeed, the left hand side of Eq.~\eqref{eq:CG_saddle} vanishes for $b\to + \infty$. Furthermore using the mass conservation, we have
\begin{equation}
\begin{split}
\int_{0}^{+\infty} \rmd b'\,\log \abs{b-b'} (\rho_*(b')-\rho_{\rm Ai}(b'))&=\int_{0}^{+\infty} \rmd b'\,\log \abs{1-\frac{b'}{b}} (\rho_*(b')-\rho_{\rm Ai}(b'))\\
&=b\int_{0}^{+\infty} \rmd z\,\log \abs{1-z} (\rho_*(bz)-\rho_{\rm Ai}(bz))\\
&\underset{b\to +\infty}{\simeq}b \int_0^{+\infty}\rmd z \, \log \abs{1-z} \frac{A}{(bz)^{3/2}}=o(\frac{1}{\sqrt{b}})
\end{split}
\end{equation}
where we used $\int_0^{+\infty}\rmd z \, \log\abs{1-z} z^{-3/2}=0$ and the fact that 
\begin{equation}
\label{eq:diff_density}
\begin{split}
\rho_*(b)-\rho_{\rm Ai}(b)&=\frac{1}{2\pi}\int_0^u \rmd x \ \left[ \frac{1}{\sqrt{(b-x-v_*(x)})_+}-\frac{1}{\sqrt{(b-x)_+}}\right]\\
&= \frac{1}{2\pi}\int_0^u\rmd x \, \frac{v_*(x)}{\sqrt{(b-x-v_*(x)})_+\sqrt{(b-x)_+}(\sqrt{(b-x-v_*(x)})_++\sqrt{(b-x)_+})}\\
&\leq  \frac{1}{2\pi (b-u)^{3/2}_+}\int_0^u\rmd x \,v_*(x) \underset{b\to +\infty}{\sim} \frac{A}{b^{3/2}}
\end{split}
\end{equation}
Hence for all real $b$, we have the inequality
\begin{equation}
\phi(u-b)- \beta \int_{-\infty}^{+\infty} \rmd b'\,\log \abs{b-b'} (\rho_*(b')-\rho_{\rm Ai}(b'))\geq 0
\end{equation}
which turns to be an equality in the support of the optimal density $\rho_*$, i.e. $b\in [u_0,+\infty[$ and therefore the saddle point {\sf SP2} identifies with the variational equation of the Coulomb-gas provided that {\sf SP1} holds. 

\section{Optimal density : SAO/WKB and electrostatic Coulomb gas methods  }
We now derive an explicit formula for the optimal density that minimizes both the SAO/WKB and the Coulomb-gas functionals. Let us start from the expression of the density of the SAO/WKB, using the correspondence $v_*(x)=\frac{4}{\beta \pi}w(u-x)$ for $0\leq x \leq u$ and the fact that $v_*(x)=0$ for $x\geq u$
\be
\rho_*(b)= \frac{1}{2\pi}  \int_0^{+\infty} 
 \frac{\mathrm{d}x}{\sqrt{(b - x -  v_*(x))_+}}  = \frac{1}{2\pi} \int_0^{u} 
 \frac{\mathrm{d}u'}{\sqrt{(b - u + u' - \frac{4}{\beta \pi} w(u') )_+}} + \frac{ \sqrt{(b - u )_+}}{ \pi} 
\ee
Using the saddle point {\sf SP1}, we have $ u' - \frac{4}{\beta \pi} w(u')=f^{-1}(w(u'))$ and we use $w'=w(u')$ as the variable of integration leading to the first formula for the density (Eq.(34) in the Letter)
\be 
\begin{split}
 \rho_*(b) &=  \frac{\sqrt{ (b - u )_+} }{ \pi} +\frac{1}{2\pi} \int_0^{w(u)} 
\mathrm{d}w'  \frac{ [f^{-1}]'(w')+ \frac{4}{\beta \pi}}{
\sqrt{(b - u + f^{-1}(w') )_+}}  \\
& = \frac{\sqrt{(b -\frac{4}{\beta \pi} w(u) )_+  }}{ \pi}  
+ \frac{2}{\beta \pi^2} \int_0^{w(u)} 
  \frac{\mathrm{d}w'}{
\sqrt{(b - u + f^{-1}(w') )_+}  } 
\end{split}
\ee
where from the first line to the second line we explicitly integrated the term involving $[f^{-1}]'(w')$. To treat the last term, we proceed to the change of variable $z=f^{-1}(w')$ so that 
\bea 
\label{eq:density_f}
&& \rho_*(b) -  \frac{\sqrt{(b -\frac{4}{\beta \pi} w(u) )_+  }}{ \pi}  =
 \frac{2}{\beta \pi^2} \int_0^{u -\frac{4}{\beta \pi} w(u)} \rmd z \frac{f'(z)}{\sqrt{(b - u + z)_+} } \nn \\
&& =  \frac{1}{\beta \pi^2} \int_{-\infty}^{+\infty} \rmd b' \int_{\max(b',u-b)}^{u -\frac{4}{\beta \pi}  w(u)} \rmd z 
\frac{ \phi''(b')}{\sqrt{(b - u + z)_+(z-b')_+} } \label{89} 
\eea
We have used the definition of the function of $f$ as a convolution, i.e. $f=\frac{1}{2} \frac{1}{\sqrt{(b)_+}} * \phi'$ to go from the first line to the second line. We note that the integral over $z$ vanishes if $b \leq u_0=\frac{4}{\beta \pi }w(u)$ which is precisely the lower edge of the support of $\rho_*$. The integration over $z$ can be done explicitly, leading to
\begin{equation}
\begin{split}
 \rho_*(b) -  &\frac{\sqrt{(b -\frac{4}{\beta \pi} w(u) )_+  }}{ \pi}   = \frac{1}{\beta \pi^2} \int_{-\infty}^{+\infty} \rmd b' \phi''(b')
\theta(u -\frac{4}{\beta \pi} w(u) - b') \theta(b -\frac{4}{\beta \pi} w(u)) \\
& \times \log \abs{ \frac{u -\frac{8}{\beta \pi}   w(u) + b -b' + 2 \sqrt{ (b -\frac{4}{\beta \pi} w(u)) (u-\frac{4}{\beta \pi} w(u) - b')}
}
{b + b' - u} }\\
& = \frac{1}{\beta \pi^2} \dashint_{-\infty}^{+\infty}\frac{ \rmd b'}{b + b' - u } \phi'(b')
\frac{ \sqrt{ ( b-\frac{4}{\beta \pi} w(u) )_+} }{\sqrt{ (u -\frac{4}{\beta \pi} w(u) - b')_+}} 
\end{split}
\end{equation}
The second line is obtained by an integration by part, which has no boundary term. Finally, the optimal density can be factorized as
\begin{equation} \label{dens} 
\rho_*(b)=\frac{\sqrt{(b-\frac{4}{\beta \pi}w(u))_+}}{\pi}\left[1+\frac{1}{\beta \pi}\dashint_{-\infty}^{+\infty}\frac{\rmd b'}{b+b'-u}\frac{ \phi'(b')}{\sqrt{(u-\frac{4}{\beta \pi}w(u)-b')_+}}\right]
\end{equation}
which shows that for fixed $u$, in the large $b$ limit, one recovers the density of the Airy process. Upon identification of the edge of the support as $u_0=\frac{4}{\beta \pi}w(u)$, this leads to the second formula for the density (Eq.(34) in the Letter).

\subsection{Deviation of the optimal density from the Airy density}
We have shown in Eq.~\eqref{eq:diff_density} that for large argument, the optimal density $\rho_*$ is close to the Airy density $\rho_{\rm Ai}$, i.e.
\begin{equation}
\rho_*(b)-\rho_{\rm Ai}(b)\underset{b\gg 1}{=} \mathcal{O}(\frac{1}{b^{3/2}})
\end{equation}
As both densities behave asymptotically as the semi-circle, i.e. $\rho(b)=\mathcal{O}(\sqrt{b})$, it is not straightforward to see that the difference between the optimal density and the Airy one is of order $1/b^{3/2}$ and not $1/b^{1/2}$. We now show that this is a consequence of the first saddle point {\sf SP1}. Indeed, using the edge notation $u_0=\frac{4}{\beta \pi}w(u)$ and	starting from Eq.~\eqref{eq:density_f} with $b\geq u \geq u_0$, 
\begin{equation}
\rho_*(b)-\rho_{\rm Ai}(b)=  \frac{\sqrt{b -u_0   }-\sqrt{b}}{ \pi}  +
 \frac{2}{\beta \pi^2} \int_0^{u -u_0} \rmd z \frac{f'(z)}{\sqrt{b - u + z} } 
\end{equation}
We now Taylor expand both terms on the right hand side for large $b$. 
The second term reads
\begin{equation}
 \frac{2}{\beta \pi^2}\int_0^{u -u_0} \rmd z \frac{f'(z)}{\sqrt{b - u + z} } =\frac{2 f(u-u_0)}{\beta \pi^2 b^{1/2}}+\frac{1}{\beta \pi^2 b^{3/2}}\int_0^{u -u_0} \rmd z f'(z)(u-z)+\mathcal{O}(\frac{1}{b^{5/2}})
\end{equation}
Adding the contribution of the first term we find 
\begin{equation}
\rho_*(b)-\rho_{\rm Ai}(b)=\frac{2\big[ f(u-u_0)-\frac{\beta \pi }{4}u_0\big]}{\beta \pi^2 b^{1/2}}+\frac{1}{\beta \pi^2 b^{3/2}}\big[\int_0^{u -u_0} \rmd z f'(z)(u-z)- \frac{\beta \pi u_0^2}{8} \big]+\mathcal{O}(\frac{1}{b^{5/2}})
\end{equation}
The term of order $1/b^{1/2}$ is exactly the saddle point {\sf SP1} defining the edge of the support $u_0$ and is therefore zero. Hence, at large $b$, the deviation of $\rho_*$  from $\rho_{\rm Ai}$ is only of order $1/b^{3/2}$.

\section{Calculation of the PDF of {\sf L}  }
We study $P({\sf L})$ the PDF of ${\sf L} = t \sum_i \phi(u + t^{-2/3} a_i)$. Since the mean value over the APP is of order $t^2$ at large $t$, more precisely 
\be \label{firstcum} 
\mathbb{E}_\beta[{\sf L}] = - t^2 \tilde \kappa_1(u) \simeq  t^2 \frac{1}{\pi} \int_0^{+\infty} \rmd b \sqrt{b} ~\phi(u-b) 
\ee 
e.g. see Eqs.~\eqref{kappa12} and \eqref{cumL1}, we anticipate that
the PDF takes the large deviation form
\be
P({\sf L}) \sim e^{- t^2 \tilde G(\tilde L) } \quad , \quad \tilde L = {\sf L}/t^2
\ee
Introducing a parameter $B$ we see that the following average is dominated
by a saddle point
\be
\mathbb{E}_\beta [ e^{- B {\sf L} } ] = \int \rmd {\sf L} \,  P({\sf L} ) e^{- t^2 [ \tilde G(\tilde L) + B \tilde L]} 
\sim \exp\big(- t^2 \min_{\tilde L} \big[ \tilde G(\tilde L) + B \tilde L\big]\big)
\ee
Since the l.h.s. corresponds to the linear statistics problem for $\phi \to B \phi$, we
see that $G$ and $\Sigma_\phi(u)$ are related by the following 
Legendre transform $\Sigma_{B \phi}(u) = \min_{\tilde L} \left[ \tilde G(\tilde L) + B \tilde L \right]$
which can be inverted as
\be \label{s2}
\tilde G(\tilde L) = \max_{B} \left[ \Sigma_{B \phi}(u) - B \tilde L \right] 
\ee
We thus have the pair of equations relating $\tilde L$ and $B$
at the optimum 
\be
\tilde G'(\tilde L) = - B  \quad , \quad \partial_B \Sigma_{B \phi}(u) = \tilde L  
\ee
The most probable value $\tilde L=\tilde L_{\rm typ}$, which satisfies 
by definition $\tilde G'(\tilde L_{\rm typ})=0$ corresponds to $B=0$. The second 
equation shows that it equals the first cumulant 
\be
\tilde L_{\rm typ} = \mathbb{E}_\beta [ L ]/t^2 
\ee
as given by Eq.~\eqref{firstcum}, since the $\mathcal{O}(B^n)$ term in the expansion at small $B$ is 
given by the $n$-th cumulant $\kappa_n$, see Eqs.~\eqref{gener},
\eqref{cumL1} and \eqref{kappa12}. It is thus convenient to define, as in the Letter,
the dimensionless ratio $\ell = \tilde L/\tilde L_{\rm typ}$ and $\tilde G(\tilde L)=G(\ell)$, which is thus given by
\be
G(\ell) = \max_{B} \left[ \Sigma_{B \phi}(u) - A B \ell \right] 
\ee 
where $A=\tilde L_{\rm typ}=\partial_B \Sigma_{B \phi}(u)|_{B=0}$. Using the cumulant expansion \eqref{gener} we see that around the most 
probable value
\be \label{probaexpand} 
 \tilde G(\tilde L) \simeq \frac{(\tilde L- \tilde L_{typ})^2}{2 \tilde \kappa_2(u)} \quad , \quad 
 G(\ell) \simeq \frac{(\ell - 1)^2}{2 \sigma} \quad , \quad 
\sigma = \frac{\tilde \kappa_2(u)}{\tilde \kappa_1(u)^2}
\ee 
where $\sigma$ is the dimensionless ratio formed with the
first cumulant \eqref{firstcum} and the second,
$\tilde \kappa_2(u)=\kappa_2/t^2 = \frac{4}{\beta} \int_0^u \rmd u' f(u')^2$
from \eqref{kappa12}.\\

We then note that $B>0$ corresponds to $\ell\leq 1$ while $B<0$ corresponds to $\ell>1$.
We thus give here only the PDF for $\ell \leq 1$. 
To treat the case $B<0$ requires to extend the methods of the present Letter.
We know from the study of Ref.~\cite{grabsch2017truncated} that in the bulk it leads to a distinct
Coulomb gas phase, with a splitted support for the optimal density: this is likely to carry to the edge and we leave its study to future work.

\section{General scaling dependence in $\beta$  }

It is easy to see, e.g from the Coulomb gas formulation Eq. (29) in the Letter, 
that for any function $\phi$
\be \label{genbetasigma}
\Sigma^{(\beta)}_{\frac{\beta}{2} \phi}(u) = \frac{\beta}{2} \Sigma^{(\beta=2)}_{\phi}(u) 
\ee 
where the dependence in $\beta$ was made explicit. The optimal density then is the same
\be \label{genbetarho}
\rho^{(\beta)}_{*,\frac{\beta}{2} \phi}(b) = \rho^{(2)}_{*,\phi}(b)
\ee
As a result setting $B=B' \frac{\beta}{2}$ in
\eqref{s2} we also have
\be \label{genbetaG}
\tilde G^{(\beta)}(\tilde L) = \frac{\beta}{2} \max_{B'} \left[ \Sigma^{(2)}_{B' \phi}(u) - B' \tilde L \right] 
= \frac{\beta}{2} \tilde G^{(\beta=2)}(\tilde L) 
\ee
and since $\tilde L_{typ}$ does not depend on $\beta$ the same holds for
$G(\ell)$, i.e. $G^{(\beta)}(\ell) =  \frac{\beta}{2}  G^{(\beta=2)}(\ell)$.
\section{Bounds on the large deviation rate function  }
\subsection{Jensen's inequality : first cumulant upper bound}
The Jensen's inequality states that $\mathbb{E}_\beta[ e^{- L } ] \geq e^{- \mathbb{E}_\beta[ L ] }$ which provides an upper bound for $\Sigma_\phi(u)$ valid for any $\phi$
\begin{equation} 
\Sigma_\phi(u)\leq  \int_0^{+\infty} \rmd b \frac{\sqrt{b}}{\pi}\phi(u-b) \label{b1} 
\end{equation}
\subsection{Bound on the comparison of linear statistics}
We compare the linear statistics involving two functions $\phi_1$ and $\phi_2$ such that $\phi_1\leq \phi_2$. Then for all $u\geq 0$,
\begin{equation}
\label{eq:sigma_comparison}
\mathbb{E}_{\beta}\left[ \prod_{i=1}^{+\infty} e^{- t \phi_2(u+t^{-2/3} a_i) }  \right]\leq \mathbb{E}_{\beta}\left[ \prod_{i=1}^{+\infty} e^{- t \phi_1(u+t^{-2/3} a_i) }  \right]
\end{equation}
In particular, this allows to compare the excess energies of both problems as 
\begin{equation}
\forall u\geq 0, \; \Sigma_{\phi_1}(u)\leq  \Sigma_{\phi_2}(u)
\end{equation}
\subsection{Upper bound from the Tracy-Widom large deviations}
Here we assume $\phi(z) \geq 0$ and $\phi(z\leq 0)=0$ and we compare the linear statistics to the function $\phi$ to the hard wall case. We also define a function $\phi_{\rm HW}$ as $\phi_{\rm HW}(z\leq 0)=0$ and $\phi_{\rm HW}(z\geq 0)=+\infty$. By construction, $\phi\leq \phi_{\rm HW}$, which leads to, using Eq.~\eqref{eq:sigma_comparison}, 
\begin{equation}
\mathbb{E}_{\beta}\left[ \prod_{i=1}^{+\infty} e^{- t \phi_{\rm HW}(u+t^{-2/3} a_i) }  \right]\leq \mathbb{E}_{\beta}\left[ \prod_{i=1}^{+\infty} e^{- t \phi(u+t^{-2/3} a_i) }  \right]
\end{equation}
Denoting $a_{\max}= \max_i \lbrace a_i \rbrace$, the left hand side of this equality gets rewritten as 
\begin{equation}
\mathbb{E}_{\beta}\left[ \prod_{i=1}^{+\infty} e^{- t \phi_{\rm HW}(u+t^{-2/3} a_i) }  \right]=\mathbb{E}_{\beta}\left[ \prod_{i=1}^{+\infty} \theta(u+t^{-2/3} a_i)   \right]=\mathbb{P}\left( a_{\rm max} < - ut^{2/3}\right)
\end{equation}
Finally, gathering both results leads to the inequality
\begin{equation}
\mathbb{P}\left( a_{\rm max} < - ut^{2/3}\right)\leq \mathbb{E}_{\beta}\left[ \prod_{i=1}^{+\infty} e^{- t \phi(u+t^{-2/3} a_i) }  \right]
\end{equation}
Using the standard result for the large deviations of the largest eigenvalue of the
$\beta$-ensemble (i.e. from Tracy Widom for $\beta=1,2,4$) leads to a second upper bound for the excess energy
\begin{equation}
\Sigma_\phi(u)\leq \frac{\beta}{24}u^3\label{b2} 
\end{equation}
The equality is saturated by the hard wall, i.e. if one multiplies $\phi$ by an amplitude $B$, then $\lim_{B \to +\infty} \Sigma_{B \phi}(u) =  \frac{\beta}{24}u^3$. We call this bound the Tracy-Widom (hard wall) bound.

\section{$\Sigma_\phi(u)$ for the case of the monomial walls $\phi(z)=  (z)_+^\gamma$  } 

We consider the monomial walls $\phi(z)=  (z)_+^\gamma$
as well as the problem $\phi\to B\phi$ with a positive amplitude $B$. 
Let us first give the associated function $f(u)$ associated to $\phi$, using the definition (14) in the Letter
we obtain
\be \label{fu} 
f(u) = C_\gamma (u)_+^{\gamma-\frac{1}{2}}  
\quad , \quad C_\gamma = \frac{\sqrt{\pi}}{2} \frac{\Gamma(\gamma+1)}{\Gamma(\gamma+ \frac{1}{2})} ~
\ee 
Hence we see that $\phi$ in $\Omega_0$ only for $\gamma > 1/2$, the case
to which we restrict here. Before giving more explicit formula 
let us discuss some general properties.

\subsection{Consequence of the bounds}

Gathering the two bounds of Eqs.~\eqref{b1} and \eqref{b2} 
brings a stronger constraint on the large deviation function $\Sigma_\phi(u)$, indeed we find
\begin{equation} \label{bound} 
\Sigma_{B\phi}(u)\leq \min\left(  \frac{\beta}{24}u^3 , \frac{\Gamma(\gamma+1)B}{\sqrt{4\pi}\Gamma(\frac{5}{2}+\gamma)}u^{\gamma+\frac{3}{2}} \right)
\end{equation}
This implies that :

\textit{(i)} for $\gamma<3/2$ the large $u$ behavior is smaller or equal to $u^{\gamma+ \frac{3}{2}}$,
hence a $u^3$ behavior is impossible for large $u$,

\textit{(ii)} for $\gamma>3/2$ the small $u$ behavior is smaller or equal to $u^{\gamma+ \frac{3}{2}}$,
hence a $u^3$ behavior is impossible for small $u$.

\subsection{Scaling of $\Sigma_\phi(u)$ with the amplitude of the soft walls and
with the Dyson index $\beta$}

We show that for $\phi(z)= (z)_+^{\gamma}$, with 
$\gamma > 1/2$ and $\gamma \neq 3/2$ we have the scaling law
\be \label{scaling1} 
\Sigma_{B\phi}(u) = B^{\frac{6}{3-2 \gamma}}  \Sigma_{\phi}(u B^{\frac{2}{2 \gamma-3}})
\ee 
where $B \phi$ is the function $z \mapsto B \phi(z)$. Similarly, indicating
explicitly the dependence in $\beta$
\be \label{scaling2} 
\Sigma^{(\beta)}_{\phi}(u) = (\frac{2}{\beta})^{\frac{3 + 2 \gamma}{3-2 \gamma}}   
\Sigma^{(2)}_{\phi}(u (\frac{\beta}{2})^{\frac{2}{3-2 \gamma}})
\ee 

\begin{proof}
Consider the
saddle point equation {\sf SP1}
\be 
 \frac{\beta}{4} v_*(x) = \frac{B}{2 \pi} \int_{-\infty}^{+\infty} \frac{\mathrm{d}b}{\sqrt{(b)_+}} \phi'(u-b-x - v_*(x)) 
\ee
We define $u=r \tilde u$, $x=r \tilde x$, $b = r \tilde b$ and $v_* = r \tilde v_*$.
In these variables the problem corresponds to $\tilde B = \frac{2}{\beta} B r^{\gamma-3/2}$, 
and $\beta=2$ i.e
$\tilde B=1$ if we choose $r^{3/2-\gamma}=B$. For $\gamma \neq 3/2$ one
can always choose $r = B^{2/(3-2 \gamma)}$. Now using the form of 
$\Sigma_{B\phi}(u)$ valid for $\phi$ in $\Omega_0$ we obtain
\bea \label{form} 
\Sigma_{B\phi}(u) =\frac{\beta}{4} \int_0^{+\infty} \rmd x\,  x v_*(x) 
=
 r^3 \Sigma_{\phi}(\frac{u}{r}) = B^{\frac{6}{3-2 \gamma}}  
 \Sigma_{\phi}(u B^{\frac{2}{2 \gamma-3}})
\eea
\end{proof}
To obtain \eqref{scaling1} we follow the same method choosing
$r= (\frac{2}{\beta})^{2/(3-2 \gamma)}$ and keep track of $\beta$ in formula \eqref{form}. 

\subsection{Consequence of the scaling: saturation of the Tracy-Widom bound and transition
at $\gamma=3/2$}

As discussed above the limit $B \to +\infty$ corresponds to a hardwall and therefore 
leads to the Tracy-Widom left large deviation result $\Sigma_{B\phi}(u) = \frac{\beta}{24}u^3$.
Combining this with the scaling form \eqref{scaling1} we deduce that the cubic Tracy Widom
upper bound is saturated in the following cases
\begin{itemize}
\item For $\gamma<3/2$, we see the cubic behavior arises from the small $u$ behavior
\be
\Sigma_\phi(u) \underset{u \to 0}{=}  \frac{\beta}{24} u^3+ o(u^3) 
\ee
\item 
For $\gamma > 3/2$, the cubic behavior arises from the large $u$ behavior 
\be
\Sigma_\phi(u) \underset{u \to +\infty}{=}  \frac{\beta}{24}u^3 + o(u^3) 
\ee
\item At the transition value $\gamma=3/2$, the saddle point equation {\sf SP1}
admits a very simple solution.
Consider the problem with an amplitude $B$, $\phi\to B\phi$. As $f(u) = \frac{3 \pi}{8} B (u)_+$, equation {\sf SP1} reads for $x \geq 0$
\be
\frac{\beta}{4} v_*(x) = \frac{3}{8} B (u - x - v_*(x))_+
\ee
whose solution is 
\begin{equation} 
v_*(x)=\frac{3B}{3B+2\beta}(u-x)_+
\end{equation}
Hence using Eq. (12) of the Letter, the large deviation function reads
\be \label{res32} 
\Sigma_{B\phi}(u) =\frac{\beta}{24}\frac{3B}{3B+2\beta} u^3
\ee
which, remarkably, is a simple cubic for any $u$, although the coefficient
depends continuously on $\beta$ and $B$ and saturates the Tracy-Widom bound for $B\to +\infty$

\end{itemize}

\subsection{Explicit solution for general $\gamma$: series expansion}

We now present the solution for general $\gamma$. Let us start with the series
expansion representation of $\Sigma_\phi(u)$ obtained from the cumulant method, as given 
in Ref.~\cite{krajenbrink2018systematic}, obtained by inserting the expression \eqref{fu} of $f(u)$ in
Eq.~(13) of the Letter
 \begin{equation}
\begin{split}
\Sigma_\phi(u)& = -\frac{\beta}{4} \sum_{n \geq 1} \frac{(-1)^n}{n!} 
\left(\frac{2}{\beta \sqrt{\pi}} \frac{\Gamma(\gamma+1)}{\Gamma(\frac{1}{2}+\gamma)}\right)^n \partial_u^{n-3} u^{n(\gamma-\frac{1}{2})}
\end{split}
\end{equation}
Explicitly performing the derivative, this reads
\begin{equation} \label{ser2} 
\Sigma_\phi(u) =-\frac{\beta}{4} \sum_{n \geq 1} \frac{(-1)^n}{n!} 
\left(\frac{2}{\beta \sqrt{\pi}} \frac{\Gamma(\gamma+1)}{\Gamma(\frac{1}{2}+\gamma)}\right)^n\frac{\Gamma(n(\gamma-\frac{1}{2})+1)}{\Gamma(4-n(\frac{3}{2}-\gamma))} u^{3-n(\frac{3}{2}-\gamma)} 
\end{equation}
We observe that:
\begin{itemize}
\item For $\gamma<3/2$ this is a series expansion in $1/u$ around large $u$, 
starting with the $n=1$ first cumulant term
\be \label{1cum} 
\Sigma_\phi(u) = \frac{\Gamma(\gamma+1)}{\sqrt{4\pi}\Gamma(\frac{5}{2}+\gamma)} u^{\gamma+\frac{3}{2}} 
+ \dots 
\ee 
which saturates at large $u$ the bound \eqref{b1}, i.e. the second term
in the r.h.s. of \eqref{bound}. 
\item For $\gamma>3/2$ this is a series expansion in $u$ around small $u$
starting with the same $n=1$ first cumulant term \eqref{1cum},
thus saturating now the bound \eqref{b1} at small $u$. 
\end{itemize}

Combining with our previous observations we thus see that for any $\gamma>1/2$
both bounds are saturated at small and large $u$, although they are interchanged
as $\gamma$ crosses $3/2$.

\subsection{Explicit solution for general $\gamma$: saddle point equation}

Following the Letter we want to solve for $w=w(u)$
\be
f(u - \frac{4}{\beta \pi} w(u)) = w(u) 
\ee
where $f(u)$ is given in \eqref{fu}.
We obtain a trinomial algebraic equation for $w$, and we must retain only the positive root (which vanishes for $u=0$)
\be \label{weq} 
u= \frac{4}{\beta \pi} w + (\frac{w}{C_\gamma})^{\frac{1}{\gamma-1/2}}
\ee
From the middle term in Eq.~\eqref{eq:triple_int} with $a=-\frac{4}{\beta \pi}$ we obtain
\bea
\Sigma_{\phi}(u)=
\frac{1}{2 \pi} \int_0^{w(u)} \rmd w' (u(w')-u)^2 
= \frac{1}{2 \pi} \int_0^{w(u)} \rmd w' 
(\frac{4}{\beta \pi} w' + (\frac{w'}{C_\gamma})^{\frac{1}{\gamma-1/2}}-u)^2
\eea
Performing the integral and replacing all $(\frac{w}{C_\gamma})^{\frac{2}{2\gamma-1}}$ factors 
by $u-\frac{4}{\beta\pi } w $, we finally obtain 
\be \label{ressigma}
\Sigma_\phi(u)= \frac{4 u^2 w(u)}{\pi  (2 \gamma +1) (2 \gamma +3)}+\frac{(2 \gamma -3) (6 \gamma +1) u w(u)^2}{\pi
   ^2 \beta  \gamma  (2 \gamma +1) (2 \gamma +3)}+\frac{4 (2 \gamma -3)^2 w(u)^3}{3 \pi ^3 \beta
   ^2 \gamma  (2 \gamma +3)}
\ee
where $w(u)$ is the unique positive solution of Eq.~\eqref{weq}. This is the result quoted in the Letter in Eq. (36)\\

For $\gamma<3/2$ we see from \eqref{weq} that at small $u$ we have 
$w \simeq \frac{\beta \pi}{4} u$, which inserted in Eq.~\eqref{ressigma} recovers  $\Sigma_\phi(u) \simeq  \frac{\beta}{24}u^3$. For large $u$, 
$w(u) \simeq C_\gamma u^{\gamma-1/2}$ and the first term in 
\eqref{ressigma} dominates, recovering the first cumulant (which is also
a bound) given in \eqref{b1}. The same holds for $\gamma>3/2$ with the role
of large and small $u$ inverted. For $\gamma=3/2$ the two last terms
vanish and using $u=(\frac{4}{\beta \pi} + \frac{8}{3 \pi}) w$ one recovers
$\Sigma_\phi(u)  = \frac{\beta u^3}{8(3+ 2\beta)}$ which is the
result obtained above in \eqref{res32}.\\

In Table~\ref{table:sigma_example} we give a few examples of closed analytic forms which
can be obtained for some values of $\gamma$. We have checked the positivity and convexity of the above expressions.
These have been obtained
by summing the cumulant series using Mathematica. In some cases (e.g. $\gamma=5/2$)
the same result (in a different, though equivalent form) can be obtained
by solving the trinomial equation. In general, for $\gamma=\frac{1}{2} + \frac{1}{p}$ with
positive integer $p$ the solutions of the trinomial equation can be expressed
using hypergeometric functions, see Ref.~\cite{perelomov2004hypergeometric}. 

\begin{table}[h!]
\begin{center}
\hspace*{-3.45cm}
\begin{tabular}{|c||c||c||c|}
\hline 
$\gamma$& $\Sigma_\phi(u)$ & Around $u=0^+$ & Around $u=+\infty$ \\ 
\hline 
\hline 
&&&\\[-2ex]
$1$ & $\dfrac{4}{15\pi^6}(1+\pi^2 u)^{5/2}-\dfrac{4}{15\pi^6}-\dfrac{2}{3\pi^4}u -\dfrac{1}{2\pi^2}u^2$  &  $\dfrac{u^3}{12}-\dfrac{\pi^2 u^4}{96} + \mathcal{O}(u^5) $&$\dfrac{4 u^{5/2}}{15 \pi }-\dfrac{u^2}{2 \pi ^2}+\mathcal{O}(u^{3/2})$\\ [2ex]
\hline 
&&&\\[-2ex]
$\dfrac{3}{2} $ & $\dfrac{u^3}{28}$ & $\dfrac{u^3}{28}$  & $\dfrac{u^3}{28}$ \\  [2ex]
\hline 
&&&\\[-2ex]
$2$ & \makecell{\hspace*{-2cm}$\dfrac{16 u^{7/2} }{105 \pi } \, _2F_1\left(\dfrac{5}{6},\dfrac{7}{6};\dfrac{9}{2};\dfrac{48 u}{\pi ^2}\right)$\\ \vspace*{-0.2cm}\\\hspace*{0.3cm}
$+\dfrac{81 \pi ^6 }{29360128} \,  _2F_1\left(-\dfrac{8}{3},-\dfrac{7}{3};-\dfrac{5}{2};\dfrac{48 u}{\pi ^2}\right)$\\\vspace*{-0.2cm}\\  \hspace*{1cm}$\dfrac{u^3}{12}-\dfrac{3 \pi ^2
   u^2}{256}+\dfrac{27 \pi ^4 u}{81920}-\dfrac{81 \pi ^6}{29360128}$} & $\dfrac{16u^{7/2}}{105\pi}-\dfrac{4 u^4}{9\pi^2}+\mathcal{O}(u^{9/2})$&$\dfrac{u^3}{12}-\dfrac{9 (3\pi)^{2/3} u^{8/3}}{320}+\mathcal{O}(u^{7/3})$\\ [2ex]
   &&&\\[-2ex]
\hline 
&&&\\[-2ex]
$\dfrac{5}{2}$ & $\dfrac{u^3}{12}+\dfrac{2 u^2}{15}+\dfrac{32 u}{675}-\dfrac{8 (4+15u)^{5/2}}{50625}+\dfrac{256}{50625}$ & $\dfrac{5u^4 }{128} - \dfrac{45 u^5}{1024} +\mathcal{O}(u^6)	 $&$\dfrac{u^3}{12}-\dfrac{8 u^{5/2}}{15 \sqrt{15}}+\mathcal{O}(u^2)$ \\  [2ex]
\hline
&&&\\[-2ex]
$\dfrac{7}{2} $ &$\dfrac{8u}{105}  \, _2F_1\left(-\dfrac{2}{3},-\dfrac{1}{3};\dfrac{3}{2};-\dfrac{945 u^2}{128}\right)+\dfrac{u^3}{12}-\dfrac{8 u}{105}$  & $ \dfrac{7u^5}{256}-\dfrac{175u^7}{4096}  +\mathcal{O}(u^{9})$&$\dfrac{u^3}{12}-\dfrac{9 u^{7/3}}{14 \times 70^{1/3}}+\mathcal{O}(u^{5/3})$\\ [3ex]
\hline 
\end{tabular} 
\hspace*{-2.5cm}
\end{center}
\caption{Excess energies $\Sigma_\phi(u)$ for $\beta=2$ for different values of $\gamma$ for $u>0$ with the two first orders of their expansion around $u=0^+$ and $u=+\infty$. The result for other values of $\beta$ can be obtained using the scaling law of Eq.~\eqref{scaling2}.}
\label{table:sigma_example}
\end{table}

\section{Optimal density for the case of the monomial walls $\phi(z)=  (z)_+^\gamma$  }
\subsection{Scaling property} 
Let us first give a general scaling property for $\phi(z)= (z)_+^{\gamma}$, with 
$\gamma > 1/2$ and $\gamma \neq 3/2$ and consider the problem with a positive amplitude $B$ such that $\phi\to B \phi$. Using the same method as
above we now obtain the following scaling properties for the optimal density
with respect to an amplitude $B$ (making the dependence in $B$ and $u$ apparent)
\be \label{scaling1rho} 
\rho_{B,u}(b) = B^{\frac{1}{3-2 \gamma}}  \rho_{1,u B^{\frac{2}{2 \gamma-3}}}(b B^{\frac{2}{2 \gamma-3}}) 
\ee 
where $\rho_{B,u}(b)$ denotes here the optimal density associated to $B \phi$ for a parameter $u$. 
Similarly, indicating
explicitly the dependence in $\beta$ we have
\be \label{scaling2rho} 
\rho^{(\beta)}_u(b) = (\frac{2}{\beta})^{\frac{1}{3-2 \gamma}}  
\rho^{(1)}_{u (\frac{2}{\beta})^{\frac{2}{2 \gamma-3}}} (b (\frac{2}{\beta})^{\frac{2}{2 \gamma-3}}) 
\ee 
From now on, in this Section we restrict to $B=1$.
\subsection{Support of the density}
\begin{figure}[h!]
\centerline{\includegraphics[scale=0.55]{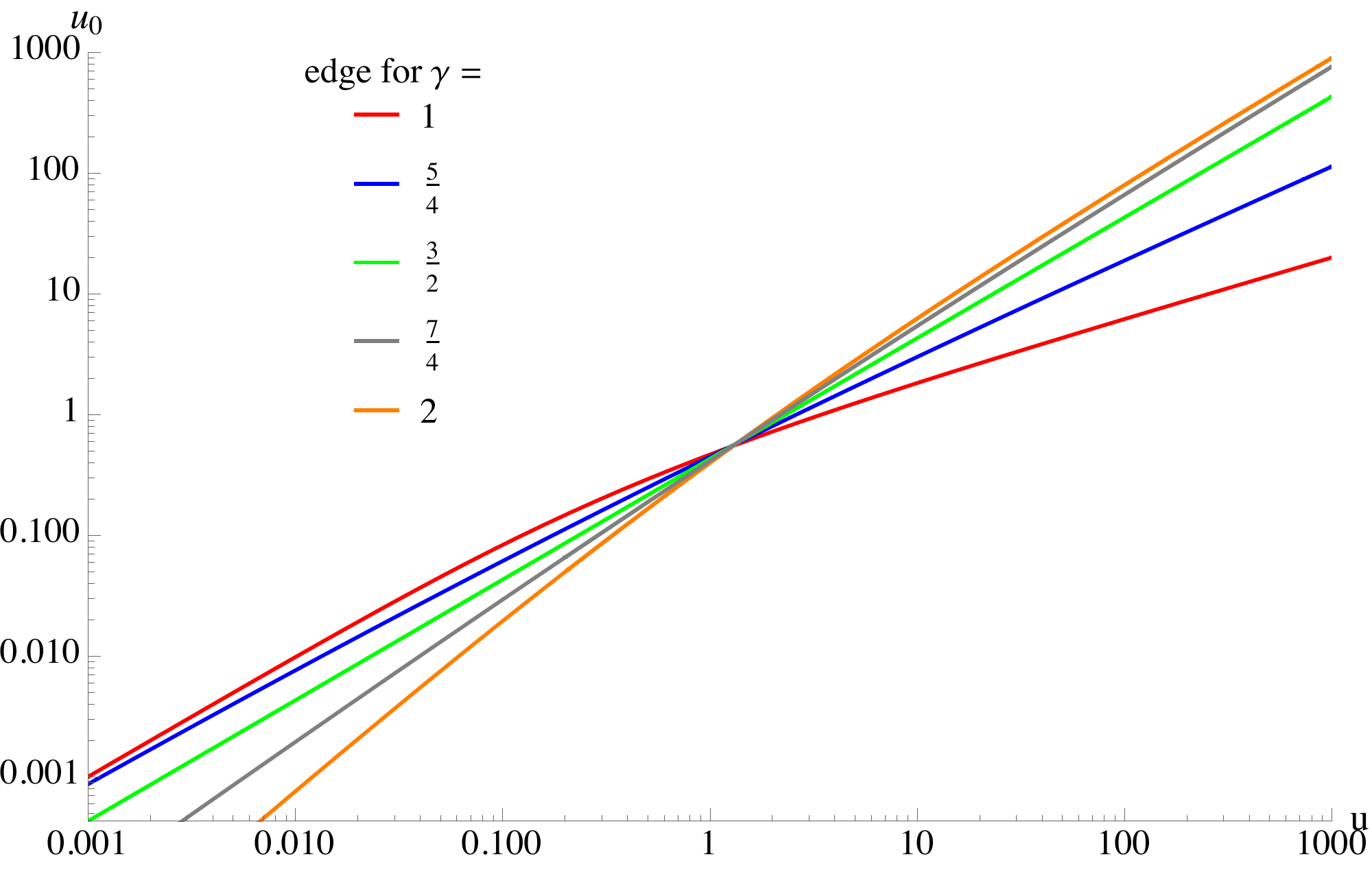}}
\caption{The value of the edge $u_0$ is plotted against $u$ in log-log space for $\beta=2$ and different values of $\gamma$.}
\label{fig:edge}
\end{figure}

As was discussed above, the support of the optimal density $\rho_*(b)$ is the interval 
$[u_0, +\infty[$ where
\be \label{edgeu0} 
u_0 = \frac{4}{\beta \pi} w(u) 
\ee 
with $0 < u_0 <u$. Note the useful relations from \eqref{weq} 
\be \label{useful} 
(u-u_0)^{\gamma- \frac{1}{2}} = \frac{\beta \pi}{4 C_\gamma} u_0 \quad , \quad 
u = u_0 + (\frac{\beta \pi}{4 C_\gamma} u_0)^{\frac{1}{\gamma-1/2}}
\ee
From the second equation in Eq.~\eqref{useful}, one sees that
\begin{itemize}
\item For $\gamma<3/2$, for $u\gg1$, $u_0\simeq \frac{4C_\gamma}{\beta \pi}u^{\gamma-1/2}$ and for $u\ll1$, $u_0\simeq u$.
\item For $\gamma<3/2$, for $u\gg1$, $u_0\simeq u$ and for $u\ll1$, $u_0\simeq \frac{4C_\gamma}{\beta \pi}u^{\gamma-1/2}$.
\end{itemize}
This behavior is summarized in Fig.~\ref{fig:edge} where the value of the edge $u_0$ is plotted against $u$ in log-log space for various values of $\gamma$ and $\beta=2$.

We now use the two equivalent formulae obtained above for $\rho_*(b)$,
where we insert $\phi(z)=(z)^\gamma_+$ which lead to equivalent forms that we give for completeness.
\subsection{First form of the density}
The optimal density reads
\begin{equation}
\begin{split}
 \label{dens11} 
&\forall b \geq u, \quad \rho_*(b) =   \frac{ \sqrt{(b - u_0 )_+  }}{ \pi}  +  \frac{1}{2 \pi} \frac{u_0}{ (b-u)^{1/2}} \, _2F_1\left(\frac{1}{2},\gamma -\frac{1}{2};\gamma +\frac{1}{2};
 \frac{u-u_0}{u-b} \right) \\
&\forall b \leq u, \quad\rho_*(b) =   \frac{ \sqrt{(b - u_0 )_+  } }{ \pi} \left[ 1
+ \frac{4 C_\gamma}{\beta \pi}  (\gamma -\frac{1}{2})  (u-b)^{\gamma-\frac{3}{2}}     \,
   _2F_1\left(\frac{1}{2},\frac{3}{2}-\gamma ;\frac{3}{2};\frac{b -u_0}{b -u}\right)    \right] 
   \end{split}
\end{equation}

\begin{proof}
From Eq. \eqref{89} we obtain using the explicit form for $f(u)$, see Eq.~\eqref{fu}
\bea \label{dens1}
&& \rho_*(b) =
  \frac{\sqrt{(b - u_0 )_+  }}{ \pi}  
+ \frac{2}{\beta \pi^2} \int_0^{w(u)} 
  \frac{\mathrm{d}w'}{
\sqrt{(b - u + (\frac{w'}{C_\gamma})^{\frac{1}{\gamma-1/2}} )_+}  }
\eea
where we recall that $w(u)$ is the unique positive root of Eq.~\eqref{weq}.
Performing the change of variable $w'=C_\gamma |b-u|^{\gamma-1/2} z$
and using Eqs.~\eqref{edgeu0} and \eqref{useful}, it can be written as (with $\epsilon={\rm sgn}(b - u)$)
\bea  \label{int1}
&& \rho_*(b) =
  \frac{\sqrt{(b - u_0 )_+  }}{ \pi}  
+ \frac{2C_\gamma}{\beta \pi^2}    |b-u|^{\gamma-1}  
\int_0^{(\frac{u-u_0}{|b-u|})^{\gamma-1/2}} 
 \frac{\mathrm{d}z}{
\sqrt{(\epsilon + z^{\frac{1}{\gamma-1/2}} )_+}  }
\eea
For $b>u$ we can perform the integral (with $\epsilon=+1$) and we find
\be \label{f1} 
 \rho_*(b) =
  \frac{ \sqrt{(b - u_0 )_+  }}{ \pi} 
+  \frac{2C_\gamma}{\beta \pi^2}   
\frac{(u-u_0)^{\gamma-1/2}}{(b-u)^{1/2}}
 \, _2F_1\left(\frac{1}{2},\gamma -\frac{1}{2};\gamma +\frac{1}{2};
 \frac{u-u_0}{u-b} \right)
\ee 
which, using \eqref{edgeu0} and \eqref{useful} can be equivalently written as
in \eqref{dens11}. For $b<u$ we calculate the same integral with $\epsilon=-1$), this gives 
\be
 \rho_*(b) =
  \frac{ \sqrt{(b - u_0 )_+  }}{ \pi}  + \frac{2C_\gamma}{\beta \pi^2}    (u-b)^{\gamma-1}  
\left( i \frac{(u-u_0)^{\gamma-1/2}}{ (u-b)^{\gamma-1/2}}
 \, _2F_1\left(\frac{1}{2},\gamma -\frac{1}{2};\gamma +\frac{1}{2};
 \frac{u-u_0}{u-b}   \right)
 -\frac{i \sqrt{\pi } \Gamma \left(\gamma +\frac{1}{2}\right)}{\Gamma (\gamma )}
 \right) 
\ee
A nicer equivalent form can be obtained by performing another equivalent integral. We rewrite the optimal density as 
\bea 
&& \rho_*(b) =
  \frac{ \sqrt{(b - u_0 )_+  }}{ \pi} 
+ \frac{2 C_\gamma}{\beta \pi^2}   |b-u|^{\gamma-1}  
(\gamma- \frac{1}{2}) \int_1^{\frac{u-u_0}{u-b}} \rmd y\, \frac{y^{\gamma-3/2}}{\sqrt{y-1}}  
\eea
which upon integration gives formula \eqref{dens11}.
\end{proof}
\subsection{Second form of the density}
Defining $\tau =  \frac{u-u_0}{b-u_0}$, the optimal density reads
\begin{equation} \label{dens21} 
\begin{split}
&\forall b \geq u, \quad \rho_*(b)=\frac{\sqrt{(b-u_0)_+} }{\pi}\left[1+\frac{u_0}{2(b-u_0)} 
~_2F_1(1,\frac{1}{2},\gamma+\frac{1}{2},\tau)\right] \\
&\forall b \leq u, \quad \rho_*(b)=\frac{\sqrt{(b -u_0)_+}}{\pi }\left[1+ \frac{(\gamma-\frac{1}{2})u_0}{u-u_0}~_2F_1(1,\frac{3}{2}-\gamma,\frac{3}{2},\frac{1}{\tau})\right]
\end{split}
\end{equation}
\begin{proof}
We now use the formula (34) given in the Letter, i.e. \eqref{dens} here. 
It leads to 
\begin{equation}
\label{eq:opt_density_pp}
\rho_*(b)=\frac{\sqrt{(b-u_0)_+}}{\pi}\left[1+\frac{\gamma}{\beta \pi}\dashint_{-\infty}^{+\infty}\frac{\rmd b'}{b+b'-u}\frac{(b')_+^{\gamma-1}}{\sqrt{(u-u_0-b')_+}}\right]
\end{equation}
We perform the change of variable $X = \sqrt{1- \frac{b'}{u-u_0}}$ and $\tau =  \frac{u-u_0}{b-u_0}$
\begin{equation}
\rho_*(b)=\frac{\sqrt{(b-u_0)_+}}{\pi}\left[1
+ \frac{2\gamma}{\beta \pi} \frac{(u-u_0)^{\gamma-\frac{1}{2}}}{b-u_0}\dashint_{0}^{1}\rmd X \,\frac{ (1-X^2)^{\gamma-1}}{1-\tau X^2} \right]
\end{equation}

For $\tau\leq 1$, i.e. $b \geq u$, the integral is given by a hypergeometric function, 
leading to formula \eqref{dens21} upon simplification using Eqs.~\eqref{fu} and \eqref{useful}. It is equivalent to Eq.~\eqref{f1} using relations between hypergeometric functions, i.e for $\tau\leq 1$ 
one has
$_2F_1(1,\frac{1}{2},\gamma+\frac{1}{2},\tau)=(1-\tau)^{-1/2} 
~_2F_1(\frac{1}{2},\gamma -\frac{1}{2};\gamma +\frac{1}{2};\frac{\tau}{\tau-1})$. For $\tau>1$, one uses the reflection formula for hypergeometric functions
\begin{equation}
\label{eq:hypergeo_continuation}
\begin{split}
~_2F_1(1,\frac{1}{2},\gamma+\frac{1}{2},\tau)&=\frac{2\gamma-1}{\tau} \! ~_2F_1(1,\frac{3}{2}-\gamma,\frac{3}{2},\frac{1}{\tau})+\frac{\Gamma(\gamma+\frac{1}{2})\sqrt{\pi}}{\Gamma(\gamma)}(-\tau)^{-\frac{1}{2}}~_2F_1(\frac{1}{2},1-\gamma,\frac{1}{2},\frac{1}{\tau})
\end{split}
\end{equation}
Bearing in mind that Eq.~\eqref{eq:opt_density_pp} contains a principal part, that $\dashint \frac{1}{x}\underset{\varepsilon\to 0}{=} \int\Re ( \frac{1}{x\pm i\varepsilon})$ and that the last term in Eq.~\eqref{eq:hypergeo_continuation} is purely imaginary, i.e. $(-\tau)^{-\frac{1}{2}} \in i \mathbb{R}$, we can discard the last term by taking the real part of Eq.~\eqref{eq:hypergeo_continuation} leading to the proposed continuation in Eq.~\eqref{dens21}. By the same transformation between hypergeometric functions evaluated at $\tau$ or $\tau/(\tau-1)$, we show the equivalence between the second equation of Eq.~\eqref{dens21} with the second equation of Eq.~\eqref{dens11}.
\end{proof}
\subsection{Singularities of the density}
The density is analytic everywhere except at $b=u_0$ (the edge)
and at $b=u$ (the first point of application of the potential). The singularity for $b=u$ is a power law divergence for $1/2 < \gamma < 1$.
From Eq.~\eqref{int1} one obtains 
\begin{equation}
\begin{split}
& \rho_*(b) \simeq 
\frac{2 C_\gamma}{\beta \pi^2} |b - u|^{\gamma-1} 
\int_0^{+\infty} \frac{\rmd z}{\sqrt{(\epsilon+ z^{\frac{1}{\gamma-1/2}})_+} } 
= D_\gamma^\epsilon |b - u|^{\gamma-1} \\
& D_\gamma^{+} = \frac{2 C_\gamma}{\beta \pi^2} \frac{\Gamma(1-\gamma) \Gamma(\frac{1}{2} + \gamma)}{\sqrt{\pi}}
= \frac{\gamma}{\pi \beta \sin(\gamma \pi)} \;  , \; D_\gamma^{-} = - \frac{2 C_\gamma}{\beta \pi^2} \frac{\sqrt{\pi } \Gamma (1-\gamma )}{\Gamma \left(\frac{1}{2}-\gamma \right)} 
= - \cos(\gamma \pi) D_\gamma^{+} 
\end{split}
\end{equation}
where $\epsilon ={\rm sgn}(b-u)$ and this power law divergence becomes 
a logarithmic singularity for $\gamma=1$.  More generally one finds, for any $\gamma>1/2$, $\gamma \neq 1,2 $, expanding \eqref{dens11} 
on both sides of $b=u$
\be
\rho_*(b) =  \frac{\sqrt{(b-u_0)_+} }{\pi}+ D_\gamma^\epsilon |b - u|^{\gamma-1}  + 
\frac{(2 \gamma -1) u_0}{4 \pi  (\gamma -1) \sqrt{u-u_0}}
+\frac{u_0 \left(2 \gamma -1 \right) (u-b)}{8 \pi  (\gamma -2) \left(u-u_0\right){}^{3/2}}+\mathcal{O}(b
   -u)^{2}
\ee 

Finally we verify that the singularity at $b=u_0$ is always of semi-circle type.
More precisely we obtain
\be
\rho_*(b) =  \frac{\sqrt{(b-u_0)_+} }{\pi}
+\frac{u_0\left(2 \gamma-1\right) \sqrt{(b -u_0)_+}}{2 \pi  (u-u_0)}+\frac{(4
   (2-\gamma) \gamma -3) u_0 \left(b -u_0\right)_+^{3/2}}{6 \pi 
   \left(u-u_0\right)^2}+\mathcal{O}\left(b -u_0\right)^{\frac{5}{2}}
\ee

\subsection{Hard wall limit for the optimal density}
In the limit $B \to +\infty$ and for any $\gamma$, one recovers the result for the hard wall
\be
\rho_{B=+\infty,u}(b) = \frac{2 b - u}{ 2 \pi \sqrt{(b-u)_+} }
\ee 
\begin{proof}
We recall that for $b \geq u$, the optimal density reads
\begin{equation}
\begin{split}
\rho_{B,u}(b) &= \frac{ \sqrt{(b - u_0 )_+  }}{ \pi}  +  \frac{1}{2 \pi} \frac{u_0}{ (b-u)^{1/2}} \, _2F_1\left(\frac{1}{2},\gamma -\frac{1}{2};\gamma +\frac{1}{2};
 \frac{u-u_0}{u-b} \right) 
 \end{split}
\end{equation}
The saddle point equation being $\frac{\beta \pi}{4} u_0= Bf(u-u_0)$, we see that the hard wall limit, $B\to +\infty$, imposes that $u_0=u$. In this case, the optimal density reads
\begin{equation}
\begin{split}
\rho_{B,u}(b) &= \frac{ \sqrt{b - u   }}{ \pi}  +  \frac{1}{2 \pi} \frac{u}{ (b-u)^{1/2}} \, _2F_1\left(\frac{1}{2},\gamma -\frac{1}{2};\gamma +\frac{1}{2};
0 \right) 
 \end{split}
\end{equation}
As $_2F_1\left(\frac{1}{2},\gamma -\frac{1}{2};\gamma +\frac{1}{2};
0 \right) =1$, we obtain the hard wall density $\rho_{B=+\infty,u}(b) = \frac{2 b - u}{ 2 \pi \sqrt{(b-u)_+} }$.
\end{proof}

\subsection{Optimal density for special values of $\gamma$ (for $B=1$)}

\begin{itemize}
\item  For $\gamma=3/2$, using $C_\gamma=3 \pi/8$, $u=(1+ \frac{2 \beta}{3}) u_0$ 
we find for all $b$ the remarkably simple expression
\be
\rho_*(b)= \left[\frac{\sqrt{b-u_0} }{\pi}+ 
 \frac{3}{2 \beta \pi} ( \sqrt{b-u_0} - \sqrt{(b-u)_+} ) \right]\theta(b-u_0)
\ee

\begin{figure}[h!]
\centerline{\includegraphics[scale=0.53]{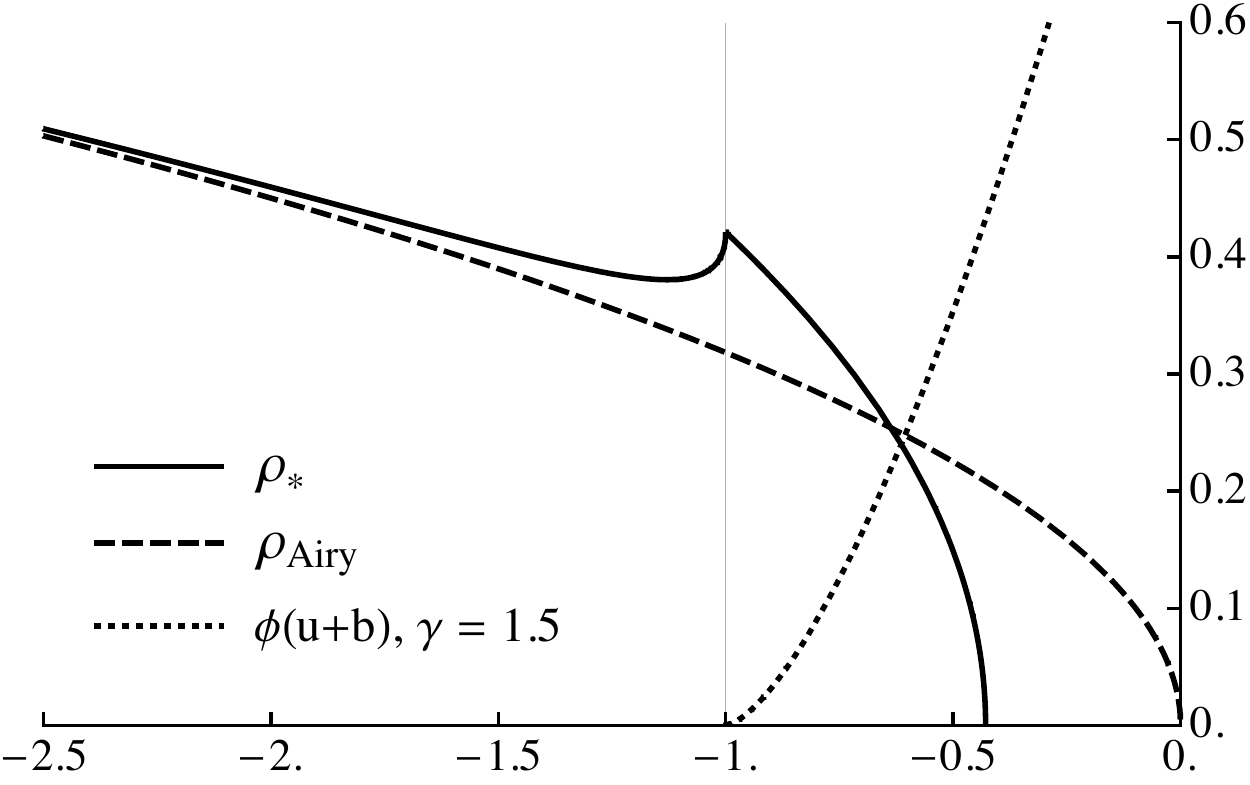}}
\caption{Optimal density $\rho_*(-b)$ for the soft wall with $\gamma=3/2$ (solid line), compared to the semi-circle density $\rho_{\rm Ai}(-b)$  (dashed line). The external potential $\phi(u+b)$ is represented on the dotted line.}
\label{gamma15}
\end{figure}

\item   For $\gamma=1$, using $C_\gamma=1$, $\frac{u_0}{\sqrt{u-u_0}}=\frac{4}{\beta \pi}$ we find for all $b$ the expression
\be
\rho_*(b) =  \left[ \frac{\sqrt{b-u_0}}{\pi} + \frac{2}{\beta \pi^2} 
\log\left( \frac{\sqrt{b - u_0} + \sqrt{u-u_0}}{\sqrt{|b - u|}}\right) \right] \theta(b-u_0)
\ee
which, upon simple manipulations, recovers the result obtained in Ref.~\cite{JointLetter} by a quite different calculation.

\begin{figure}[ht!]
  \centering
  \subfloat[$u=25$]{\label{fig:edge-a}\includegraphics[scale=0.55]{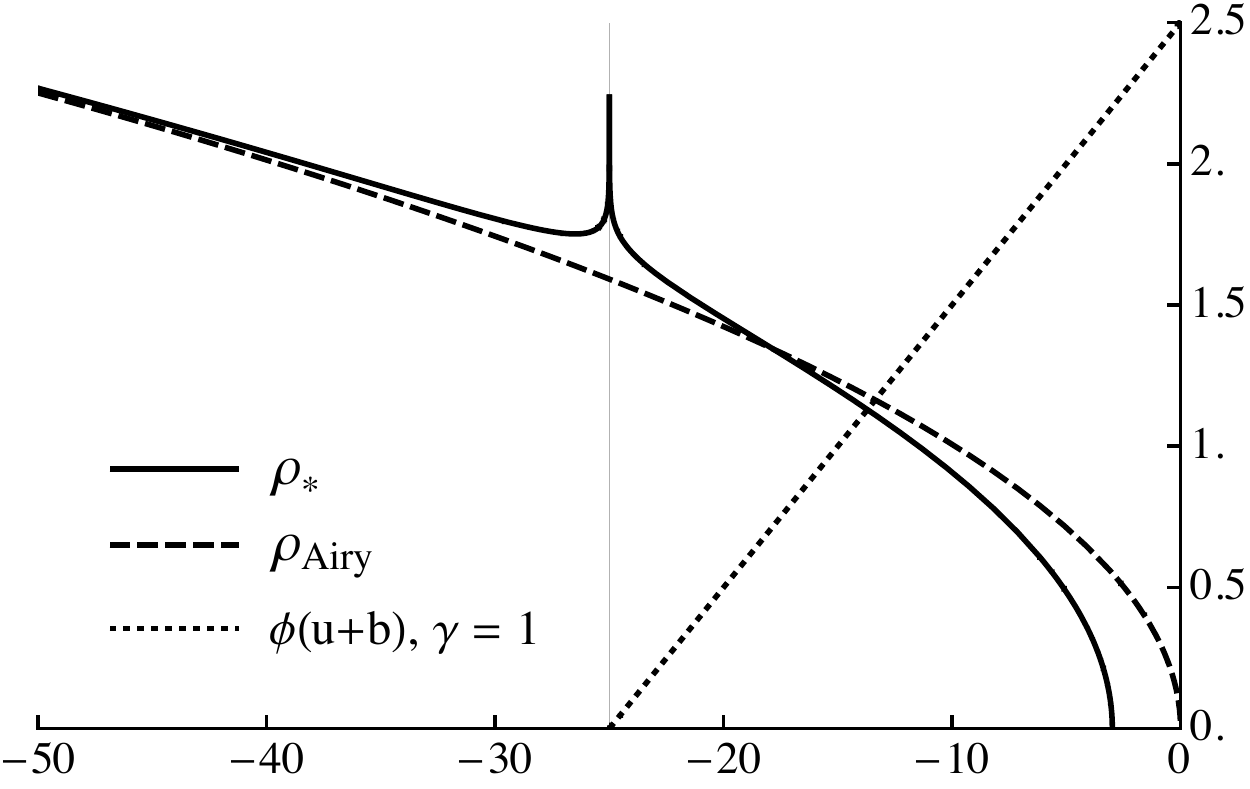}}
  \hspace{-5pt}
  \subfloat[$u=0.05$]{\label{fig:contour-b}\includegraphics[scale=0.55]{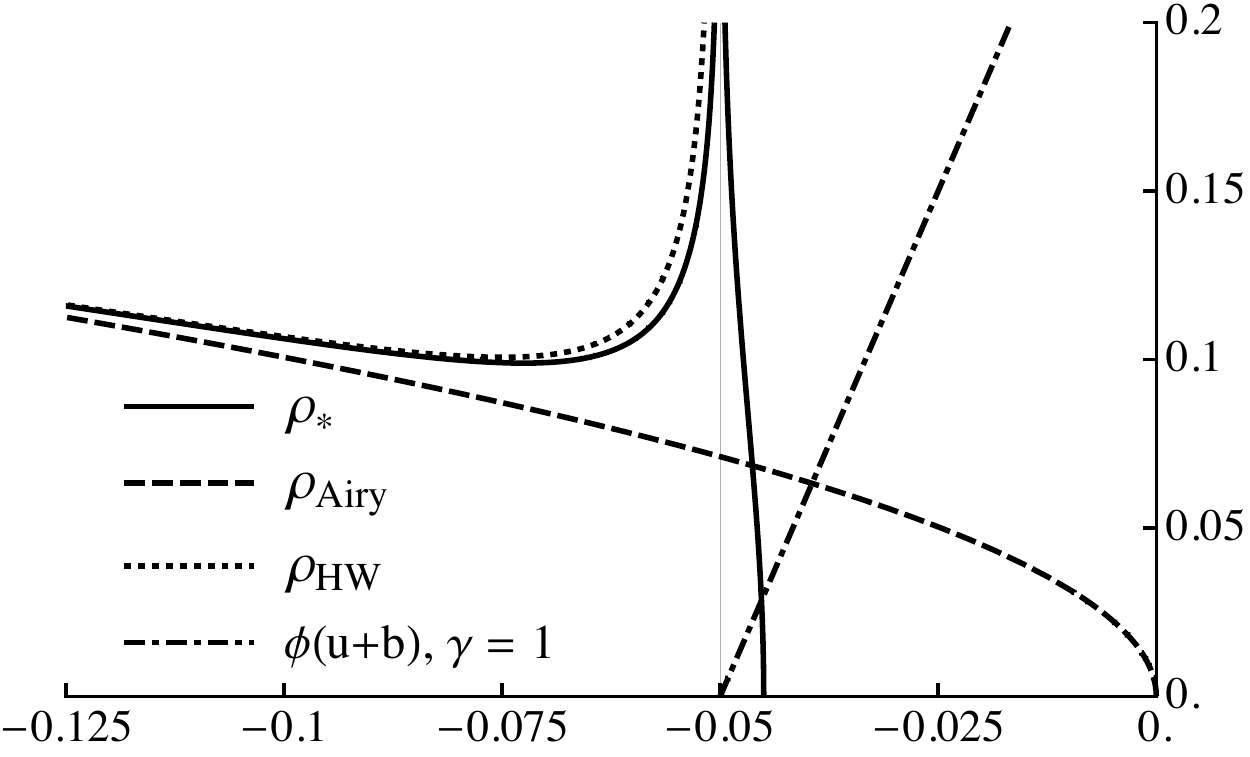}}
  \caption{Same as Fig. \ref{gamma15} for the linear wall $\gamma=1$ with $u=25$ and $u=0.05$. The optimal density for $u=0.05$ is also compared to the  infinite hard-wall $\rho_{HW}(-b)$ showing a good agreement.}
  \label{fig:gamma1}
\end{figure}

\item  For $\gamma=2$, we find
\be
\hspace*{-1.3cm}\rho_*(b)=\left[ \frac{\sqrt{b-u_0}}{\pi} +\frac{2}{\pi^2\beta }\bigg(2 \sqrt{b-u_0} \sqrt{u-u_0}+(b-u) \log
\abs{\frac{\sqrt{b-u_0}-\sqrt{u-u_0}}{\sqrt{b-u_0}+\sqrt{u-u_0}}}\bigg) \right] \theta(b-u_0)
\ee 
\begin{figure}[ht!]
  \centering
  \subfloat[$u=100$]{\label{fig:edge-a}\includegraphics[scale=0.55]{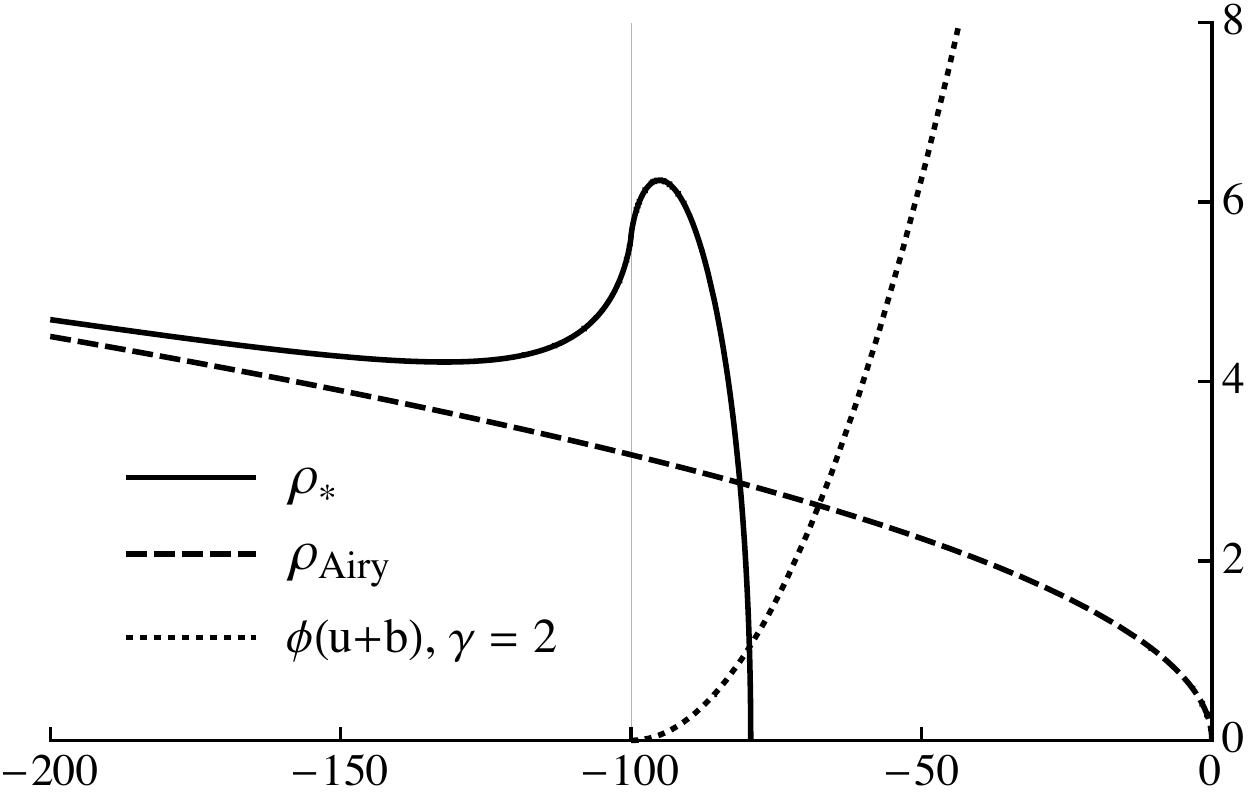}}
  \hspace{-5pt}
  \subfloat[$u=1$]{\label{fig:contour-b}\includegraphics[scale=0.55]{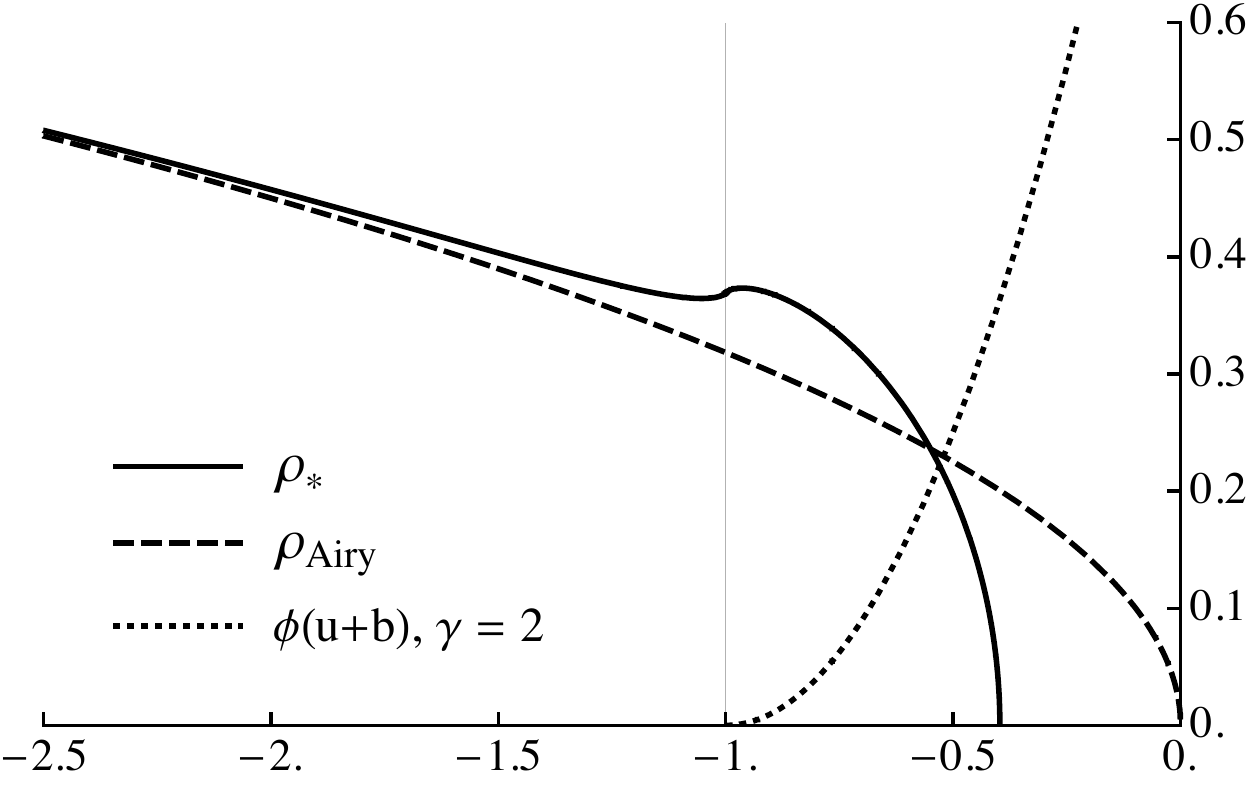}}
  \caption{Same as Fig. \ref{gamma15} for the quadratic wall $\gamma=2$ with $u=100$ and $u=1$.}
  \label{fig:gamma2}
\end{figure}

\item For $\gamma=5/2$ we find
\be
\rho_*(b)=\left[ \frac{\sqrt{b-u_0}}{\pi} + 
\frac{5}{4 \beta \pi} ( \sqrt{b-u_0}  (3 (u-b) + (b-u_0)) 
+ 2 (b - u)_+^{3/2} )\right] \theta(b-u_0)
\ee 

\end{itemize}

\section{PDF $P({\sf L})$ for monomial walls} 
We study the PDF $P({\sf L})$ for monomial walls, $\phi(z)=  (z)_+^\gamma$ and we first start with the case $\gamma=3/2$ where calculations are simple. Inserting the 
result for $\Sigma_{B \phi}$ of Eq.~\eqref{res32} into Eq.~\eqref{s2} we have 
\be \label{s22}
\tilde G(\tilde L) = \max_{B} \left[ \frac{\beta}{24} \frac{3 B}{3 B + 2 \beta} u^3  - B \tilde L \right] 
\ee
The optimal $B$ is given by $B = \dfrac{\beta}{6} (\dfrac{u^{3/2}}{\sqrt{\tilde L}} - 4)$, which
inserted into Eq.~\eqref{s22} gives
\be
G(\tilde L) = \frac{\beta}{24} (u^{3/2} - 4 \sqrt{\tilde L})^2 \quad 0 < \tilde L < \mathbb{E}_\beta[{\sf L}]=\frac{u^3}{16} 
\ee
Hence, as given in the Letter $G(\ell) = \frac{\beta u^3}{24} (1- \sqrt{\ell})^2$. 
Near the typical value $\ell=1$, the rate function takes the form
\be
G(\ell) \simeq \frac{(\ell - 1)^2}{2 \sigma} \quad , \quad 
\sigma = \frac{48}{\beta u^3}
\ee
which consistent with the first two cumulants,
$\kappa_1(u)=- \frac{1}{16} u^3$ and $\kappa_2(u)=\frac{3}{16 \beta} u^3$. \\

For $\gamma \neq 3/2$, let us first recall the cumulants
$\mathbb{E}_\beta[{\sf L}^n]^c=t^2 (-1)^n \tilde \kappa_n(u)$, for $n=1,2$
\bea
\label{eq:cum_L_gamma11}
&& \mathbb{E}_\beta[{\sf L}] = t^2 k_1 u^{\gamma+ \frac{3}{2}}  \quad , \quad k_1= \frac{\Gamma(\gamma+1)}{2 \sqrt{\pi} \Gamma(\gamma+\frac{5}{2})} = - \frac{1}{\pi} \frac{C_\gamma}{(\gamma+ \frac{3}{2})(\gamma+ \frac{1}{2})}  \\
&& \mathbb{E}_\beta[{\sf L}^2]^c = t^2 k_2 u^{2 \gamma} \quad , \quad k_2=  \frac{2}{\beta \pi^2 \gamma} C_\gamma^2
\eea
We can now use the scaling relations \eqref{scaling1} and \eqref{scaling2} to write
\be
\tilde G(\tilde L) = \max_{B} \left[  (\frac{2}{\beta})^{\frac{3 + 2 \gamma}{3-2 \gamma}}  
B^{\frac{6}{3-2 \gamma}}  \Sigma_{\phi}\left(u B^{- \frac{2}{3-2 \gamma}}
(\frac{\beta}{2})^{\frac{2}{3-2 \gamma}} \right) 
- B \tilde L \right] 
\ee
Let us denote $U=u (\frac{\beta}{2})^{\frac{2}{3-2 \gamma}}B^{- \frac{2}{3-2 \gamma}}$ and insert $\tilde L = k_1 u^{\gamma+ \frac{3}{2}} \ell$ so that
\be
G(\ell) = \frac{\beta}{2} u^3 \max_{U} [  U^{-3} \Sigma_{\phi}(U)- k_1 U^{\gamma- 3/2} \ell ]
\ee
where here and below $\Sigma_\phi=\Sigma_\phi^{\beta=2}$. The rate function is determined by the parametric system
\bea
&& G(\ell) =  \frac{\beta}{2}  u^3 (U^{-3} \Sigma_\phi(U)- k_1 U^{\gamma- 3/2} \ell) \\
&&  3 \Sigma_{\phi}(U) - U \Sigma'_{\phi}(U)  = (\frac{3}{2} - \gamma) k_1 U^{\gamma+ 3/2} \ell \\
&& G'(\ell) = - k_1 U^{\gamma-3/2} 
\eea
Note that, remarkably, for any $\gamma$ the only dependence in $u$ and $\beta$ 
is in the cubic prefactor $\beta  u^3$ as noted in the Letter. In the vicinity of the typical value, the fluctuations are Gaussian and given by
\be \label{probaexpand2} 
 G(\ell) \simeq \frac{(\ell - 1)^2}{2 \sigma} \quad , \quad 
\sigma = \frac{\tilde \kappa_2(u)}{\tilde \kappa_1(u)^2} = \frac{2}{\beta} \frac{(\gamma+ \frac{3}{2})^2 
(\gamma+ \frac{1}{2})^2}{u^3} 
\ee 

\section{Exponential walls $\phi(z)=e^{ z}$   } 

Until now we considered $\phi$ in $\Omega_0$, with $\phi(z\leq 0)=0$.
It is possible to extend our formula to a larger class of soft walls, $\Omega_1$,
such that $\phi$ is still positive and increasing but does not vanish on $\mathbb{R}^-$,
instead it vanishes smoothly as $z \to -\infty$. The bounds of the integrals
over $u$ have to be taken at $u=-\infty$ and the formula go through.

\subsection{$\Sigma_\phi(u)$ for the exponential wall}

Consider the following linear statistics 
\be
{\sf L}_{c,B}(t,u) = Bt \sum_i  e^{c (u + t^{-2/3} a_i)}
\ee
we first note that by rescaling of $t$ and $u$ we have ${\sf L}_{c,B}(t,u) = {\sf L}_{1,1}(c^{-3/2} t, c u + \log( B c^{3/2}) ) $, hence it is sufficient to study the case $c=1,B=1$ since the $c,B$ dependence can be restored easily.
The function $f(u)$ and the saddle point {\sf SP1} are then
\be
f(u) = \frac{1}{2} \int_0^{+\infty} \frac{\rmd b}{\sqrt{b}} e^{u-b} = \frac{\sqrt{\pi }}{2} e^{ u}  \qquad, \qquad w = f(u - \frac{4}{\beta \pi} w) = \frac{\sqrt{\pi }}{2}   e^{u-  \frac{4}{\beta \pi} w}
\ee 
We obtain the solution of {\sf SP1} in terms of the principal branch of the Lambert function $W_0$ \cite{corless1996lambertw}
\begin{equation}
w(u) = \frac{\beta \pi}{4 } W_0(\frac{2e^{ u} }{\beta \sqrt{\pi}} ) 
\end{equation}
We calculate the excess energy using the formula $ \Sigma_\phi(u) = \frac{1}{\pi} \int_{-\infty}^u \rmd u' w(u') [u-u'] $
\be
\Sigma_\phi(u)  = \frac{\beta}{48 } (2 W^3+9 W^2+12 W) \quad , \quad W 
:= W_0\left(\frac{2  e^{ u}}{\beta \sqrt{\pi }  }\right) \label{SigExp}
\ee 
It is useful to note that the derivative of the express energy reads
\be
\Sigma'_\phi(u) = \frac{1}{\pi} \int_{-\infty}^u \rmd u' \, w(u')  
= \frac{\beta}{8 } (2 W+W^2) \label{SigPExp}
\ee 
The different asymptotics are
\begin{equation}
\begin{split}
&w(u)\underset{u \to -\infty} {\simeq} \frac{\sqrt{\pi  } }{2} \, e^{ u}+\mathcal{O}(e^{2u}) \quad , \quad w(u)\underset{u \to +\infty}{\simeq}\frac{\pi  \beta  }{4}\left[u +
 \log( \frac{2 }{\sqrt{\pi} \beta u })+\mathcal{O}(\frac{\log u}{u})\right]\\
&\Sigma_\phi(u)\underset{u \to -\infty} {\simeq} \frac{e^{u}}{2\sqrt{\pi}} +\mathcal{O}(e^{2u}) \; , \quad \Sigma_\phi(u)\underset{u \to +\infty}{\simeq}\frac{\beta  u^3}{24}+
\frac{\beta  u^2 }{16 } \log \left(\frac{4  e^{2/3}}{\pi \beta^2 u^2 }\right)+\mathcal{O}(u\log(u)^2)
\end{split}
\end{equation}

\subsection{Optimal density for the exponential wall} 

\begin{figure}[ht!]
  \centering
  \subfloat[$u=20$]{\label{fig:edge-a}\includegraphics[scale=0.55]{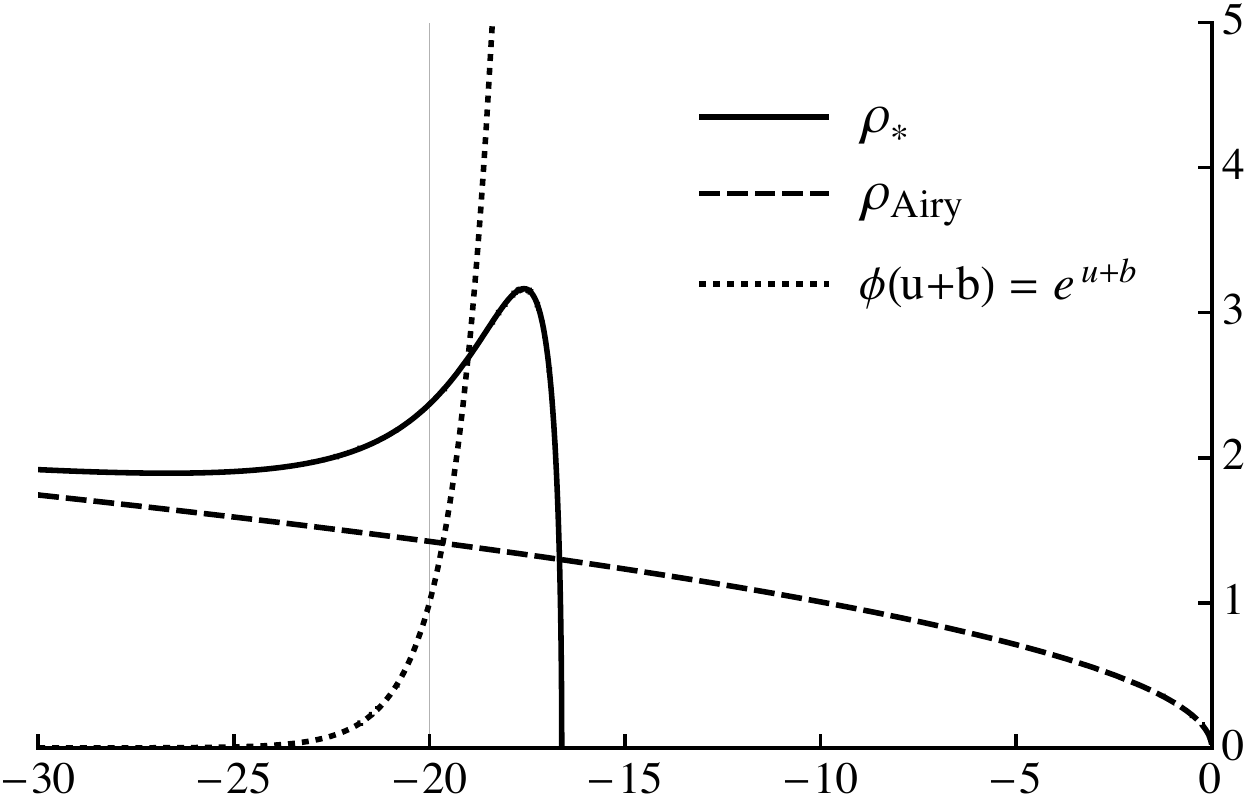}}
  \hspace{-3pt}
  \subfloat[$u=1$]{\label{fig:contour-b}\includegraphics[scale=0.55]{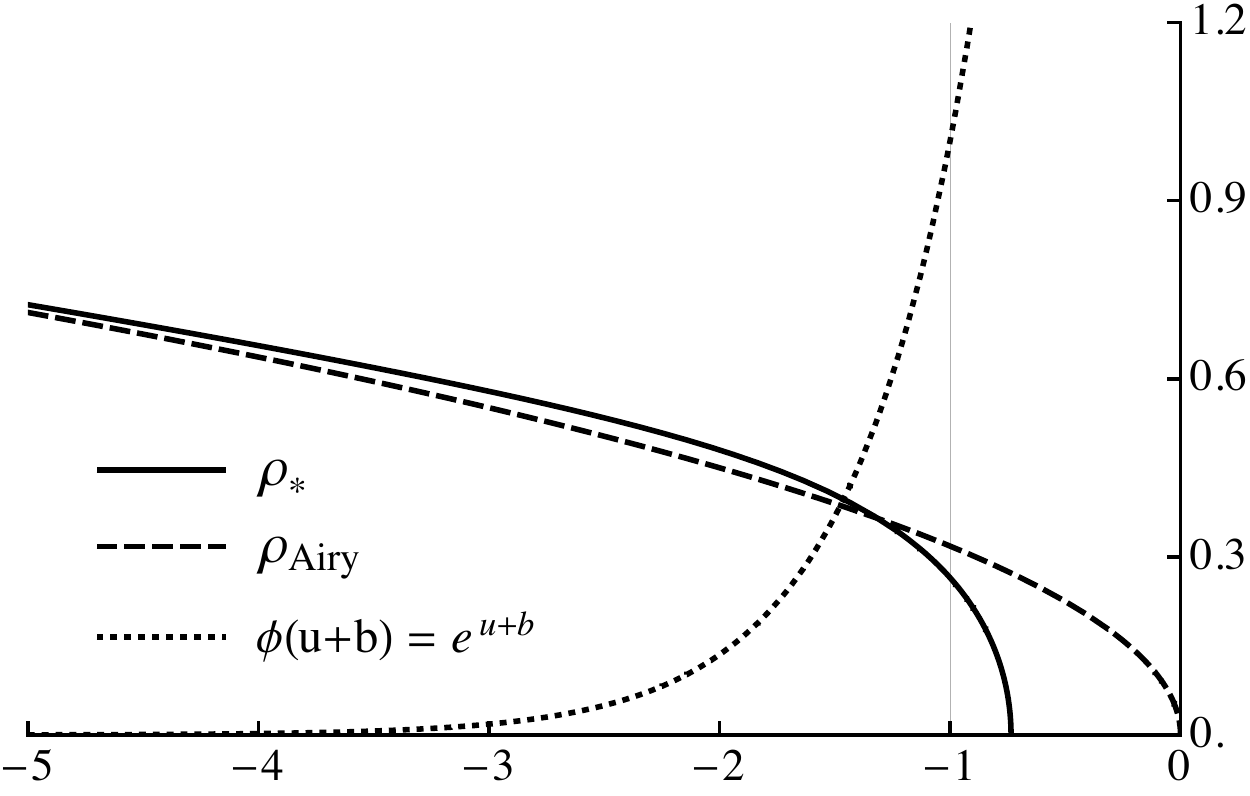}}
  \caption{Optimal density $\rho_*(-b)$ for the exponential wall (solid line), compared to
  the semi-circle density $\rho_{\rm Ai}(-b)$
  (dashed line) for $u=20$ and $u=1$. The potential $\phi(u+b)$ is represented with the dotted line.}
  \label{fig:exp1}
\end{figure}

The associated optimal density has a support $[u_0,+\infty[$
with $u_0>0$ given by $ u_0 = \frac{4}{\beta \pi} w(u) =  W_0(\frac{2 }{\beta \sqrt{\pi}} e^{ u}) $ and is given by
\be
 \rho_*(b)  = \frac{ \sqrt{(b - u_0 )_+  }}{ \pi} + \frac{2}{\beta \pi^2} \int_0^{w(u)}    \frac{\mathrm{d}w'}{\sqrt{(b - u + f^{-1}(w') )_+}  }
\ee
Let us calculate the second term
\bea
&& \int_0^{w(u)} 
 \frac{\mathrm{d}w' }{\sqrt{(b - u + f^{-1}(w') )_+}  } = \int_0^{w(u)} 
  \frac{\mathrm{d}w'}{\sqrt{(b - u +  \log \frac{2 w'}{\sqrt{\pi }} )_+}  }
\\
&& = \frac{ \sqrt{\pi }}{2} 
\int_{u-b}^{u-u_0} 
\mathrm{d}u'  \frac{e^{ u'}}{\sqrt{b - u +u'}  }
=  \frac{ \pi e^{u-b } }{2}    \text{Erfi}\left(\sqrt{ b-u_0}\right)
\eea
where we have defined
$u'=\log \frac{2 w'}{\sqrt{\pi }}$ and used that $ \log \frac{2 w(u)}{\sqrt{\pi }} = u - \frac{4}{\beta \pi} w(u) = u - u_0$. Since there is a relation between $u$ and $u_0$, $e^{ u} =  \frac{\beta \sqrt{\pi}}{2} u_0  \, e^{ u_0} $, we can express the optimal density $\rho_*(b)$ only in terms of $u_0$ leading to
\begin{equation}
\begin{split}
\rho_*(b)  
  &= \frac{ \sqrt{(b - u_0 )_+  }}{ \pi} 
+  \frac{u_0\,   e^{u_0-  b}}{2 \sqrt{\pi}}  \text{Erfi}\left(\sqrt{b   -u_0}\right)
   \end{split}
\end{equation}
The optimal density is plotted in Fig.~\ref{fig:exp1} for $u=20$ and $u=1$. We see that for large $u$, the density becomes close to the hard wall one, while for small $u$ the reorganization of the density is perturbative.

\begin{figure}[ht!]
\centerline{\includegraphics[scale=0.45]{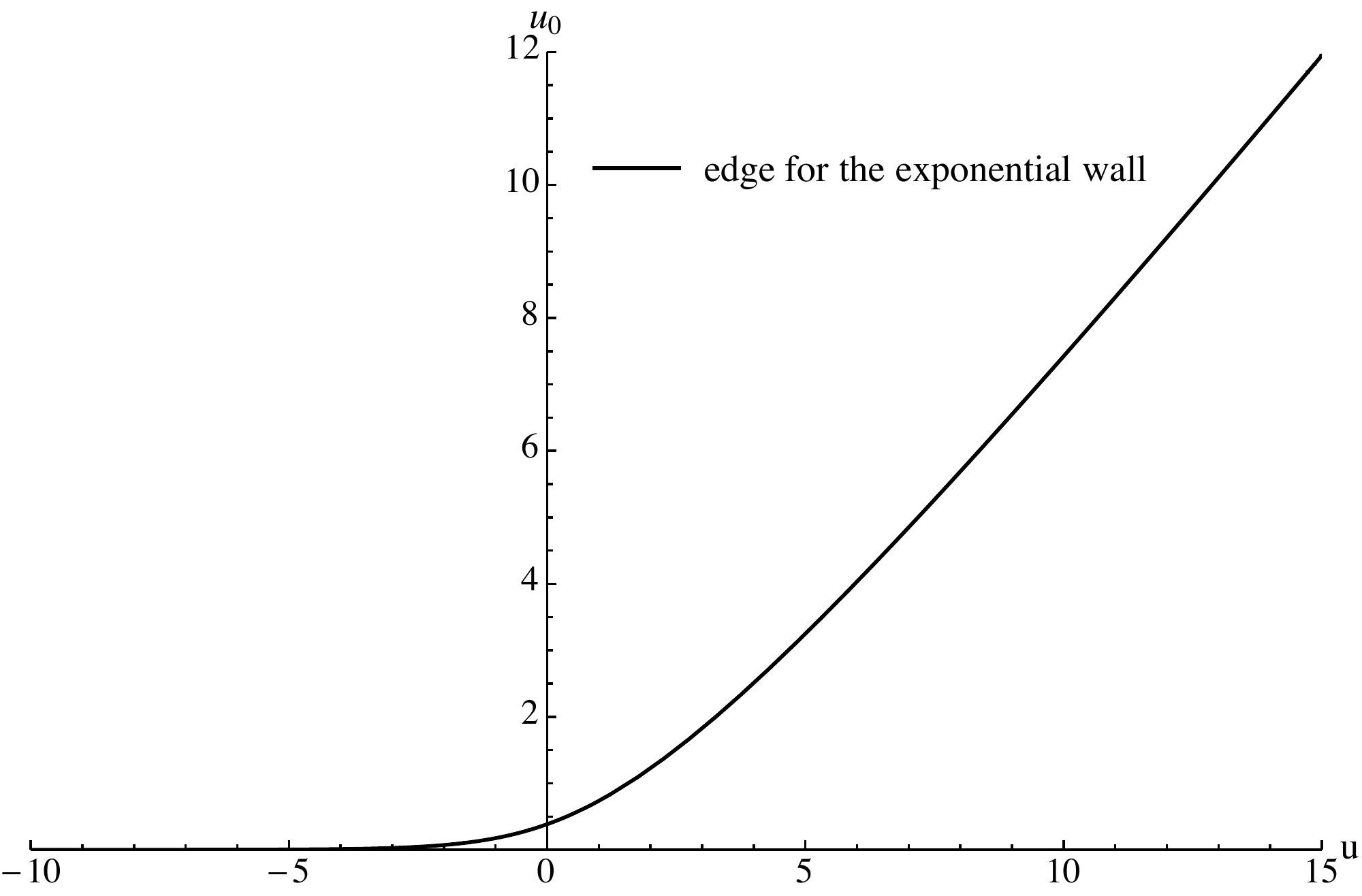}}
\caption{Edge of the exponential wall for $\beta=2$ as a function of $u$. For large $u$, $u_0\simeq u$.}
\label{fig:edge_exp}
\end{figure}

\subsection{PDF $P({\sf L})$ for the exponential wall}

Let us first note that for the exponential wall the first cumulant is\footnote{Note that this result is equivalent to the Okounkov formula for the 
average, see Proposition 2.13 of \cite{GorinShkolnikov2016}, setting
$t^{-2/3}=T/2$ and $u=0$.}
\be
\label{avexp}
\mathbb{E}_\beta[L]= \frac{t^2}{\pi} \int_0^{+\infty} \rmd b\,  \sqrt{b}
e^{u-b} = \frac{e^{ u}}{2  \sqrt{\pi}} t^2
\ee

Now, it is easy to see that for the exponential wall $\Sigma_{B \phi}(u) = \Sigma_\phi(u + \log B) $
which allows to calculate easily the PDF of ${\sf L}$, indeed defining $\hat u = u+ \log B$, we have
\be
 G(\ell) = \max_{B} \left[\Sigma_\phi(u + \log B)  -  \frac{e^{ u}}{2  \sqrt{\pi}} B \ell \right]  
= \max_{\hat u} \left[ \Sigma_{\phi}(\hat u) -  \frac{e^{ \hat u}}{2  \sqrt{\pi}} \ell  \right]
\ee
$G(\ell)$ is independent of $u$ and is given by the parametric system of equations
\begin{equation}
\begin{split}
& G(\ell) = \Sigma_{\phi}(\hat u) -  \frac{e^{ \hat u}}{2  \sqrt{\pi}} \ell 
\qquad , \qquad    \Sigma'_{\phi}(\hat u) = \frac{e^{ \hat u}}{2 \sqrt{\pi}} \ell
\end{split}
\end{equation}

Using the results of the previous subsection (Eqs.~\eqref{SigExp} and \eqref{SigPExp}) for $\Sigma_{\phi}$ and $\Sigma'_{\phi}$ in terms of the variable
\be
W := W_0(\frac{2 e^{ \hat u}}{\beta \sqrt{\pi }  }) \quad \Longleftrightarrow \quad 
\frac{2 e^{ \hat u}}{\beta \sqrt{\pi }}  = W e^W 
\ee
we obtain the system
\begin{equation}
\begin{split}
 G(\ell) = \frac{\beta}{48 } W^2 (2 W+3) \qquad , \qquad   \ell = \frac{1}{2} (2+W) e^{-W}  \label{sec} 
\end{split}
\end{equation}
and $G'(\ell) = -  \frac{\beta}{4 } W e^W  $. The typical value therefore corresponds to $W=0$ and $\ell=1$. We solve the second equation in \eqref{sec} by writing it as $- (2+W) e^{-(2+W)} = -2 \ell e^{-2}  $ so that
\be
-(2+W) = W_{-1}(-2 \ell e^{-2}) \label{lamb2}
\ee
where one needs to take the second branch of the Lambert function 
to recover that $\ell=1$ is realized for $W=0$. This leads to our final result, for $\ell \leq 1$
\be \label{Gexp} 
  G(\ell) = - \frac{\beta}{48 } (2 + W_{-1}(-2 \ell e^{-2}))^2 (1+ 2 W_{-1}(-2 \ell e^{-2}))
\ee
which is positive, as required, with its minimum at $\ell=1$.
\begin{itemize}
\item  Near $\ell=1$,
\be
 G(\ell) = \beta  \left[ \frac{1}{4} (\ell-1)^2 - \frac{1}{3} (\ell-1)^3 + \frac{1}{3} (\ell-1)^4 + \mathcal{O}((\ell-1)^5)\right]
\ee
\item Near $\ell=0$,
\be \label{tailH} 
 G(\ell) =\frac{\beta}{24}\left[-\log ^3(\ell )+ \frac{  \log ^2(\ell ) (3+6 \log (-\frac{\log \ell }{\log 2})}{2 }+\mathcal{O}(\log (\ell) \log(-\log \ell)^2	)\right]
\ee
\end{itemize}
Again, here we have access only to the side $\ell \leq 1$, the side $\ell > 1$
requires to be able to treat the case of negative $B$, which goes beyond this Letter.

\section{Inverse monomial walls $\phi(z)=(- z)^{-\delta}$, $\delta>3/2$   } 
\begin{figure}[ht!]
\centerline{\includegraphics[scale=0.64]{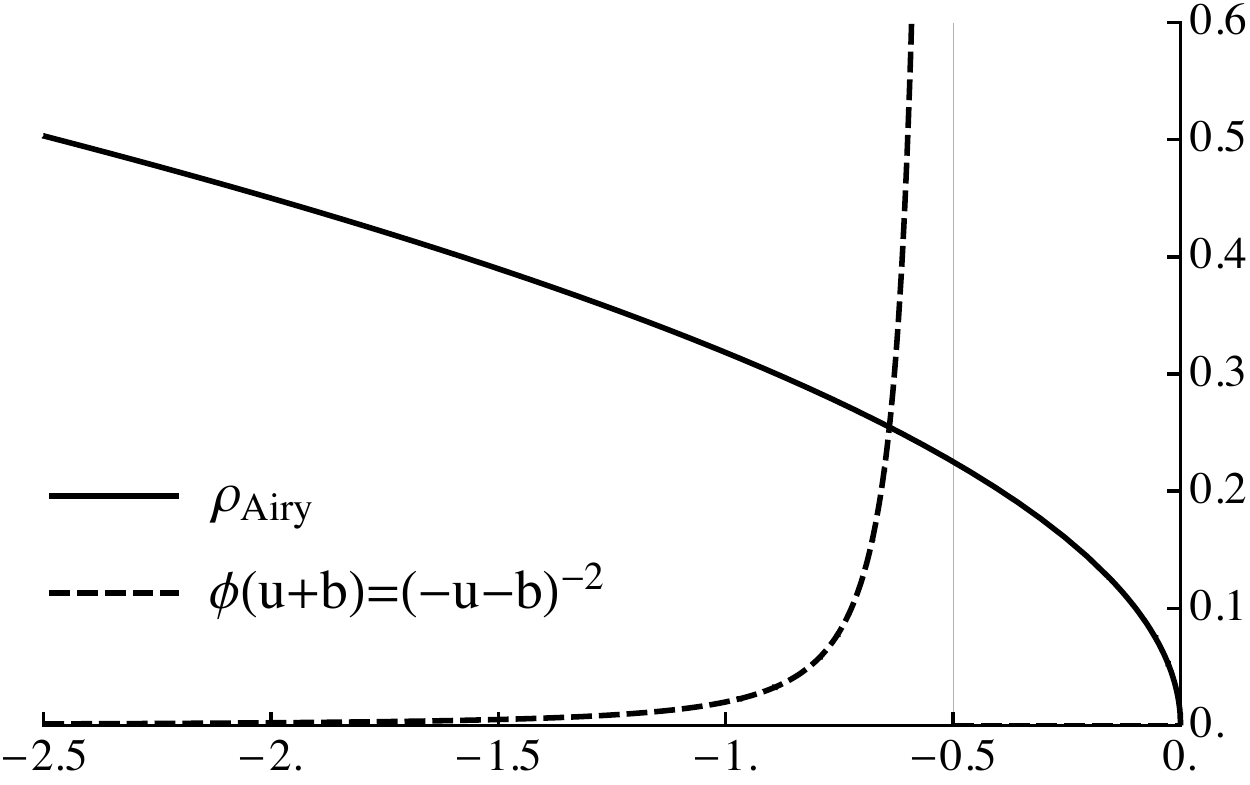}}
\caption{Representation of an inverse monomial wall with $\delta=2$ and of the density of the unperturbed Coulomb gas $\rho_{\rm Ai}$.}
\label{fig:inverse_exp}
\end{figure}

Another example of functions in the set $\Omega_1$ are the inverse monomial walls 
\be
\phi(z)= (-z)^{-\delta} \quad b<0 \quad , \quad \phi(z)=+\infty \quad z>0
\ee
such that $\phi(u + t^{-2/3} a_i)$ has an infinite hard wall for $t^{-2/3} a_i>-u$,
which penetrates the semi-circle as a power law for $t^{-2/3} a_i < -u$. For $u>0$
the infinite hard wall part penetrates the semi-circle, while for $u<0$ it does not. The influence
of the wall can be felt for any $u \in \mathbb{R}$ although it becomes weaker and
weaker for a distant wall at large negative $u$ (see Fig.~\ref{fig:inverse_exp}).\\

Since $a_i \sim - i^{2/3}$ for large $i$, we see that one must take $\delta > 3/2$ for
$L$ to be a convergent sum. The function $f(u)$ is given for $u<0$ as
\be
f(u) = \frac{1}{2} \int_0^{+\infty} \frac{\rmd b}{\sqrt{b}} \frac{\delta}{(b-u)^{\delta+1}} 
= \frac{D_\delta}{(-u)^{\delta+ \frac{1}{2} }} 
\quad , \quad D_\delta = \frac{\sqrt{\pi} \Gamma(\delta+\frac{1}{2})}{2 \Gamma(\delta)} 
\ee 
with $f(u>0)=+\infty$. We must thus solve
\be
w = \frac{D_\delta}{(-u + \frac{4}{\beta \pi} w)^{\delta+ \frac{1}{2} }} \Longleftrightarrow 
u = \frac{4}{\beta \pi} w - (\frac{D_\delta}{w})^{\frac{1}{\delta + 1/2}}
\ee 
There is a unique positive solution $w=w(u)$ (since $\phi \in \Omega_1 \subset \Omega_2$)
increasing function of $u \in \mathbb{R}$, with 

\begin{itemize}
\item $w(u) \simeq \frac{D_\delta}{(-u)^{\delta+ \frac{1}{2} }}$ for large negative $u$ (distant wall),
\item $w(u) \simeq \frac{\beta \pi}{4} u$ for large
positive $u$.
\end{itemize}

The SAO/WKB {\sf SP1} saddle point is then $v(x)=\frac{4}{\beta \pi} w(u-x)$ for $x>0$
and is now non-zero and decreasing on $\mathbb{R}^+$ and decaying at large $x$
as $v(x) \simeq \frac{4}{\beta \pi}  \frac{C_\delta}{x^{\delta+ \frac{1}{2} }}$.
For the excess energy $\Sigma_\phi(u)=\frac{\beta}{4} \int_0^{+\infty} \rmd x\, x v(x)$ to be finite, 
we need $\delta > 3/2$ as anticipated above. We can use the following formula to obtain the expression of the excess energy
\be
\Sigma_{\gamma}(u)=
\frac{1}{2 \pi} \int_0^{w(u)} \rmd w' [u(w')-u]^2 
= \frac{1}{2 \pi} \int_0^{w(u)} \rmd w' 
[\frac{4}{\beta \pi} w' - (\frac{D_\delta}{w'})^{\frac{1}{\delta + 1/2}} - u]^2
\ee
Performing the integral and replacing all  $(\frac{D_\delta}{w})^{\frac{2}{2\delta+1}}$ factors 
by $\frac{4}{\beta\pi } w-u $, we find
\begin{equation}
\begin{split}
\Sigma_{\gamma}(u) &=\frac{w(u) \left(12 \pi ^2 \beta ^2 \delta  u^2-3 \pi  \beta  (2 \delta +3) (6 \delta -1) u w(u)+4 (2 \delta -1) (2 \delta +3)^2 w^2(u)\right)}{3 \pi ^3 \beta ^2 \delta 
   (2 \delta -3) (2 \delta -1)}\\
   \end{split}
\end{equation}
The asymptotic behavior of the excess energy is
\begin{itemize}
\item $\Sigma_\phi(u)=\frac{\Gamma \left(\delta -\frac{3}{2}\right) }{2 \sqrt{\pi } \Gamma (\delta )} \frac{1}{(-u)^{\delta-\frac{3}{2} }}$ for large negative $u$ (distant wall),
\item $\Sigma_\phi(u)=\frac{\beta}{24}u^3$ for large positive $u$.
\end{itemize}

\subsection{Upper and lower bounds on the excess energy }
The inverse monomial walls $\phi(z)$ are larger then the hard wall potential $\phi_{\rm HW}$ implying the lower bound on the excess free energy
\begin{equation}
\forall u\geq 0, \qquad   \Sigma_\phi(u) \geq \frac{\beta}{24}u^3
\end{equation}
Besides, by the Jensen (first cumulant) inequality, we have
\begin{equation} 
\forall u\leq 0, \quad \Sigma_\phi(u)\leq  \frac{\Gamma \left(\delta  -\frac{3}{2}\right) }{2 \sqrt{\pi } \Gamma ( \delta )}\frac{1}{(-u)^{\delta-\frac{3}{2} }}
\end{equation}

\section{Relation to truncated linear statistics: matching bulk and edge  }

In this Section we show that there is a smooth matching between the 
results of Ref. \cite{grabsch2017truncated} for truncated linear
statistics in the bulk and our results at the edge for the linear wall $\gamma=1$. The details of the matching are non-trivial and instructive. 
Furthemore we show universality, i.e. for any linear statistics smooth at
a soft edge we obtain, up to coefficients, the same results
corresponding to ours for $\gamma=1$.

\subsection{Summary of results in the bulk}
Let us first recall briefly the results of Ref. \cite{grabsch2017truncated}. We use most of 
their notations. They study,
for large $N$
\be
{\cal L} = \sum_{i=1}^{N_1} f(\lambda_i) \quad , \quad f(\lambda)=\sqrt{\lambda}  \quad , \quad \kappa=\frac{N_1}{N}
\ee 
where the sum is over the $N_1$ largest eigenvalues of the Laguerre ensemble,
which can be written as $\lambda_i = N x_i$. Let us define the general density
of eigenvalues in the bulk as $\hat \rho(x)=N^{-1} \sum_{i=1}^N \delta(x-x_i)$,
i.e. with unit normalization. 
For the Laguerre ensemble at large $N$ this density converges to
the Marchenko-Pastur distribution $\hat \rho(x)=\rho_{\rm MP}(x)=\frac{1}{2 \pi} \sqrt{\frac{4-x}{x}}$,
which has a soft edge at $x=4$ with locally a semi-circle form. In Ref \cite{grabsch2017truncated} 
the scaling $N_1/N=\kappa$ fixed was studied. The question is whether for 
small $\kappa$ one is able to match to the edge problem studied here.\\

From the replacement $\lambda_i = N x_i$ one sees that for $f(\lambda)=\sqrt{\lambda}$ the typical
size of the sum is of order $ N^{3/2}$ (for $\kappa>0$), hence the authors introduced 
the random variable\footnote{For simplicity we abuse notations
by denoting with the same letter $s$ the random variable and its value.} $s={\cal L}/N^{3/2}$ 
and obtained the following large deviation form 
for the probability of $s$, at fixed $\kappa$ 
\be \label{probform}
P_\kappa(s) \sim \exp\left(- { \frac{\beta N^2}{2}}  \Phi_\kappa(s) \right)
\ee 
The typical (and mean) value for $s$, denoted $s_0(\kappa)$, such that
$\Phi_\kappa(s_0(\kappa))=0$, is obtained 
simply by noting that for large $N$ there is a well defined 
level $c_0$ in $x$ which corresponds to $\kappa$, and by
eliminating $c_0$ in the system
\be
s_0= \int_{c_0}^4 \sqrt{x} \rho_{\rm MP}(x) 
\quad , \quad 
\kappa = \int_{c_0}^4 \rho_{\rm MP}(x) 
\ee 
Let us recall here the result for small $\kappa$, to the order relevant for us
\be \label{exps0} 
s_0(\kappa) = 2 \kappa - \frac{3 (3 \pi)^{2/3}}{10 \times 2^{1/3}} \kappa^{5/3} + \dots
\ee 
where the $2 \kappa$ comes simply from $\sqrt{x_{\rm edge}}=2$.\\

The authors of Ref. \cite{grabsch2017truncated} write the JPDF of the eigenvalues $x_i$ 
as $\sim \exp(- { \frac{\beta N^2}{2}} {\cal E}[\hat \rho])$ where ${\cal E}[\hat \rho]$ is the standard logarithmic Coulomb Gas in the
bulk. We have here generalized their calculation to arbitrary $\beta$, which is immediate.
In addition to the usual constraint $\int \rmd x \hat \rho(x)=1$, they impose
\be \label{constraints} 
\int_{c}^d \rmd x \hat \rho(x) = \kappa \quad , \quad \int_{c}^d \rmd x \sqrt{x} \hat \rho(x) = s
\ee 
where $d$ is the upper edge of the support of $\rho$. They add Lagrange multiplier
terms, ${\cal E}[\hat \rho] \to {\cal E}[\hat \rho] + \mu_1(\int_{c}^d \rmd x \sqrt{x} \hat \rho(x) -s)$
and similarly for the two other conditions (see their equation (3.8)). They then look for the  minimal energy configuration, in the ensemble with fixed $\kappa,s$, which we denote $\hat \rho(x) = \rho_{\kappa,s}(x)$.\\

Here we discuss only the side $s<s_0(\kappa)$, relevant for us, and where the
density has a single interval support. They find the optimal
density
\be \label{rho0} 
\rho_{\kappa,s}(x) = \frac{1}{2 \pi} \sqrt{\frac{d-x}{x}} + \frac{4-d}{8\pi \sqrt{d-c} \sqrt{x}} 
\log \frac{\sqrt{d-c} + \sqrt{d-x}}{|\sqrt{d-c} - \sqrt{d-x}|}
\ee
where the three parameters $c,d,\mu_1$ are determined at the optimum by the three
equations (3.35), (3.36) and (3.37) there, as a function of $\kappa,s$. 
The last two equations simply express the constraints \eqref{constraints}.
The large deviation function $\Phi_\kappa(s)$ is determined from 
integrating the relation 
\be
\Phi'_\kappa(s) = - \mu_1
\ee 
where $\mu_1=\mu_1(s,\kappa)$. No closed form was found but $\Phi_\kappa(s)$
was determined perturbatively near $s_0(\kappa)$ and for $s \to 0$. The optimal density \eqref{rho0} is strikingly similar to the one obtained here for the linear wall
$\gamma=1$ (i.e. the result related to the KPZ large deviations first obtained in \cite{JointLetter}). 
We now explain why, and give the connection between the two sets of results. 

\subsection{Connecting bulk truncated linear statistics and the soft wall at the edge}

Let us start from the bulk, and consider a general linear statistics in the limit to the edge, $\kappa=N_1/N \to 0$.
From the universality of the soft edge,
the eigenvalues very near the edge take the form $\lambda_i = 4 N + \alpha a_i N^{1/3}$
for some constant $\alpha$, where the $\lbrace a_i \rbrace $  forms the $\beta$ Airy point process. For the case studied in Ref \cite{grabsch2017truncated} it is easy to see
that $\alpha=4^{2/3}$. We can thus write heuristically, for large $N$
\be
{\cal L }= \sum_{i=1}^{N_1} f(4 N + \alpha a_i N^{1/3}) =
N_1 f(4 N) + \alpha N^{1/3} f'(4 N) \sum_{i=1}^{N_1} a_i 
+ \frac{1}{2} \alpha^2  N^{2/3} f''(4 N) \sum_{i=1}^{N_1} a_i^2  + \dots
\ee 
while the first two terms are certainly present, the last term (neglected below)
may not be the only subdominant one, but it is sufficient 
to illustrate our argument. 
For $f(\lambda)=\sqrt{\lambda}$ we see that 
\be
{\cal L }= 2 N^{1/2} N_1 + \frac{\alpha}{4 N^{1/6}} \sum_{i=1}^{N_1} a_i  - \frac{\alpha^2}{64 N^{5/6}}  \sum_{i=1}^{N_1} a_i^2  + \dots
\ee
For $N_1$ large we can use that the ordered $a_i \simeq - (3 \pi/2)^{2/3} i^{2/3}$ 
at large $i$, and obtain the typical value
\be
{\cal L }_{typ}= 2 N^{1/2} N_1 - \frac{3 \alpha (3 \pi/2)^{2/3}}{20 N^{1/6}} N_1^{5/3}  
+ \mathcal{O}( N^{-5/6} N_1^{7/3}) 
\ee
Inserting $N_1=\kappa N$ and $\alpha=4^{2/3}$ it correctly reproduces the first two terms in the expansion \eqref{exps0}
of $s_0(\kappa)$ at small $\kappa$ of \cite{grabsch2017truncated}. This already
suggests a smooth matching to the edge, since here we used the Airy point process, at least
at the level of typical fluctuations. Note however that
all terms are of the same order $N^{3/2}$ and the only perturbative parameter is thus 
small $\kappa$, which suggests that the higher order terms can be dropped. \\

On the other hand for $N_1$ large but fixed, taking $N$ to infinity first, we see that
the successive terms in the expansion become smaller and smaller, 
and that the only remaining fluctuating term is linear in the Airy point process
\be
\alpha N^{1/3} f'(4 N) \sum_{i=1}^{N_1} a_i 
\ee
which is similar to the one which we studied for the monomial wall 
with $\gamma=1$, i.e. $\phi(z)=(z)_+$ 
with the correspondence indicated in the Letter, that {\it the typical} 
$N_1 \simeq \frac{2 u^{3/2}}{3 \pi} t$ (both $N_1$ and $t$ large). Let us now make this more precise. We want to compare the 
bulk random variable $s$ studied in \cite{grabsch2017truncated} (first line) 
and the edge random variable $\tilde L$ studied here (second line) 
\bea \label{precise} 
&& s= 2 \kappa + 4^{-1/3} N^{-5/3} \sum_{i=1}^{N_1} a_i  \\
&& \tilde L = t^{-2} {\sf L} = t^{-1} \sum_i (u + t^{-2/3} a_i)_+ \simeq 
u \frac{K}{t} + t^{-5/3} \sum_{i=1}^{K} a_i 
\eea 
We must be careful that the first problem was studied at fixed $s, \kappa=\frac{N_1}{N}$,
while the second is studied at fixed $u$, and that here $K$ is {the cut-off} defined by $a_K \approx - t^{2/3} u$, {i.e. the largest index for which $u+t^{-2/3}a_i>0$},
so it is a priori  a fluctuating quantity. To connect the two, we want to identify
$K=N_1$ although the ensembles may be different. It turns out that it works,
and that for large $N_1,t$ at fixed $N_1/t$ one can perform this identification,
in the way described below.\\

Thus, to summarize, we want to identify , at small $\kappa$
\be \label{id} 
s-s_0(\kappa) = 4^{-1/3} (\frac{t}{N})^{5/3} (\tilde L - \tilde L_{typ}) 
\ee 
A first check is to calculate the variance of both sides. Using the
result for the variance of $s$ in (4.9) of \cite{grabsch2017truncated}
we obtain for small $\kappa$ 
\be \label{var} 
{\rm Var}(s) \simeq   \frac{2}{\beta} \frac{(6 \pi \kappa)^{4/3}}{16 \pi^2 N^2}  = 4^{-2/3} 
\frac{2}{\beta}  (\frac{t}{N})^{10/3}  \frac{u^2}{\pi^2 t^2} 
= 4^{-2/3} (\frac{t}{N})^{10/3} {\rm Var}(\tilde L) 
\ee
consistent with Eqs.~\eqref{cumL1} and \eqref{eq:cum_L_gamma11}. In the middle equality we have replaced $N_1$ by $K_{typ}$ where $K_{typ}$ is determined by 
\be
\int_0^{u} \rmd b \, \rho_{\rm Ai}(b) = \frac{K_{typ}}{t} \quad \Leftrightarrow \quad \frac{K_{typ}}{t} = \frac{2}{3 \pi} u^{3/2}
\ee
This is consistent with \eqref{id}.\\

However, if one now calculates the third cumulant on both sides one finds {\it that it fails}
by a factor of $4$. The reason is that the two ensembles (fixed $\kappa$ and fixed
$u$) are not identical, and must be related by a transformation which we describe below.
Hence the identification \eqref{id} is subtle, although the final correct version
is quite natural, as shown below. In a nutshell the idea is that, conditionned to
a very atypical value of $s$, or of $\tilde L$, the optimal density deviates from the unperturbed semi-circle 
$\rho_{\rm Ai}(b)$, hence the relation of $N_1/t$ to $u$ changes.

\subsection{Recalling results for $\gamma=1$} 
Before working it out, it is useful to recall the full set of results for the monomial wall $\gamma=1$, $\phi(z)= (z)_+$. We will indicate by an index $B$ the result for
$\phi \to B \phi$ since we need the amplitude $B$ to probe all values for $\tilde L$ by Legendre transform.
The associated functions $f_B(u)$ and $w(u)=w_B(u)$ solving Eq.~(15) of the Letter are
\be
f_B(u) = B \sqrt{(u)_+} \quad , \quad u^B_0={ \frac{4}{\beta \pi} 
w_B(u)=\frac{8 B^2}{\pi^2 \beta^2} (\sqrt{1+ \frac{\beta^2 \pi^2}{4 B^2} u}-1)}
\ee 
leading to the excess energy which we call $\Sigma_B(u)$, and its scaling form
\be \label{sigexpl}
\Sigma^{(\beta=2)}_{B=1}(u):=\Sigma(u)=\frac{4 \left(\pi ^2
   u+1\right)^{5/2}}{15 \pi ^6}-\frac{u^2}{2 \pi ^2}-\frac{2 u}{3 \pi
   ^4}-\frac{4}{15 \pi ^6} \;\;  , \;\; { \Sigma^{(\beta)}_{B}(u)= B^6 (\frac{2}{\beta})^5 \Sigma(u (\frac{\beta}{2B})^2) }
\ee 
The optimal density and its scaling property read 
\bea \label{opt1} 
&& \rho^{(\beta)}_{*,B,u}(b)=
\frac{\sqrt{b-u_0} }{\pi} 
+ \frac{B}{\beta  \pi^2} \log \frac{\sqrt{u-u_0} + \sqrt{b-u_0}}{|\sqrt{u-u_0} - \sqrt{b-u_0}|} 
\quad , \quad u_0=u_0^B
\\
&&   \frac{\beta \pi}{4 B} u^B_0 = \sqrt{u-u^B_0}  \quad , \quad \rho^{(\beta)}_{*,B,u}(b)=
 \frac{2B}{\beta} \rho^{(2)}_{*,1,u (\frac{\beta }{2B})^{2}}(b (\frac{\beta }{2B})^2)  
\eea 

\subsection{Matching bulk and edge}

The optimal edge density \eqref{opt1} is strikingly similar to the optimal bulk density \eqref{rho0}
near the soft edge, $x,d,c \approx 4$. More precisely if we write
\bea \label{id2} 
d=4 - \epsilon u^B_0 \quad , \quad x=4 - \epsilon b \quad , \quad c=4 - \epsilon u \quad , \quad \sqrt{d-c} = \sqrt{\epsilon} 
\sqrt{u-u^B_0} 
\eea
then \eqref{rho0} becomes, using {$\sqrt{u-u_0^B}=\frac{\beta \pi}{4 B} u_0^B$}, for small $\epsilon$
\be \label{corrdens} 
\rho_{\kappa,s}(x) \simeq \frac{\sqrt{\epsilon}}{4} \rho^*_{B,u}(b)
\ee
The factor $\epsilon$ which relates the volume elements, $\rmd x=\epsilon \rmd b$, can
be predicted from the edge behavior of the rescaled eigenvalues {$x_i=\lambda_i /N$}
\be
\rmd x_i = \frac{\alpha}{N^{2/3}} \rmd a_i = \alpha (\frac{t}{N})^{2/3} \rmd b_i 
\ee 
with here $\alpha=4^{2/3}$ hence
\be \label{epsilon}
\epsilon= (\frac{4 t}{N})^{2/3} = (\frac{4 t}{N_1})^{2/3} \kappa^{2/3} 
\ee 
The number of eigenvalues in interval $\rmd x$ is equal to the one in $\rmd b$, hence, from our
definitions, $N \hat \rho(x) \rmd x = t \rho(b) \rmd b$. Using $\rmd x=\epsilon \rmd b$ and \eqref{epsilon}
explains the prefactor in the correspondence \eqref{corrdens}.\\

We now justify the equations \eqref{id2}, and give the correspondence between parameters
$\kappa,s$ and $B,u$. We also derive the relation between our function $\Sigma_{\phi}(u)$
and the results of \cite{grabsch2017truncated}. Let us now write the equations (3.35), (3.36) and (3.37) in the limit $\kappa \to 0$. 
Let us define $d= 4 - \epsilon d_1$, $c=4 - \epsilon c_1$, then these equations
become, discarding terms of higher orders in $\epsilon$
\bea \label{equations} 
&& \kappa = \frac{\sqrt{c_1-d_1}}{24 \pi} (4 c_1-d_1) \epsilon^{3/2} \\
&& s - 2 \kappa = \frac{\sqrt{c_1-d_1} \left(d_1^2 - 2 c_1 d_1- 4
   c_1^2\right)}{160 \pi } \epsilon^{5/2} \\
&& \mu_1= \frac{\pi}{2} \frac{d_1}{\sqrt{c_1-d_1}} \epsilon^{1/2} 
\eea 
So one must calculate the $\mathcal{O}(1)$ parameters $c_1$ and $d_1$ as functions of $s$ and $\kappa$, obtain $\mu_1(\kappa,s)$ from the last equation, and finally obtain the large deviation for the probability 
by integration
\be
\Phi_\kappa(s) = - \int_{s_0(\kappa)}^s \mu_1(\kappa,s) \rmd s 
\ee 
at fixed $\kappa$. The typical value $s_0(\kappa)$ in \eqref{exps0} is recovered and corresponds
to $d_1=\mu_1=0$. It is easy to see that $\Phi_\kappa(s)$ then takes the scaling form in the small $\kappa$
limit (i.e. small $\epsilon$) 
\be \label{scalingsmall} 
\Phi_\kappa(s) = \kappa^2 \hat \Phi(S) \quad , \quad S=\kappa^{-5/3} (s-s_0(\kappa))
\ee 

To calculate $\hat \Phi(S)$ we proceed as follows. Define $\tilde \kappa=\kappa \epsilon^{-3/2}$. Set
$\sqrt{c_1-d_1}= y (6 \pi \tilde \kappa)^{1/3}$, substitute for $c_1$ as a function of
$d_1$ and $y$ in the first equation in \eqref{equations} then solve the resulting linear equation for for $d_1$. 
Report $c_1$ and $d_1$ in the second equation. It reads now
\be
S = \pi^{2/3} 6^{-1/3} ( \frac{y^5}{10} + y^2 - \frac{2}{y} + \frac{9}{10} ) 
\ee 
We can also report $c_1$, $d_1$ in $\mu_1$ and obtain
\be
\frac{\mu_1}{\kappa^{1/3}} = \frac{(2 \pi)^{4/3}}{3^{2/3}}  (\frac{1}{y^2}-y)
\ee 
Hence $y=1$ corresponds to the typical value. 
Then we have
\be
\Phi_\kappa =  \frac{2 \pi^2}{3} \kappa^2 \int_1^y (y'-\frac{1}{y'^2}) \rmd\big[ \frac{y'^5}{10} + y'^2 - \frac{2}{y'} \big]
\ee 
leading to the parametric formula for $\hat \Phi(S)$
\bea \label{PhiS}
&& \hat \Phi(S)=  \frac{2 \pi^2}{3} ( \frac{y^6}{12}+\frac{y^3}{2}+\frac{2}{3 y^3}-\frac{5}{4}) \\
&& S = \pi^{2/3} 6^{-1/3} ( \frac{y^5}{10} + y^2 - \frac{2}{y} + \frac{9}{10} ) 
\eea 
The side $S<0$ corresponds to $y<1$.\\

 On this parametric form it is easy to generate the series in $S$ around $S=0$ by
expanding around $y=1^-$. We put it in the following convenient form
\bea \label{PhiSsmall}
&& \hat \Phi(S) = (\frac{3 \pi}{2})^2 \tilde \Phi(2^{7/3} 3^{-5/3} \pi^{-2/3} S) \\
&& \tilde \Phi(S) = \frac{S^2}{2}-\frac{S^3}{6}+\frac{S^4}{24}-\frac{7
   S^6}{1440}+\frac{S^7}{1440}+\frac{7 S^8}{5760}-\frac{5
   S^9}{10368}+\mathcal{O}(S^{10})
\eea
where the $S^2$ term is compatible with the variance \eqref{var}.\\

The series for $S \to - \infty$ corresponds to $y \to 0^+$ and we obtain
\bea \label{PhiSlarge}
&& \hat \Phi(S) = 
\frac{2 \pi^2}{3} \bar \Phi(6^{1/3} \pi^{-2/3} S) \\
&& \bar \Phi(S) = -\frac{S^3}{12}+\frac{9 S^2}{40}-\frac{81
   S}{400}-\frac{757}{4000}-\frac{4}{5 S^3}-\frac{54}{25
   S^4}-\frac{486}{125
   S^5}+\mathcal{O}(S^{-6}) \eea 

To compare with our edge results it is more convenient to consider the 
generating function, $H_\kappa(\mu_1)$, of the cumulants of $s$
\be
\int \rmd s P_\kappa(s) e^{- { \frac{\beta N^2}{2}} \mu_1 s}  \sim e^{- { \frac{\beta N^2}{2}} H_\kappa(\mu_1)} 
\ee 
Inserting \eqref{probform} we see that it is given by the Legendre transform 
$H_\kappa(\mu_1) = \min_s [ \Phi_\kappa(s) + \mu_1 s]$. Since it satisfies $H'_\kappa(\mu_1) = s$ it is also easy to calculate
from an expansion of \eqref{equations} in powers of $\mu_1$
and integration $H_\kappa(\mu_1)= \int_0^{\mu_1} s(\mu_1) \rmd \mu_1$.
We obtain that for $\kappa \to 0$, $H_\kappa(\mu_1)$ takes
the scaling form
\be
H_\kappa(\mu_1) \simeq 2 \kappa \mu_1 + \kappa^2 \hat H(\mu_1/\kappa^{1/3}) 
\ee
and one finds the expansion
\bea \label{reshatH} 
&& \hat H(\mu) =  (\frac{3 \pi}{2})^2 \tilde H(\frac{\mu}{6^{1/3} \pi^{4/3}}) \\
&& \tilde H(\mu) = - \frac{2}{5} \mu -\frac{\mu ^2}{2}+\frac{\mu ^3}{6}-\frac{\mu
   ^4}{12}+\frac{\mu ^5}{24}-\frac{3 \mu
   ^6}{160}+\frac{\mu ^7}{144}-\frac{\mu ^8}{576}+\frac{11
   \mu ^{10}}{41472}+\mathcal{O}(\mu ^{11}) \nonumber 
\eea
The cumulants of $s$ can be extracted from $H_\kappa(\mu_1) = \sum_{n\geq 2} \frac{(-1)^{n+1}}{n!}
\mathbb{E}[s^n]^c N^{2 n-2} \mu_1^n$, and the variance agrees with \eqref{var}.
The two scaling forms obey the scaled Legendre transform relation
\be
\hat H(\mu) = \min_\sigma [ \hat \Phi(\sigma - s_1) + \mu \sigma ]
\ee 
where $s_1 = - \frac{3 (3 \pi)^{2/3}}{10 \times 2^{1/3}}$.\\

To identify with the edge problem we can compare the Coulomb gas free energy 
(3.8) in \cite{grabsch2017truncated} with our expression (29) in the Letter.
We note the equivalent roles of the terms
\be
{ \frac{\beta N^2}{2}} \mu_1 s \equiv t^2 B \tilde L
\ee
(both $s$ and $\tilde L$ denote here the fluctuating random variable),
which leads to the correspondence
\be \label{mu1B}
\mu_1 ={ \frac{2B}{\beta} }  (\frac{4 t}{N})^{1/3}  =  { \frac{2B}{\beta} }\sqrt{\epsilon}
\ee 
We thus now want to match the fixed $\mu_1$, fixed (and small) $\kappa$,
i.e. fixed $N_1$ bulk problem, with fixed $B$, fixed $K/t=N_1/t$ edge problem. 
In that edge problem $u$ is not fixed, and will be determined by optimisation.
Schematically we write (where here $\langle \dots \rangle$ denote averages over
the Coulomb Gas measure at fixed values of the parameters indicated in subscripts)
\bea
&& e^{- { \frac{\beta N^2}{2}} (H_\kappa(\mu_1)-2 \kappa \mu_1)} = \langle e^{- { \frac{\beta N^2}{2}} \mu_1 (s-2 \kappa)}  \rangle_{\kappa,\mu_1}
\\
&& \equiv  \langle e^{- { \frac{\beta N^2}{2}} 4^{-1/3} (\frac{t}{N})^{5/3} \mu_1 (\tilde L  - u \frac{N_1}{t}   )} 
\rangle_{\frac{N_1}{t},\mu_1} = \langle e^{- B t^2 (\tilde L- u \frac{N_1}{t}) } \rangle_{\frac{N_1}{t},B} 
\eea 
using \eqref{mu1B}. We can use that 
\be
\langle e^{- B t^2 \tilde L} \rangle_{u,B}  := \mathbb{E}_\beta[e^{- B t^2 \tilde L} ]= e^{- t^2 \Sigma_B(u)} 
\ee 
The question is now to fix $u$. It is natural to set $u$ to $u_*$ determined
by the fact that the number of eigenvalues below the level $u_*$ is precisely
equal to $N_1$, which leads to the condition
\be \label{condu} 
\frac{N_1}{t} = \int_{u_0^B(u_*)}^{u_*} \rmd b \, \rho^*_{B,u_*}(b)
\ee
where $\rho^*_{B,u}(b)$ is the optimal density for the edge problem at
fixed $B,u_*$. We can now write 
\be
\langle e^{- B t^2 (\tilde L- u \frac{N_1}{t}) } \rangle_{\frac{N_1}{t},B} 
= \exp\big( - t^2  \big[\Sigma_B(u_*) - u_* \frac{N_1}{t} B\big] \big)  
\ee 
We thus identify
\be
{ \frac{ \beta N^2}{2}} H_\kappa(\mu_1) = t^2  [ \Sigma_B(u_*) - u_* \frac{N_1}{t} B ] \quad , \quad \mu_1  = { \frac{2B}{\beta} }  \sqrt{\epsilon} 
\ee
Since ${ \frac{ \beta N^2}{2}} H_\kappa(\mu_1) = { \frac{\beta N_1^2}{2}} \hat H(\mu_1/\kappa^{1/3})$
and $\sqrt{\epsilon}= \kappa^{1/3} 4^{1/3} (\frac{t}{N_1})^{1/3}$ we obtain the
identification 
\be \label{idH} 
{ \frac{\beta}{2}}\hat H(\frac{2}{\beta} 4^{1/3} (\frac{t}{N_1})^{1/3} B) = (\frac{t}{N_1})^2  \big[ \Sigma_B(u_*) - u_* \frac{N_1}{t} B \big] 
\ee 
which should be valid for any (positive) values of the parameters $B, N_1/t$. Interestingly
one can show that the condition \eqref{condu} is equivalent to the condition 
\be \label{sad} 
\frac{N_1}{t} = \frac{\Sigma'_B(u_*)}{B} = B^3 \Sigma'_{1}(u_* B^{-2}) 
\ee 
where in the last equality we used the scaling property \eqref{sigexpl} of $\Sigma$.
Hence it shows that one can also write 
\be \label{idH2} 
{ \frac{\beta}{2}}\hat H(\frac{2}{\beta} 4^{1/3} (\frac{t}{N_1})^{1/3} B) = (\frac{t}{N_1})^2  \min_u[ \Sigma_B(u) - u \frac{N_1}{t} B \big] 
\ee 
To show that Eqs.~\eqref{condu} and \eqref{sad} are equivalent 
we rewrite 
\be
\int_{u_0^B}^u \rmd b \, \rho^*_{B,u}(b)  =
\frac{1}{\pi} \int_0^{+\infty} \rmd x \sqrt{(u-x-v_*(x))_+} = \frac{\beta}{4 B} \int_0^u \rmd x\,  v_*(x) 
= \frac{\Sigma'_{B}(u)}{B}  
\ee 
which is valid for any $u,B$ and 
where we used the SAO/WKB formula for the density together with the saddle point equation ${\sf SP1}$ i.e.
$\frac{\beta \pi}{4} v_*(x) = B \sqrt{ (u-x-v_*(x))_+}$ for the
linear soft wall $\gamma=1$. This is a remarkably simple formula for the total
number of eigenvalues which belong to the support of the linear soft wall. \\

Note that rewriting $B=\frac{\beta}{2} B'$ and using the general $\beta$ dependence 
\eqref{genbetasigma} we can express the relation independently of $\beta$ 
(since $\hat H$ is $\beta$-independent)
\be \label{idH3} 
\hat H(4^{1/3} (\frac{t}{N_1})^{1/3} B') 
=  (\frac{t}{N_1})^2  \min_u[ \Sigma^{\beta=2}_{B'}(u) - u \frac{N_1}{t} B' \big] 
\ee 

Introducing the variables $U=u B'^{-2}$ and $y=  (\frac{t}{N_1})^{1/3} B'$, Eq. \eqref{idH} becomes 

\be
\hat H(4^{1/3} y) = y^6  \min_U[ \Sigma(U) - U y^{-3}] 
\ee 
We can now check this prediction, which comes from the identification described above.
Inserting on the r.h.s. the function $\Sigma$ from \eqref{sigexpl} and
performing the Legendre transform we obtain a series at small $y$ which
perfectly matches the result \eqref{reshatH} from the bulk calculation.  \\

Finally, we can check that the parameters $c_1$ and $d_1$ solutions 
of the first and last equations in \eqref{equations} at fixed $\kappa$ and
$\mu_1$ coincide with $u_*$ and $u_0$ the endpoint of the edge density for
the corresponding value of $N_1/t$ and $B=\mu_1/\sqrt{\epsilon}$
\be \label{idc1} 
c_1(\frac{N_1}{t},\mu_1)=u_* \quad , \quad d_1(\frac{N_1}{t},\mu_1)=u_0^B(u_*) 
\ee
i.e. the endpoint of the density predicted from the edge 
coincides with the one predicted of the bulk, which provides
another consistency check.\\

In summary, we have shown the perfect matching of 
(i) {\it truncated} linear statistics in the bulk for large $N$, at fixed $\kappa=N_1/N$,
in the limit of small $\kappa$ (ii) our results for the linear soft wall $\gamma=1$ at the edge
for large $t$ and $N_1$, at fixed ratio $N_1/t$, with a wall parameter $u=u_*$ determined
by the condition \eqref{condu}. This equation simply expresses that there are $\simeq N_1$ 
eigenvalues below the level $u_*$ 
in the optimal density calculated for that value of $u_*$. Furthermore one can
identify this condition with the first of the equations \eqref{equations}. This is the meaning
of \eqref{idc1} and it
is quite natural since it is the small $\kappa$ limit of the 
same bulk condition (3.36) and (3.16) in \cite{grabsch2017truncated}. 
So the endpoint of the support of the optimal density matches smoothly. 
The identification
was performed here at fixed chemical potential $\mu_1$ which
corresponds to fixed wall amplitude $B$. The equation for $\mu_1$ 
(last of \eqref{equations}) can be seen as our central equation (15) in the Letter.
The matching can also be performed in different ensembles, and the PDF of $s$ and
$\tilde L$ can be similarly related using the appropriate Legendre 
transforms. \\

It should be possible to perform a similar limit on the solution of
Ref.~\cite{grabsch2017truncated} with splitted supports, i.e. $s>s_0(\kappa)$, which should then provide a solution to the edge problem for $\gamma=1$ but
for $B<0$, the pulled Coulomb Gas, this is left for future study. 

It would also be interesting to be able to treat more general cases of functions $\phi$
by taking the limit from the bulk linear statistics. Preliminary calculations show that
it requires to linear statistics functions $f(\lambda)$ singular at the edge since 
truncated linear statistics with smooth functions always lead to $\gamma=1$, and
work in that direction is in progress. 

\section{Various applications of the exponential soft wall  } 
Because of its special form the exponential wall $\phi(z)=e^{z}$  enjoys a number of applications.

\subsection{Exponential linear statistics in the bulk}
A simple way to generate {the linear statistics with an exponential wall }${\sf L}$ with the exponential wall from a bulk linear statistics ${\cal L}$ is to consider
sums of the kind  

\be
{\cal L} = t  \sum_i e^{(N/t)^{2/3} (\lambda_i-2)} \simeq t \sum_i e^{t^{-2/3} a_i} 
\ee 
which for $t/N \ll 1$ are dominated by the edge, and to which our results apply for large $t$.

\subsection{Large power of a random matrix}
A concrete realization {of the exponential linear statistics} is given by the large powers of a random matrix, in the spirit of 
Refs. \cite{gorin2018kpz,GorinShkolnikov2016}. Consider $\beta=2$ for simplicity and a standard $N\times N$ GUE random matrix $M_N$ 
with the measure such that the support is $[-2,2]$ at large $N$ (as in Eq. (1) in the Letter).
Define the matrix $M'_N=\frac{1}{2} M_N$, then the quantity 
\be \label{gorin} 
{\cal L} =    \frac{t}{2} {\rm Tr}  \left[  (M'_N)^{[2  (N/t)^{2/3}]} + (M'_N)^{[2  (N/t)^{2/3}]+1} \right]
\ee
can be written as 
\be 
{\cal L} =   \frac{t}{2}  \sum_i  \left[  (\lambda'_i)^{[2  (N/t)^{2/3}]} + (\lambda'_i)^{[2  (N/t)^{2/3}]+1}  \right]
\ee 
where $\lambda'_i$ are the eigenvalues of $M'_N$.
For $N \to +\infty$ at fixed $t$ these sums are clearly dominated by the two edges. If one considers
the contribution from the right edge
\be
(\lambda'_i)^{[2  (N/t)^{2/3}]}  \simeq e^{[2  (N/t)^{2/3}] \log(1+ \frac{a_i}{2 N^{2/3}})}
\simeq e^{t^{-2/3} a_i} 
\ee
The contribution of the left edge will be cancelled by the second term in \eqref{gorin}. 
Hence for large $N$
\be 
{\cal L} \simeq {\sf L} = t   \sum_i e^{t^{-2/3} a_i} 
\ee
the limit being in law. Hence our results for the exponential wall readily apply to traces of a large power of
a GUE matrix,  and of a $\beta$ tridiagonal random matrix. Note that the
power is $ (N/t)^{2/3}$ where $N$ is taken large first (the needed condition is $t \ll N$).
While the scaling in $N$ is similar to the quantities considered in Ref.~\cite{gorin2018kpz}
the $t$ dependence is quite different (there a matrix element was considered instead
of a trace).

\subsection{Quantum particle and polymer in linear plus random potential at high temperature}

There are several problems related to the canonical partition sum 
\be
Z(T):=Z_{\rm SAO}(T,\beta) = \sum_i e^{ - \epsilon_i/T} \label{ZSAO} 
\ee 
where the $\epsilon_i=-a_i$ are the eigenenergies of the SAO Hamiltonian, 
${\cal H}_{SAO}$, given by (7) in the Letter, equivalently the reversed Airy$_\beta$ point process.\\

\begin{itemize} 

\item One is a quantum particle in a linear plus random potential described by the Hamiltonian
\be \label{calH} 
{\cal H} = - D  \partial_z^2 + h z + \sqrt{g} W(z)
\ee 
with $\overline{W(z) W(z')}=\delta(z-z')$ and vanishing
wavefunction at $z=0$ (Dirichlet boundary condition). Defining $z=(D/h)^{1/3} y$ we obtain
\be
{\cal H} = h^{2/3} D^{1/3} {\cal H}_{\rm SAO}|_{\beta=4 D h/g}
\ee
in the notations of (7) in the Letter. The energy levels are thus
$h^{2/3} D^{1/3} \epsilon_i$. We study the canonical partition sum for this particle at temperature $T'$. 
\be \label{part1} 
Z_{h,g}(T') = \sum_i e^{ - h^{2/3} D^{1/3} \epsilon_i/T'} = Z_{\rm SAO}(T=h^{-2/3} D^{-1/3} T',\beta=4 D h/g)
\ee

\item An equivalent realization is a continuum polymer in $d=1+1$, directed along $\tau$, in presence of columnar disorder and of a linear binding potential to the wall, of length $1/T'$, described by the partition sum
\be
\!\!\!\! {\cal Z}(z_1,z_0,\frac{1}{T'}) = \int_{z(0)=z_0}^{z(\tau)=z_1}\!\!\! \mathcal{D}z(\tau) \exp\left(- \int_0^{1/T'} \rmd \tau \big[ \frac{1}{4 D} (\frac{\rmd z(\tau)}{\rmd \tau})^2 + h z(\tau) + \sqrt{g} W(z(\tau))\big]\right)
\ee 
where the sum is over paths $z(\tau) \geq 0$, i.e. there is an impenetrable hard
wall at $z=0$. Introducing the eigenstates $\psi_i$ of ${\cal H}$ one has
\be
{\cal Z}(z_1,z_0,\frac{1}{T'}) = \sum_i \psi_i^*(z_1) \psi_i(z_0) e^{- h^{2/3} D^{1/3}  \epsilon_i/T'} 
\ee
It was shown in Refs.~\cite{gorin2018kpz,GorinShkolnikov2016} that the partition sum
{\it with fixed enpoints near the wall} for $\beta=2$ is equal in law to the one point
partition sum associated to the KPZ problem.  More precisely, for $D=1/2$, $h=1/2$, $1/T'=2 b$ and $g=1/2$
one finds
\be \label{equivKPZ} 
\lim_{\varepsilon \to 0} \varepsilon^{-2} {\cal Z}(\varepsilon,\varepsilon,\frac{1}{T'}) \simeq 
\sum_i \psi^{\prime *}_i(0) \psi'_i(0) e^{- h^{2/3} D^{1/3} \epsilon_i/T'} \equiv
Z_{\rm KPZ}(0,b^3) e^{b^3/12} 
\ee
where $Z_{\rm KPZ}(z,t)= e^{{\sf H}_{\rm KPZ}(z,t)}$ with ${\sf H}_{\rm KPZ}(z,t)$ being the solution for the height field of the full
space KPZ equation with droplet initial conditions at $z=0$ in the units and conventions of Ref.~\cite{KrajLedou2018}.
To obtain Eq.~\eqref{equivKPZ} from the identity between (1.9) and (1.12) 
in \cite{gorin2018kpz} we note that the latter contains an expectation over Brownian excursions,
thus a ratio of partition sums whose denominator is simply the free Brownian propagator in presence of the absorbing wall. Hence (1.12) reads
\be
\frac{1}{\sqrt{4 \pi b^3}} \lim_{z_1,z_2 \to 0}
\frac{\langle z_1 |e^{-2 b {\cal H}}|z_2 \rangle}{\langle z_1 |e^{-2 b {\cal H}_{g=0,h=0}}|z_2 \rangle}
\simeq \lim_{z_1,z_2 \to 0} \frac{\langle z_1 |e^{-2 b {\cal H}}|z_2 \rangle}{z_1 z_2} 
\ee 
identical to \eqref{equivKPZ}. One checks that the small $b$ behavior,
$Z_{\rm KPZ}(0,b^3) \simeq_{b \to 0} (4 \pi b^3)^{-1/2}$, matches.

Here instead we are interested in identifying and summing over the endpoints 
\be
\int_0^{+\infty} \rmd z {\cal Z}(z,z,\frac{1}{T'}) = \sum_i e^{- h^{2/3} D^{1/3} \epsilon_i/T'}
= Z_{\rm SAO}(T=h^{-2/3} D^{-1/3} T',\beta=4 D h/g)
\ee 
as in Eq.~\eqref{part1}. Note that the inverse temperature $1/T'$ of the 
particle problem plays the role of the polymer length. 

\end{itemize}

We thus study the partition sum $Z(T)$ defined by Eq.~\eqref{ZSAO}.
Consider the high temperature limit, and write $T=t^{2/3} \gg 1$.
The reduced partition sum $\tilde Z(T)= T^{3/2} Z(T)$ is then 
exactly the linear statistics with the exponential wall
\be
\tilde Z(T) = {\sf L} = t \sum_i \phi(u + t^{-2/3} a_i)
\quad , \quad  \phi(z)=e^{z} \quad , \quad u=0
\quad , \quad t = T^{3/2}
\ee 
The average partition sum is, from Eq.~\eqref{avexp} (recall that $\beta$ here is not the
inverse temperature, but 
the Dyson index equal to the inverse random potential strength) 
\be 
\mathbb{E}_\beta[\tilde Z(T)]= \mathbb{E}_\beta[{\sf L}] = T^3 \mathbb{E}_\beta[\tilde L] 
\quad , \quad   \mathbb{E}_\beta[\tilde L] = \frac{1}{2 \sqrt{\pi}}
\ee
We have, for typical fluctuations
{
\bea
&& {\sf L} = T^3 \mathbb{E}_\beta[\tilde L] + T^{3/2} \sqrt{\tilde \kappa_2(0)} \, \omega + \dots \\
&& \log Z(T) = - \frac{3}{2} \log T + \log {\sf L} = \log (\frac{T^{3/2}}{2 \sqrt{\pi}})
+ \frac{1}{T^{3/2}} \sqrt{\frac{2}{\beta}} \omega + \mathcal{O}(T^{-3}) 
\eea
} 
where $\omega$ is a unit Gaussian random variable and higher order terms 
are subdominant. Hence we see that the free energy $F = - T \log Z(T)$
has fluctuations of variance of order $ 1/T$ at high temperature. \\

We now wonder about the large deviations and from our study we know that 
\be
 {\sf L} = t^2 \mathbb{E}_\beta[\tilde L]  \ell \quad \text{with probability} \quad e^{- t^2 G(\ell)} 
\ee 
Let us define by analogy with \eqref{equivKPZ} a "height field" 
${\sf H} = \log Z(T)$. Taking the logarithm, we have
\be
{\sf H} = \log Z(T) = \mathbb{E}_\beta[{\sf H}]  + \ln \ell \quad \text{with probability} \quad e^{- t^2 G(\ell)} 
\ee 
hence the PDF of ${\sf H}$ takes the form
\be
P({\sf H}) \sim \exp(- T^3 G( e^{{\sf H} - \mathbb{E}_\beta[{\sf H}]}))  
\ee 
with $\mathbb{E}_\beta[{\sf H}] = \log (\frac{T^{3/2}}{2 \sqrt{\pi}})$. The function $G$ to be used here is the one of the exponential wall, given in \eqref{Gexp},
and the formula is valid for ${\sf H} < \mathbb{E}_\beta[{\sf H}]$. In particular
from Eq.~\eqref{tailH} we find that for ${\sf H}  \to - \infty$
\be
P({\sf H}) \sim \exp(- T^3 |{\sf H}|^3 )
\ee 
i.e. a cubic tail. Since we are looking at the high $T$ limit here, from the
relations described above this should be rather compared with the
"short time" (i.e. small $b$) large deviation form for ${\sf H}_{\rm KPZ}$, which has
instead a $5/2$ exponent (see \cite{KrajLedou2018} and references therein).

\section{Ground state energy of non-interacting fermions in a linear plus random potential}

Consider $N_1$ non-interacting fermions with single particle Hamiltonian ${\cal H}_{SAO}$ and let us consider the ground state energy
\be
E_0(N_1) = \sum_{i=1}^{N_1} \epsilon_i 
\ee 
obtained using the Pauli exclusion principle by filling the $N_1$ lowest energy levels. It follows from our analysis of matching in the previous section that the PDF of
$E_0(N_1)=E_0$ takes the form for $1 \ll N_1 \ll N$
\be
 P(E_0) \sim \exp\left(  - \frac{\beta  N^2}{2} \kappa^2 \hat \Phi\big( \frac{s-s_0(\kappa)}{\kappa^{5/3}} \big) \right) = \exp\left( - \frac{\beta N_1^2}{2}  \hat \Phi \big( \frac{\mathbb{E}_\beta[E_0]- E_0}{4^{1/3} N_1^{5/3}} \big) \right)
\label{PE0} 
\ee
where the function $\hat \Phi(S)$ is given in a parametric form
in \eqref{PhiS}, and its small $S$ expansion was given in \eqref{PhiSsmall}.
In the first line we have used \eqref{probform} (formally since we are now at the edge)
and its small $\kappa$ limit form 
\eqref{scalingsmall}, and then identified $E_0(N_1) \equiv - 4^{1/3} (s-2 \kappa) N^{5/3}$
from \eqref{precise} with $\epsilon_i = - a_i$ to obtain the second line. This formal
limit was shown to be correct in the previous section by introducing the parameter
$t$ and working at fixed $N_1/t$, however $t$ drops from the final formula \eqref{PE0}.
The side studied here corresponds to 
$E_0 > \mathbb{E}_\beta[E_0]$ so the {\it right tail} of the PDF of $E_0$. The average ground state energy which appears in \eqref{PE0} is
\be \label{avE0} 
\mathbb{E}_\beta[E_0] \simeq \frac{3}{5} (\frac{3 \pi}{2})^{2/3} 
N_1^{5/3}
\ee
which is independent of $\beta$, i.e. of the random potential. 
To make contact with our edge problem, we note that it can also be obtained from eliminating $u$ in the system
\bea
&& \frac{N_1}{t} = \int_0^u \rmd b \, \rho_{\rm Ai}(b) \quad , \quad  \rho_{\rm Ai}(b)=\frac{\sqrt{b}}{\pi} \\
&& \frac{E_0(N_1)}{t^{5/3}} =  \int_0^u \rmd b \,b\rho_{\rm Ai}(b)
\eea 
and one checks that $t$ drops out. 
The large deviations for $E_0$ then corresponds to replacing $\rho_{\rm Ai}(b)$
by some optimal density, leading to some optimal $u^*$, as explained
in the previous Section.\\

It is interesting to indicate the tail of the PDF of the ground state energy  for large positive $E_0$, more precisely for 
$(E_0-\mathbb{E}_\beta[E_0])/N_1^{3/5} \gg 1$. Using \eqref{PhiSlarge} we obtain a
cubic far tail 
\be
 P(E_0)  \sim \exp\left( - \frac{\beta}{2}  \frac{(E_0-\mathbb{E}_\beta[E_0])^3}{12 N_1^3} \right)
\label{PE0large} 
\ee

We can also obtain the Laplace transform of the PDF of $E_0(N_1)$.
Using the matching detailed in the previous Section, and the above arguments, we identify
\be
\langle e^{- \frac{\beta N^2}{2}  \mu_1 (s-2 \kappa)} \rangle_{\mu_1,\kappa}  \; 
\Longleftrightarrow \; \mathbb{E}_\beta[ e^{ \frac{\beta}{2} 4^{-1/3} \mu N_1^{1/3} E_0(N_1)} ]
\ee
where $\mu=\mu_1/\kappa^{1/3}$,
which leads to the Laplace transform 
\be
\mathbb{E}_\beta[ e^{- \frac{\beta}{2} q N_1^{1/3} E_0(N_1)} ] \sim \exp\left(-  \frac{\beta N_1^2}{2}  \hat H(-4^{1/3}q) \right)
\ee
where the function $\hat H$ was obtained in \eqref{reshatH}. From it
one obtains the cumulants of $E_0(N_1)$. It reproduces the average 
\eqref{avE0} and gives a variance
\be
{\rm Var}_\beta [ E_0(N_1) ] \simeq \frac{2}{\beta} (\frac{3}{2 \sqrt{\pi}})^{4/3} N_1^{4/3} 
\ee 
which is proportional to the variance of the random potential term
in the SAO Hamiltonian ${\cal H}_{\rm SAO}$. 

If we now consider instead the Hamiltonian ${\cal H}$ in \eqref{calH} we obtain 
\bea
&& \mathbb{E}_\beta[E_0(N_1)] \simeq \frac{3}{5} (\frac{3 \pi}{2})^{2/3} h^{2/3} D^{1/3}
N_1^{5/3} \\
&& {\rm Var}_\beta [ E_0(N_1) ] \simeq \frac{g}{2}(\frac{h}{D})^{1/3}  (\frac{3}{2 \sqrt{\pi}})^{4/3} N_1^{4/3} 
\eea
The limit $h \to 0$ ($\beta \to 0$) is of particular interest as the problem becomes the
usual Anderson model for localization in 1D (up to the hard wall at $z=0$). It is
known that the bottom of the spectrum of the SAO in that limit becomes Poisson distributed
\cite{AllezDumaz2013,DumazLabbe2017}. Hence it would be interesting to
compare this limit to the results obtained in \cite{HartmannMajumdar2018}
for the same problem with independent
random energy levels chosen with a PDF $p(\epsilon) \sim \epsilon^\alpha$.
For instance
choosing $\alpha=1/2$ leads to the scaling 
$\mathbb{E}[E_0(N_1)] \sim N^{-2/3} N_1^{5/3}$. Clearly small $h$ provides
a cutoff scale (i.e. an effective system size) but the precise study 
of this limit is left for the future.

Finally, one could study the same problem in the grand canonical ensemble
at fixed chemical potential $\mu$. At $T=0$ the mean energy 
$E_0 = \sum_i \epsilon_i \theta(\mu-\epsilon_i)$ and the mean number
$N_1= \sum_i \theta(\mu-\epsilon_i)$ are both fluctuating. Note however that the
fluctuations of the $T=0$ grand potential 
\be
\Omega = E_0 - \mu N  = - \sum_i (\mu - \epsilon_i) \theta(\mu-\epsilon_i)
\ee 
is readily obtained, in the large $\mu= t^{2/3} u$ limit,
as $- t^{1/3} \Omega \equiv {\sf L}$ from our results on the linear monomial wall $\phi(z)=(z)_+$. Its
finite temperature fluctuations are similarly described by a mixed linear-exponential
wall $\phi(z)=\frac{z}{1+ e^{-z/T}}$.

\section{Fermions with $1/r^2$ interaction in an harmonic trap  }
Consider the Calogero model \cite{Calogero1969,forrester2010log} for 
$N$ spinless fermions in a one dimensional harmonic trap, with a $1/r^2$ mutual interaction,
described by the Hamiltonian 
\bea
H=H_{N,\omega,\beta,\{x_i\}} = - \frac{1}{2} \sum_{i=1}^N \partial^2_{x_i} + \sum_{1 \leq i < j \leq N}
\frac{g}{(x_i-x_j)^2} + \frac{\omega^2}{2}  \sum_{i=1}^N x_i^2
\eea 
One must have $g > -1/4$ to avoid the fall to the center. Parameterizing the coupling constant \cite{Calogero1969}
\be
g=\frac{\beta}{2} (\frac{\beta}{2}-1)
\ee
for $\beta>1$, the ground state wave function reads, in the sector $x_1>x_2> \dots >x_N$,
\be
\Psi_0(x_1,\dots,x_N) = A_N \prod_{1 \leq i < j \leq N} |x_i-x_j|^{\beta/2} e^{- \frac{\omega}{2} \sum_{i=1}^N x_i^2}
\ee
and its value in the other sectors has the same expression up to a sign determined by 
the antisymmetry of the wavefunction. Here $A_N$ is a normalizing constant.\\

The JPDF $P[\lambda]$ of the eigenvalues of the Gaussian $\beta$ ensemble,
given by Eq. (1) in the Letter, is thus the quantum probability 
$|\Psi_0(\lambda_1,\dots,\lambda_N)|^2$ of a Hamiltonian $H_{N,\omega_N,\beta,\{\lambda_i\}}$
with $\omega_N=\frac{\beta N}{4}$. If we define a fermion problem with
$x_i = \lambda_i \sqrt{N/2}$ it will be described by the Hamiltonian 
$\frac{N}{2} H_{N, \frac{\beta}{2},\beta,\{x_i\}}$. For $\beta=2$ this is the problem 
for non-interacting fermions studied in \cite{dean2015finite,dean2016noninteracting} and for general $\beta>1$, $\beta\neq 2$ there is an interaction. In all cases the density of the 
fermions is the semi-circle with support $[-\sqrt{2 N}, \sqrt{2 N}]$.\\

The results of the present Letter thus apply to describe linear statistics at the edge of the Fermi gas
in the ground state. If one considers the rescaled positions 
$\xi_i = (x_i - \sqrt{2 N})/w_N$ with $w_N= N^{-1/6}/\sqrt{2}$ the width of
the edge regime, these for large $N$ behave jointly as the Airy$_\beta$ process
$\xi_i \equiv a_i$. One example is the fluctuations of the center of mass position 
of the $N_1$ right most fermions
\be
X(N_1) = \frac{1}{N_1} \sum_{i=1}^{N_1} \xi_i 
\ee 
Since it identifies with the ground state energy of $N_1$ fermions in the
SAO Hamiltonian, via $X(N_1) = - E_0(N_1)/N_1$, its PDF is 
given by Eq.~\eqref{PE0}.

\section{Non-intersecting Brownian interfaces subject to a needle potential}

The results presented in the Letter additionally apply to non-intersecting Brownian interfaces representing elastic domain walls between different surface phases adsorbed on a crystalline substrate and perturbed
by a soft, needle like potential. These provide
a natural classical statistical mechanics analog of the trapped fermions studied in previous Sections.
Here we will heavily borrow from the very elegant presentation given in Refs.~\cite{ grabsch2017truncated, nadal2009nonintersecting}. There is a related extensive work on the fluctuations of vicinal
surfaces, including experiments, and we refer to \cite{EinsteinVicinal2003} for an introduction.\\

Consider $N$ non-intersecting ordered interfaces at heights $h_1(x)<\dots<h_N(x)$ that live around a cylinder of radius $L/(2\pi)$, they can be thought as random walkers with periodic boundary condition. Add a hard wall at $h=0$ (so that $h_i(x)>0$ for all $i$) induced some effective potential for each interface and consider the large system limit, i.e. $L\to \infty$, where the interfaces reach equilibrium.\\

\begin{figure}[h!]
\centerline{\includegraphics[scale=0.75]{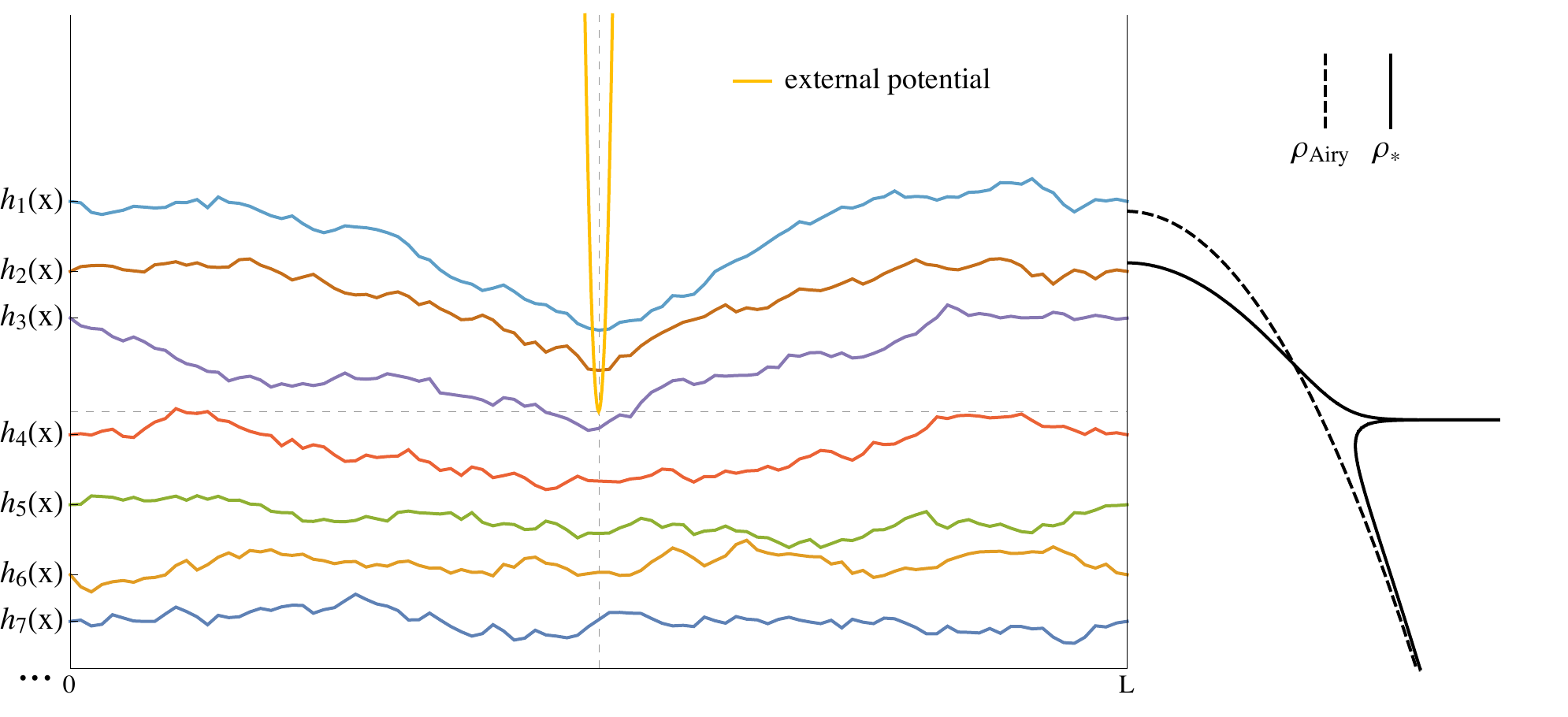}}
\caption{ Representation of the seven top Brownian lines subject to the needle external potential. In absence of the potential, the density of the top lines as a function of the depth is described by the edge of the semi-circle $\rho_{\rm Airy}$ (dashed lined on the right) and in presence of a smooth potential, the reorganization of the interfaces imposes a new optimal density $\rho_*$ (black line on the right).}
\label{brownian}
\end{figure}

We introduce four contributions to the energy of the interfaces :
\begin{enumerate}
\item An elastic energy $E_{\rm elastic}(h)=\frac{1}{2}(\frac{\rmd h}{\rmd x})^2$,
\item A confining energy $V(h)=\frac{b^2h^2}{2}+\frac{\alpha(\alpha-1)}{2h^2}$ with $b>0$ and $\alpha>1$,
\item A pairwise interaction between interfaces $V_{\rm pair}(h_i,h_j)=\frac{\beta}{2}(\frac{\beta}{2}-1)\left[ \frac{1}{(h_i+h_j)^2}+\frac{1}{(h_i-h_j)^2} \right]$ with $\beta>0$,
\item An external \textit{needle} soft potential probing the interfaces at the position $x=0$ on the cylinder, $V_{\rm needle}(h,x,U)=\delta(x)W(h(x)-U)$ (see Fig. \ref{brownian}). The parameter $U>0$ controls the depth of the probe and the exact form of $W$ controls the type of measurement on the interfaces. The $\delta$ function indicates that the probe is 
sufficiently local in space. It could be realized in practice as an STM tip. 
\end{enumerate}

The choice of the confining energy comes from the fact that confinement is necessary not to have a zero mode, so for simplicity we consider a quadratic one, plus a repulsive inverse square potential natural from entropic considerations as shown by Fisher in Ref.~\cite{fisher1984walks}. By a path integral calculation, it was proved in Ref.~\cite{nadal2009nonintersecting} that the equilibrium joint distribution of heights at a fixed space point can be obtained from the spectral properties of the quantum Calogero-Moser Hamiltonian.
However, concerning the edge properties that we are probing here, these are not important
details. From the universality of the soft edge, a purely quadratic confining potential with no hard wall at $h=0$, as considered in \cite{EinsteinVicinal2003}, would
do as well. 
\\

Indeed, at equilibrium, the probability to observe a particular realization of $N$ lines is given by the Boltzmann weight of the problem (in units where temperature is unity)
\begin{equation}
P\left[\lbrace h_i(x) \rbrace_{i\in [1,N]}\right] \propto \exp \left(-\sum_{i=1}^N E \left[  h_i(x)\right] -\sum_{1\leq i<j \leq N} V_{\rm pair}(h_i,h_j)\right) \mathds{1}_{h_i(x)>0}
\end{equation}
with $E[ h]=E_{\rm elastic}(h)+V(h)+V_{\rm needle}(h,x,u)$.
The joint probability to see interfaces at positions $\lbrace h_1,\dots,h_N\rbrace$ at $x=0$ and $x=L$  (because of the periodic boundary condition) is given by the path integral
\begin{equation}
P(h)\propto \prod_{i=1}^N \int_{h_i(0)=h_i}^{h_i(L)=h_i} \mathcal{D}h_i(x) \mathds{1}_{h_i (x)>0} \; e^{- E\left[  h_i (x) \right]} \prod_{1\leq i <j \leq N} e^{- V_{\rm pair}(h_i,h_j)}
\end{equation}
which in turn can be seen as a propagator of $N$ quantum particles
\begin{equation}
P(h_1,\dots, h_N)\propto  e^{-\sum_{i=1}^N W(h_i-U)} \bra{h_1,\dots, h_N} e^{-L\mathcal{H}_{\rm interface}}\ket{h_1,\dots , h_N}
\end{equation}
subject to the many-body Hamiltonian 
\begin{equation}
\mathcal{H}_{\rm interface}=\sum_{i=1}^N \left[-\frac{1}{2}\frac{\rmd^2}{\rmd h_i^2}+V(h_i)\right]+\sum_{1\leq i <j \leq N} V_{\rm pair}(h_i,h_j)
\end{equation}
In the large $L$ limit, the marginal PDF is given by the $N$-body ground state of $\mathcal{H}_{\rm interface}$ which is exactly the Calogero-Moser model \cite{moser1976three}. As the brownian interfaces are non-intersecting, the corresponding quantum particles are fermionic and the ground state is formed by filing the first $N$ eigenstates of the Hamiltonian and given by the Slater determinant of the first $N$ eigenfunctions $\lbrace{\psi_i \rbrace}_{i\in \mathbb{N}}$. This determinant was computed \cite{nadal2009nonintersecting} using exacts results on the Calogero-Moser Hamiltonian eigenstates.
\begin{equation}
\begin{split}
P(h_1,\dots,h_N)&\propto \prod_{i=1}^N  e^{ -W(h_i-U)}\abs{\det[\psi_i(h_j)]_{i,j\in [1,N]}}^2\\
&\propto  \prod_{i=1}^N  h_i^{2\alpha} e^{ -W(h_i-U)-bh_i^2} \prod_{1\leq i<j\leq N} (h_i^2-h_j^2)^\beta\\
\end{split}
\end{equation}
After the change of variable $bh_i^2=\lambda_i$, this PDF corresponds to the general Wishart ensemble with arbitrary $\beta\geq 0$ and an external potential $W$. In the large $N$ limit and in the absence of the potential $W$, the arrangement of the top brownian lines is described by the soft edge of the Marcenko-Pastur distribution around $\lambda\sim 4N$ or equivalently $h\sim 2\sqrt{N}$.\\

The results of the Letter readily apply to describe the linear statistics of the top non-intersecting Brownian interfaces in the ground state in a region of width $N^{-1/6}$ around the top line located at a height $\sim 2\sqrt{N}$. Indeed, if one considers the rescaled heights
$\tilde{h}_i = (\sqrt{b} h_i  - 2\sqrt{N})4^{1/3}N^{1/6}$, these behave for large $N$ jointly as the Airy$_\beta$ process $\tilde{h}_i \equiv a_i$. One observable studied in 
\cite{grabsch2017truncated} in the bulk is the center of mass position of the $N_1$ top interfaces
$H(N_1) = \frac{1}{N_1} \sum_{i=1}^{N_1} h_i$. As we see, at the edge, in the absence of a
potential $W$ it is distributed (up to a scale factor) as the variable $L_{N_1}$ in Eq. (44) of the Letter.
In presence of the needle potential $W$, parameters can be adjusted so that the
soft potential $W$ translate into the soft potential $\phi$ in our units, using
the correspondence $W(h_i-U) \equiv t \phi(u+ t^{-2/3} a_i)$ with 
$\sqrt{b} h_i \equiv 2 \sqrt{N} + 4^{-1/3} N^{-1/6} a_i$. A practical
way to measure the value of $u$ is to measure the position of the center of mass $H(N_1)$,
from which we can determine the optimal density of the first $N_1$ brownian lines yielding this specific position. Finally, we represent in Fig.~\ref{brownian} the top interfaces (at a distance of order $N^{-1/6}$ to the first line) subject to an external potential and the optimal density for the first top lines.

\section{Appendix: Mellin-Barnes summation  }
Here we perform the summation of the series which appears in \eqref{3der}. 
We use a Mellin-Barnes summation method inspired from Lemma 6 of Ref. \cite{imamura2013stationary} which was introduced to calculate the sum over replicas in the context of the KPZ equation. For sufficiently nice real test functions $f$, 
assumed to be positive, the following series admits a closed algebraic form 
\begin{equation}
\label{eq:mellin}
\mathcal{S}(u)=\sum_{n \geq 1} \frac{a^n}{n!} (\partial_u)^n f(u)^n=\sum_i \frac{1}{\abs{a f'(u+aw_i)-1}}  -1
\end{equation}
where the $\lbrace w_i \rbrace$'s are the positive solutions of the equation $f(u+aw)=w$.
We use this formula in the Letter only in the case of a unique solution. The present Mellin-Barnes method proposes a formula in the case of multiple solutions. Testing
that formula for the present problem is work in progress, we will not use it here.

\begin{proof}
Let us start by manipulating the summand
\begin{equation}
 \frac{a^n}{n!} (\partial_u)^n f(u)^n=\int_{\mathbb{R}}\mathrm{d}y \, \delta(y) \frac{a^n}{n!} (\partial_u)^n f(u)^{n+iy}
\end{equation}
Let us express the delta in Fourier space and proceed to the change of variable $z=n+iy$,
\begin{equation}
 \frac{a^n}{n!} (\partial_u)^n f(u)^n=\int_{n+i\mathbb{R}}\mathrm{d}z\int_{\mathbb{R}}\frac{\mathrm{d}r}{2i\pi} \, e^{-r(z-n)}\frac{a^n}{n!} (\partial_u)^n f(u)^{z}
\end{equation}
Let us suppose that we can shift the contour of integration of $z$ such that there is no $n$ dependency anymore. Let us call $\Gamma$ the new shifted contour. 
\begin{equation}
 \frac{a^n}{n!} (\partial_u)^n f(u)^n=\int_{\Gamma}\frac{\mathrm{d}z}{2i \pi}\int_{\mathbb{R}}\mathrm{d}r \, e^{-r(z-n)} \frac{a^n}{n!} (\partial_u)^n f(u)^{z}
\end{equation}
Let us choose the contour $\Gamma=a+i\mathbb{R}$ for some $a\in\, ]0,1[$ so that $\Gamma$ is parallel to the imaginary axis and let us proceed to the summation over $n$.
\begin{equation}
\sum_{n \geq 1} \frac{a^n}{n!} (\partial_u)^n f(u)^n=\int_\Gamma \frac{\mathrm{d}z}{2i\pi}\int_\mathbb{R}\mathrm{d}r \, e^{-rz} \left(\sum_{n \geq 1} \frac{e^{rn} a^n}{n!} (\partial_u)^n  \right) f(u)^z
\end{equation}
One recognizes an exponential series, and more particularly, the series of a translation operator.
\begin{equation}
\begin{split}
\sum_{n \geq 1} \frac{a^n}{n!} (\partial_u)^n f(u)^n&=\int_\Gamma \frac{\mathrm{d}z}{2i\pi}\int_\mathbb{R}\mathrm{d}r \, e^{-rz} \left[e^{ae^r \partial_u}-1 \right] f(u)^z\\
&=\int_\Gamma \frac{\mathrm{d}z}{2i\pi}\int_\mathbb{R}\mathrm{d}r \, e^{-rz} \left[f(u+ae^r)^z -f(u)^z \right] 
\end{split}
\end{equation}
As $\Gamma$ is parallel to the imaginary axis and as both $r$ and $f$ are real valued, one recognizes the integral over $z$ as Fourier transform and we therefore have
\begin{equation}
\begin{split}
\sum_{n \geq 1} \frac{a^n}{n!} (\partial_u)^n f(u)^n&=\int_\mathbb{R}\mathrm{d}r \,\left[\delta(\log f(u+ae^r)-r)  -\delta(r)\right]\\
&=\sum_i \frac{1}{\abs{a f'(u+ae^{r_i})-1}}  -1\\
\end{split}
\end{equation}
where $r_i$ are the real solutions of the equation $ f(u+ae^r)=e^r$. As $r$ is real, we define $w=e^r>0$ which concludes the derivation. 
\end{proof}
Furthermore suppose now, as in the Letter, that there exists a unique real solution $w=w(u)$ to the equation $f(u+aw)=w$ and that $1-af'(u+aw)>0$ for this solution. It is possible to further simplify the series. Indeed, differentiating the equation $f(u+aw)=w$ leads to
\begin{equation}
\left[ 1-af'(u+aw)\right]\mathrm{d}w=f'(u+aw)\mathrm{d}u
\end{equation}
Inserting this differential relation into Eq.~\eqref{eq:mellin} yields
\begin{equation}
\label{eq:Su}
\mathcal{S}(u)=\sum_{n \geq 1} \frac{a^n}{n!} (\partial_u)^n f(u)^n=\frac{a f'(u+aw)}{1-a f'(u+aw)} =a  \frac{\mathrm{d}w}{\mathrm{d}u}
\end{equation}